\documentclass[aps,prx,twocolumn,floatfix,longbibliography,nofootinbib,tbtags,10pt]{revtex4-2}
\usepackage[utf8]{inputenc}
\usepackage[english]{babel} 

\usepackage{siunitx}
\usepackage{natbib}
\usepackage{graphicx} 
\usepackage{xcolor}
\usepackage{mathtools}
\usepackage{amsmath}
\usepackage[colorlinks, linkcolor=blue]{hyperref}
\hypersetup{colorlinks,allcolors=black}
\usepackage{amssymb}
\usepackage{gensymb}
\usepackage{float}
\usepackage{tabularx}
\usepackage{epstopdf}
\usepackage{latexsym}
\usepackage{color, colortbl}
\usepackage{psfrag}
\usepackage{bbm}
\usepackage{bm}
\usepackage{titlesec}
\usepackage{dsfont}
\usepackage{feynmp}
\usepackage{slashed}
\usepackage{multirow}
\usepackage{comment}
\usepackage[normalem]{ulem}
\renewcommand{\vec}[1]{\boldsymbol{#1}}
\def \nn {\nonumber}

\def \k {{\vec k}}

\def \ve {\varepsilon}
\def \r {{\vec r}}

\def \q {{\vec q}}

\def \l {\ell}
\def \ve {\varepsilon}

\def \l {\ell}

\def \beq {\begin{eqnarray}}
\def \eeq {\end{eqnarray}}
\def \hi {H_{\tn{int}}}
\def \hk {H_{\tn{kin}}}
\def \tn {\textnormal}

\def \sxx {\sigma_{\rm{xx}}}

\def \hk {H_{\tn{kin}}}
\def \hi {H_{\tn{int}}}

\def \l {\ell}

\def \heff {\mathcal{H}_{\tn{eff}}}

\begin{document}
\title{Low-energy optical absorption in correlated insulators: \\Projected sum
    rules and the role of quantum geometry}
\author{Dan Mao}\thanks{Current address: Department of Physics, University of Z\"urich, 8057 Z\"urich, Switzerland}
\author{Juan Felipe Mendez-Valderrama}\thanks{Current address: Department of Physics, Princeton University, Princeton, New Jersey 08544, USA}
\author{Debanjan Chowdhury}\email{debanjanchowdhury@cornell.edu}
\affiliation{Department of Physics, Cornell University, Ithaca, New York 14853, USA.}

\begin{abstract}
Inspired by the discovery of a variety of correlated insulators in the moir\'e universe, controlled by interactions projected to a set of isolated bands with a narrow bandwidth, we examine here a partial sum-rule associated with the inverse frequency-weighted optical conductivity restricted to low-energies. Unlike standard sum-rules that extend out to {\it infinite} frequencies, which include contributions from {\it all} inter-band transitions, we focus here on transitions associated {\it only} with the {\it projected} degrees of freedom. We analyze the partial sum-rule in a non-perturbative but ``solvable" limit for a variety of correlation-induced insulators. This includes (i) magic-angle twisted bilayer graphene at integer-filling with projected Coulomb interactions, starting from the chiral flat-band limit and including realistic perturbations, (ii) fractional fillings of Chern-bands which support generalized Laughlin-like states, starting from a Landau-level and including a periodic potential and magnetic-field, respectively, drawing connections to twisted MoTe$_2$, and (iii) integer filling in toy-models of non-topological flat-bands with a tunable quantum geometry in the presence of repulsive interactions. The partial sum-rule in all of these examples is implicitly constrained by the form of the band quantum geometry via the low-lying excitation spectrum, but is not related to it explicitly. For interacting Slater-determinant insulators, the partial sum-rule is related to a new quantity --- ``many-body projected quantum geometry" --- obtained from the interaction-renormalized electronic bands. We also point out an intriguing connection between the partial sum-rule and the quantum Fisher information associated with the projected many-body position operator. Finding new experimental routes that enable a direct measurement of the quantum geometry in correlated insulators remains an exciting and uncharted territory. 
\end{abstract}

\maketitle

\section{Introduction}
  {Electronic solids can display insulating behavior due to a variety of non-trivial quantum mechanical reasons, that range from robust band-filling constraints \cite{martin2020electronic}, strong interactions at commensurate lattice fillings \cite{revMIT}, large disorder \cite{LeeRama}, and a non-trivial band-topology \cite{revTI}.} Regardless of the underlying mechanism, a universal feature shared by all insulators is their vanishing longitudinal conductivity at zero temperature in the limit of a vanishing frequency (i.e. the Drude weight) \cite{kohn,scalapino}:
\beq
\tn{Re}[\sxx(\omega\rightarrow0)]=0,~~\tn{at}~T=0.
\label{ins:def}
\eeq  
In generic quantum many-body systems, computing the detailed frequency dependence of $\tn{Re}[\sxx(\omega)]$ can be a priori a non-trivial task. However, in an electrical insulator with a set of fully filled valence bands, the ultraviolet (UV) dynamical response extending out to $\omega\rightarrow\infty$ satisfies an elegant sum-rule:
\beq
\mathbb{O} \equiv \int_0^\infty d\omega \frac{\delta_{\mu\nu}\tn{Re}[\sigma_{\mu\nu}^L(\omega)]}{\omega} = \frac{\pi e^2}{\hbar} \delta_{\mu\nu} \tn{Tr}[g_{\mu\nu}],
\label{eq:SWM}
\eeq
where $L$ denotes the longitudinal response, respectively. This is the classic result by Souza-Wilkens-Martin (SWM) \cite{SWM}, described in terms of the trace over the occupied valence states and their quantum geometry, $g_{\mu\nu}$, respectively. Quantum geometry \cite{torma}, while being an old concept \cite{provost}, has found a resurgence in the field especially with the advent of moir\'e materials \cite{QGreview}. Specifically, it is now well established that quantum geometry and the so called ``ideal-droplet" condition plays an important role in stabilizing Laughlin-like states at fractional fillings of Chern bands \cite{Roy14,crphys,ledwith2020fractional,andrews24}. In terms of establishing a theoretical connection to experiments, a number of works have investigated the relationship between the quantum geometric properties of the electronic Bloch wavefunctions and their (non-)linear optical response \cite{ahn20,holder20,Ma2021,sentef21,annrev21,Moore21,Ahn2022,CL22,holder22,MC23,deshmukh,bouhon2023quantum, jankowski2023optical, Bradlyn24a,bradlyn24b,jankowski2024quantized, antebi2024drude, avdoshkin2024multi}. Connections between quantum geometry and entanglement have also been pointed out via different means \cite{NP24,PMT,XCW,SR24}.

Various complementary aspects associated with the above sum-rule, including its relationship to polarization fluctuations in insulators have been analyzed previously \cite{RestaSorella,Resta}; an earlier version of a related sum-rule also appeared in Ref.~\cite{Kudinov}. Recent works have used the above bounds in combination with the analogous $f-$sum-rule to put bounds on the insulating gaps and related quantities \cite{OnishiFu24a,souza24,feldman}, and also highlighted a complementary interpretation of these sum-rules in the time-domain \cite{VQ24a,VQ24b}. The above integrated dynamical response is related to a specific coefficient associated with the long-wavelength $(\q\rightarrow0)$ limit of the static momentum-dependent density structure-factor, $\mathcal{S}(\q) = Kq^2 + ...$, where
\beq
\mathbb{O} = \frac{e^2}{2\hbar} K,
\label{eq:weight}
\eeq
and $K$ is denoted the ``quantum weight" \cite{OnishiFu24b}. Charge continuity associated with the globally conserved $U(1)$ density and the insulating property of the ground-state ensures the relationship in Eq.~\ref{eq:weight}. 
\begin{figure}[pth!]
\centering
\includegraphics[width=\linewidth]{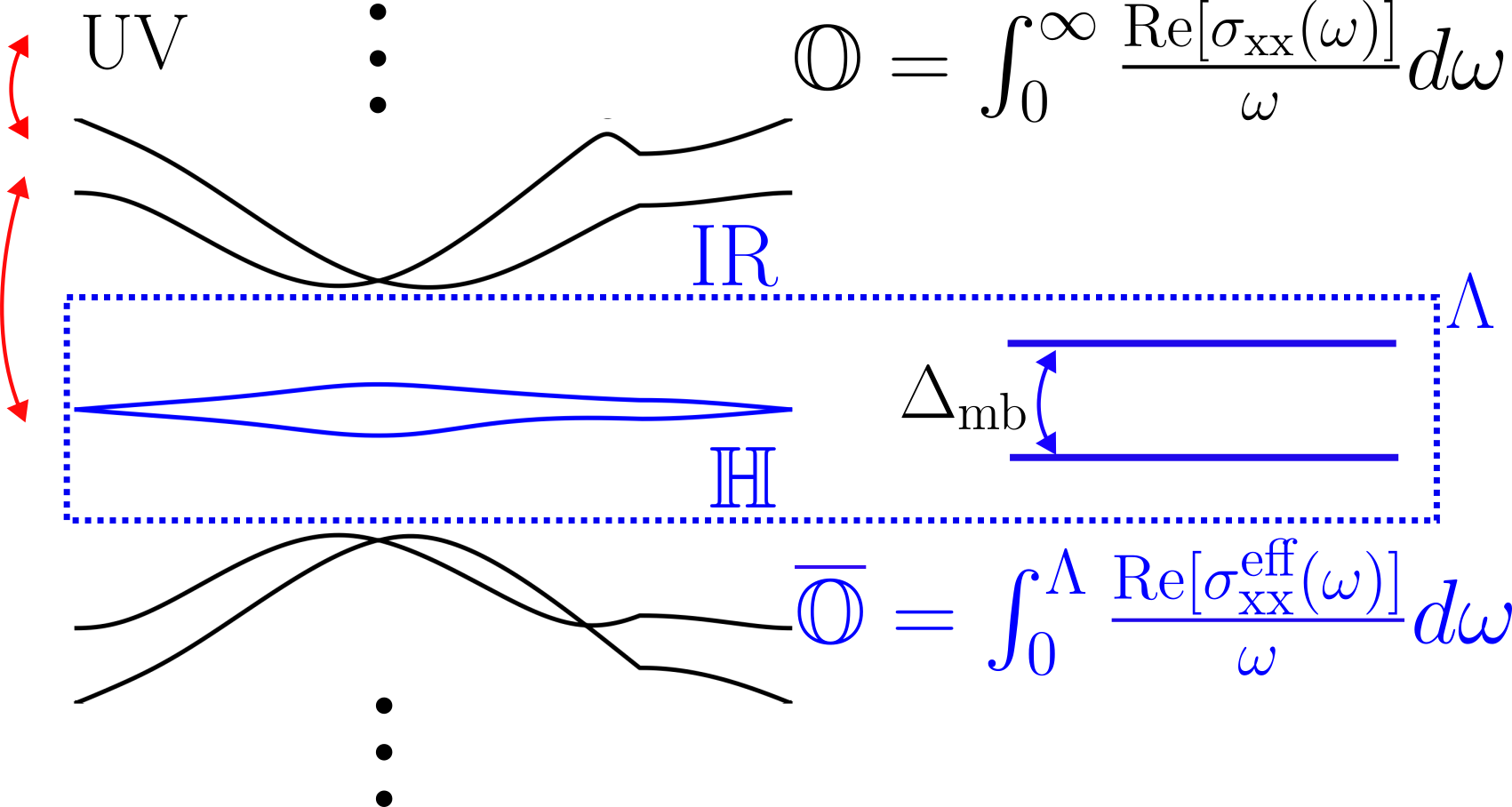} 
\caption{The non-interacting continuum model dispersion of magic-angle twisted bilayer graphene (MATBG) near $\theta=1.08^0$. The UV sum-rule, $\mathbb{O}$ in Eq.~\ref{eq:SWM} includes all of the inter-band transitions (red arrows). The low energy sum-rule, $\overline{\mathbb{O}}$ in Eq.~\ref{PSR} only includes the transitions within the projected Hilbert space, $\mathbb{H}$, restricted to the energy window $<\Lambda$ (marked in blue). We will compute $\mathbb{O}$ for correlation induced insulators at integer filling in MATBG, in solvable models of fractional Chern insulators, and in various insulating states associated with a trivial flat-band but tunable quantum metric.} 
\label{fig:model}
\end{figure}

 The UV sum-rule in Eq.~\ref{eq:SWM} extends out to $\omega\rightarrow\infty$, and includes all interband transitions (red arrows in Fig.~\ref{fig:model}). From an experimental point of view, it is impossible to obtain the weight in $\mathbb{O}$ directly without further assumptions about the high-frequency tails  of $\tn{Re}[\sigma^L(\omega)]$, and (possibly) uncontrolled extrapolations. From a purely theoretical perspective of modeling {\it correlated} electronic solids, it is rare to have a description that accurately captures the low-energy physics of interest, as well as the high-energy UV degrees of freedom extending out to infinity. Instead, given an interacting effective theory, $\heff(\Lambda)$, below some physically motivated cutoff, $\Lambda$, we are often interested in the low-energy response associated with its ground-state and low-lying excited states. For instance, in the context of magic-angle twisted bilayer graphene (MATBG) \cite{Bistritzer2011}, the relevant effective theory is typically described in terms of the interactions projected to only a subset of the low-energy bands (Fig.~\ref{fig:model}). Since the quantum metric controls the minimal spread of the maximally localizable Wannier wavefunctions \cite{MarzariVanderbilt}, it is natural to ask the extent to which $\mathbb{O}$ is related to the notion of a possibly interaction-renormalized quantum geometry tied to $\heff(\Lambda)$.   {Our goal here will be to instead analyze {\it only} the low-energy optical absorption spectral weight $\overline{\mathbb{O}}$ associated with the IR transitions (blue arrows in Fig.~\ref{fig:model}) in solvable, strongly correlated examples of many-body insulators in the isolated ``flat-band" setting, and make connections to the underlying quantum geometry. \footnote{Here the UV versus IR is formulated in terms of the frequency cutoff ($\Lambda$) in the integrated optical spectral weight, to be clarified below, and not to be confused with the UV versus IR distinction in terms of the optical transition discussed in the band electron context \cite{Barun}, where UV deals with dipole transition due to the spreading of the Wannier orbitals and IR deals with the contribution from electron hopping.}} 

  {
Our main results can be summarized as follows: We have expressed the projected sum-rule in terms of quantum many-body operators acting only within the low-energy Hilbert space. We find that unlike the full SWM sum-rule, there is no universal relationship between the projected sum-rule and the band quantum geometrical quantities. In the case where the correlated insulating state can be viewed as a Slater-determinant state, obtained e.g. using Hartree-Fock calculations, the projected sum-rule is related to a ``many-body projected quantum geometry" (MBPQG) associated with the Bloch wavefunction of the {\it interaction-renormalized} bands. For fractional Chern insulators that are similar to fractional quantum Hall insulators in the lowest Landau level, the projected sum-rule is controlled by the fluctuations in the quantum metric around the uniformly distributed value in momentum space, as opposed to the quantum metric itself. The drastic difference between the full sum-rule and projected one indicates that the low-energy optical absorption can fail to directly reveal the band Bloch geometry, but is controlled by geometry within the projected band manifold instead, which may or may not admit a single-particle description.
}

The remainder of this manuscript is organized as follows: In Sec.~\ref{sec:PSR} we introduce the central quantity of interest, the projected sum-rule, from a variety of complementary perspectives. In Sec.~\ref{sec:examples} we discuss results for the projected optical absorption sum-rule for distinct examples of interaction-induced insulators in (non-)topological flat-band problems. Specifically, in Sec.~\ref{sec:TBG}, starting from the chiral flat-band limit of magic-angle twisted bilayer graphene with projected Coulomb interactions, where the spectral weight vanishes, we discuss the effects of a heterostrain-induced band renormalization on the sum-rule. We place these results in the context of the interaction-renormalized quantum geometry of the isolated bands. In Sec.~\ref{sec:FCI}, we turn to an evaluation of the projected sum-rule in generalized Laughlin-like states obtained at partial fillings of Landau-level in a periodic potential or periodic magnetic-field, respectively, and comment on their relationship to the fluctuations in quantum geometry. We make direct connection to the fractional quantum anomalous Hall states in minimal models of twisted MoTe$_2$. Finally in Sec.~\ref{sec:chiral}, we turn to the evaluation of the low-energy sum-rule in trivial symmetry-broken many-body insulating states obtained in topologically trivial flat-bands with a tunable quantum metric. We end with an outlook towards other future directions in Sec.~\ref{sec:outlook}. The appendices include a number of technical details.

\section{``Projected" sum-rule}
\label{sec:PSR}
This manuscript will be concerned with {\it only} the ``low-energy" contribution to the optical spectral weight, associated with many-body insulating phases that satisfy Eq.~\ref{ins:def}. Consider for instance the schematic Fig.~\ref{fig:model}, where the focus is on the low-energy subspace denoted $\mathbb{H}$ (shown in blue), that is separated from the high-energy (``remote") subspace by a bandgap. The effective theory defined in $\mathbb{H}$ drives the interaction-induced insulating phase, with the fermi-energy ($\ve_F$) placed in the many-body gap ($\Delta_{\rm{mb}}$). Let us consider the optical absorption spectral weight in the {\it projected} subspace ($\in\mathbb{H}$),
\beq
{\overline{\mathbb{O}}}\equiv\int_0^\Lambda ~{\cal{I}}(\omega)~d\omega,~\tn{where}~~{\cal{I}}(\omega)=\frac{\tn{Re}[\sigma_{\rm{xx}}^{\tn{eff}}(\omega)]}{\omega},
\label{PSR}
\eeq
where $\Lambda$ is the UV cutoff associated with $\mathbb{H}$ and $\sxx^{\tn{eff}}$ represents the longitudinal conductivity obtained from $\heff$. Our modified sum-rule of interest includes all of the inter-band transitions within this low-energy subspace, but not to the high-energy subspace. The central question addressed in this manuscript can also be rephrased differently as follows: {\it What fraction of the total absorption spectral weight, $\mathbb{O}$, is contained in the low-energy theory, $\heff(\Lambda)$, below an energy scale, $\Lambda$?} 

Given that $\mathbb{O}$ is dominated by an integrand that is large at small $\omega$, one might expect the absorption spectral weight is dominated by the low-energy contribution, with most of it contained in $\overline{\mathbb{O}}$. It is with this perspective that we revisit the many-body sum-rule for generic insulating ground-states of $\heff$, restricted to an energy window up to the scale $\Lambda$. In this manuscript, we will examine and comment on the low-energy optical absorption spectral weight for a wide class of correlated insulators: (i) flavor symmetry-broken insulators in MATBG at and away from the chiral flat-band limit at integer fillings, (ii) fractional Chern insulators (FCI) due to Coulomb interactions projected to a set of topological vortexable bands inspired by the recent observation of such states in twisted MoTe$_2$, and (iii) an interaction-induced charge-density wave insulator in topologically trivial flat-bands with a tunable quantum metric. Interestingly, we will demonstrate that a variety of insulating ground states with a non-vanishing 
quantum geometry can remain optically dark when restricting the sum-rule to only low frequencies.

As a result of the projection to the low-energy subspace, it is important to realize that $\sxx^{\tn{eff}}$ is not necessarily due only to a ``free-fermion" current. Instead it can include many-body contributions, as is the case for models projected to Landau-levels (LL) or low-energy isolated bands of moir\'e materials, which we discuss below. We will  focus only on many-body phases of matter that satisfy Eq.~\ref{ins:def} for the effective conductivity. In the Lehmann representation with a many-body energy eigenbasis $\{E_n,~|n\rangle \}$, we can rewrite (see Appendix~\ref{sec:Kubo})
\beq
{\overline{\mathbb{O}}} &=& \pi e^2 \sum_{m,n\in \mathbb{H}} (p_n-p_m) ~ |\langle n|\hat{\overline{X}}|m\rangle|^2 ~\theta(E_m-E_n),\label{eq:proj}
\eeq
where $p_n=e^{-\beta E_n}/Z$, with $Z$ the partition function, and $\hat{\overline{X}}=\mathbb{P}\hat{X}\mathbb{P}$ is the projected many-body position operator ($\mathbb{P}$ is the many-body projector to $\mathbb{H}$). The many-body eigenstates are denoted $|m\rangle\equiv|\Psi_m\rangle$, with $m=0$ labeling the ground-state. At zero temperature, the above simplifies, 
\begin{subequations}
\beq
{\overline{\mathbb{O}}}\bigg|_{T=0} &=& \pi e^2 \sum_{m\in \mathbb{H}, m\neq 0} |\langle \Psi_m|\hat{\overline{X}}|\Psi_0\rangle|^2 
\label{eq:projT=0}
\eeq
\end{subequations}
where $\hat\rho_G\equiv|\Psi_0\rangle\langle\Psi_0|$ is the density-matrix (projector) associated with the ground-state wavefunction. When $\Lambda\rightarrow\infty$, these reduce to the SWM result in Eq.~\ref{eq:SWM}. 

The projected theory defined in $\mathbb{H}$ has an associated {\it emergent} $U(1)$ conserved density. Therefore, the arguments leading up to Eq.~\ref{eq:weight} can be extended to relate the partial optical absorption spectral weight to the structure-factor associated with the projected density,
\begin{subequations}
\beq
\overline{\mathbb{O}} &=& \frac{e^2}{2\hbar}\lim_{q\rightarrow0}\frac{\overline{\mathcal{S}}(\q)}{q^2}~ (\equiv \overline{K}),\label{PQW}\\
\tn{where}&~&\overline{\mathcal{S}}(\q) \equiv \frac{1}{N}\langle \overline{\rho}_{\q}~\overline{\rho}_{-\q}\rangle - \frac{1}{N}\left\langle \overline{\rho}_{\q}\right\rangle \left\langle \overline{\rho}_{-\q}\right\rangle ,\label{eq:Struct}\\
\tn{and}~ \overline{\rho}_{\q} &=& \sum_{\alpha,\beta \in \mathbb{H}} \lambda_{\alpha \beta}(\k,\q) c_{\k,\alpha}^\dag c_{\k-\q,\beta}, 
\label{eq:QW}
\eeq
\end{subequations}
with $\lambda_{\alpha \beta}(\k,\q) = \langle u_{\k, \alpha}|u_{\k-\q, \beta}\rangle$ being the form factor of the Bloch-states associated with the bands in the projected subspace and $\alpha$, $\beta$ denote generic indices including band, orbital and spin, respectively. The quantity $\overline{K}$ is now the ``projected" quantum weight.  
All of the expectation values $\langle ...\rangle$ in Eq.~\ref{eq:proj} and \ref{PQW} are meant to be evaluated with respect to the density matrix associated with the many-body insulating state. 

While we will mostly focus on results at $T=0$ below, it is interesting to also directly evaluate $\overline{\mathbb{O}}$ in a ``high" temperature limit in the same projected subspace, where $T$ is the largest energy scale in the problem (but implicitly smaller than the bandgap to remote bands). At the leading order in a small$-\beta$, 
\beq
{\overline{\mathbb{O}}}(\beta\rightarrow0) = \frac{\pi e^2\beta}{{\cal{D}}_{\mathbb{H}}}\sum_{m,n\in \mathbb{H}} (E_m-E_n) ~ |\langle n|\hat{\overline{X}}|m\rangle|^2 ~\theta(E_m-E_n),\nn\\
\label{highT}
\eeq
where ${\cal{D}}_{\mathbb{H}}$ represents the Hilbert-space dimension,   {arising from evaluating the partition function in the limit of small $\beta$, $Z = \lim_{\beta \rightarrow 0} \rm{Tr}\left[e^{-\beta H}\right] = \rm{Tr} \left[\mathbb{I}\right] = \cal{D}_{\mathbb{H}}$.}
Note that in this high-temperature limit the notion of an ``insulating-like" response becomes irrelevant, but we will nevertheless be able to make some rigorous remarks for specific many-body problems. At finite temperature, we can also consider a modified absorption spectral weight, 
\begin{subequations}
\beq
\widetilde{\overline{\mathbb{O}}} &=& \int_0^\Lambda {\cal{I}}(\omega) \tanh\bigg(\frac{\omega}{2T}\bigg)~d\omega,\\
 &=& \frac{\pi e^2}{2} \sum_{m,n\in\mathbb{H}} \frac{(p_n-p_m)^2}{p_n+p_m}  |\langle n|\hat{\overline{X}}|m\rangle|^2.
\eeq
\end{subequations}
  {Note that as $T\rightarrow0$, $\widetilde{\overline{\mathbb{O}}}\rightarrow\overline{\mathbb{O}}$, implying no fundamental distinction between the two quantities. Interestingly, $\widetilde{\overline{\mathbb{O}}}$ represents the quantum Fisher information (QFI) associated with the operator $\hat{\overline{X}}$. The QFI is a well-known object in quantum metrology with connections to entanglement \cite{QFI12a,QFI12b,QFI14,hauke2016measuring}, and in the above context helps distinguish between the density matrix $\hat\rho$ and its unitarily transformed version, $\hat{\rho}(\theta)=e^{-i\theta\hat{\overline{X}}}\hat\rho e^{i\theta\hat{\overline{X}}}$; note that $\hat\rho(\theta)$ is effectively the density matrix associated with a ``projected" gauge-transformation. We return to this quantity in the outlook, after working our way through all of the examples in the next section. }

\section{Correlated insulators at partial band-filling}
\label{sec:examples}
In the following subsections, we will evaluate the partial sum-rule associated with a variety of correlated insulators inspired by the physics of moir\'e materials, that range from the spin and/or valley polarized generalized ferromagnets in twisted bilayer graphene, fractional Chern insulators in partially filled Chern bands, and charge-density wave insulators appearing in topologically trivial flat-bands with a tunable quantum metric.

\subsection{Magic-angle twisted bilayer graphene}
\label{sec:TBG}
MATBG is a fascinating experimental platform for studying the phenomenology of correlated electrons in a topological nearly flat-band, which has revealed an interesting interplay of correlated insulators and superconductors across its phase-diagram \cite{Cao2018,Cao2018b,Yankowitz_2019,Lu2019,Park_2021,Zondiner_2020,uri2020mapping,saito2020independent,saito2021isospin,Cao_2021,liu2021tuning,Das_2021,rozen2021entropic,Serlin900,Sharpe_2019,Stepanov_2020,Wu_2021,Saito_2021Hofstadter,nuckolls2023quantum,Grover2022mosaic,Yu2022hofstadter,Yu2022skyrmion,Morisette2022Hunds,Tseng2022nu2QAH,Choi2019,Oh2021unconventional,Nuckolls2020strongly,Xie2021fractional,Diez-Merida2021diode,Jiang2019charge,Arora2020SC,Kerelsky2019,choi2021correlation,Xie2019stm,Wong_2020,pierce2021unconventional}. The isolated bands can accommodate 8 electrons per unit cell, usually denoted $\nu\in[-4,4]$ on either side of charge neutrality. At all integer fillings, there is an increased tendency to form interaction-induced insulators. The insulating states at $\nu=\pm2$ are robust and omnipresent, and with doping they show the most pronounced tendency towards superconductivity. Theoretically, ignoring the bare single-particle dispersion, in the ``chiral'' limit \cite{tarnopolsky_origin_2019}, the strongly-coupled interacting problem effectively maps to the quantum Hall ferromagnet with an enlarged Hilbert space \cite{bultinck_ground_2020,TBG4}. The continuum effective Hamiltonian for the isolated bands in momentum-space with $\Lambda$ placed in the bandgap to the remote bands is given by (Fig.~\ref{fig:model}), $H_{\rm{TBG}} = \hk + \hi$, where $\hi = \frac{1}{2A}\sum_{\q} V_\q ~\bar\rho(\q)\bar\rho(-\q)$ and $\bar\rho(\q) = \sum_{\k}\lambda^{\alpha\beta}_\mu(\k,\q) c_{\k,\alpha, \mu}^\dag c^{\phantom\dagger}_{\k - \q, \beta, \mu}$ is the projected density operator. 
Here $\lambda^{\alpha\beta}_\mu(\k,\q) = \langle u_{\k,\alpha,\mu}|u_{\k-\q,\beta,\mu}\rangle$ is the form-factor constructed out of the Bloch functions, and $\mu$ denote the valley/spin quantum numbers while $\alpha, \beta$ denote the sub-lattice indices, and $A$ represents the area. We consider the double-gated Coulomb interaction given by $V_\q= \,V_{0}d\,\tanh\left(qd\right)/q$, with  $V_0=e^2/2\epsilon\varepsilon_0 d$. For later computations, we will take the dielectric constant $\epsilon = 10$ and the screening length $d=25$ nm, with $V_0 =~18.1$ meV. A standard starting point for $\hk$ is the Bistritzer-MacDonald (BM) model \cite{Bistritzer2011}.  

\begin{figure}[pth!]
\centering
\includegraphics[width=\linewidth]{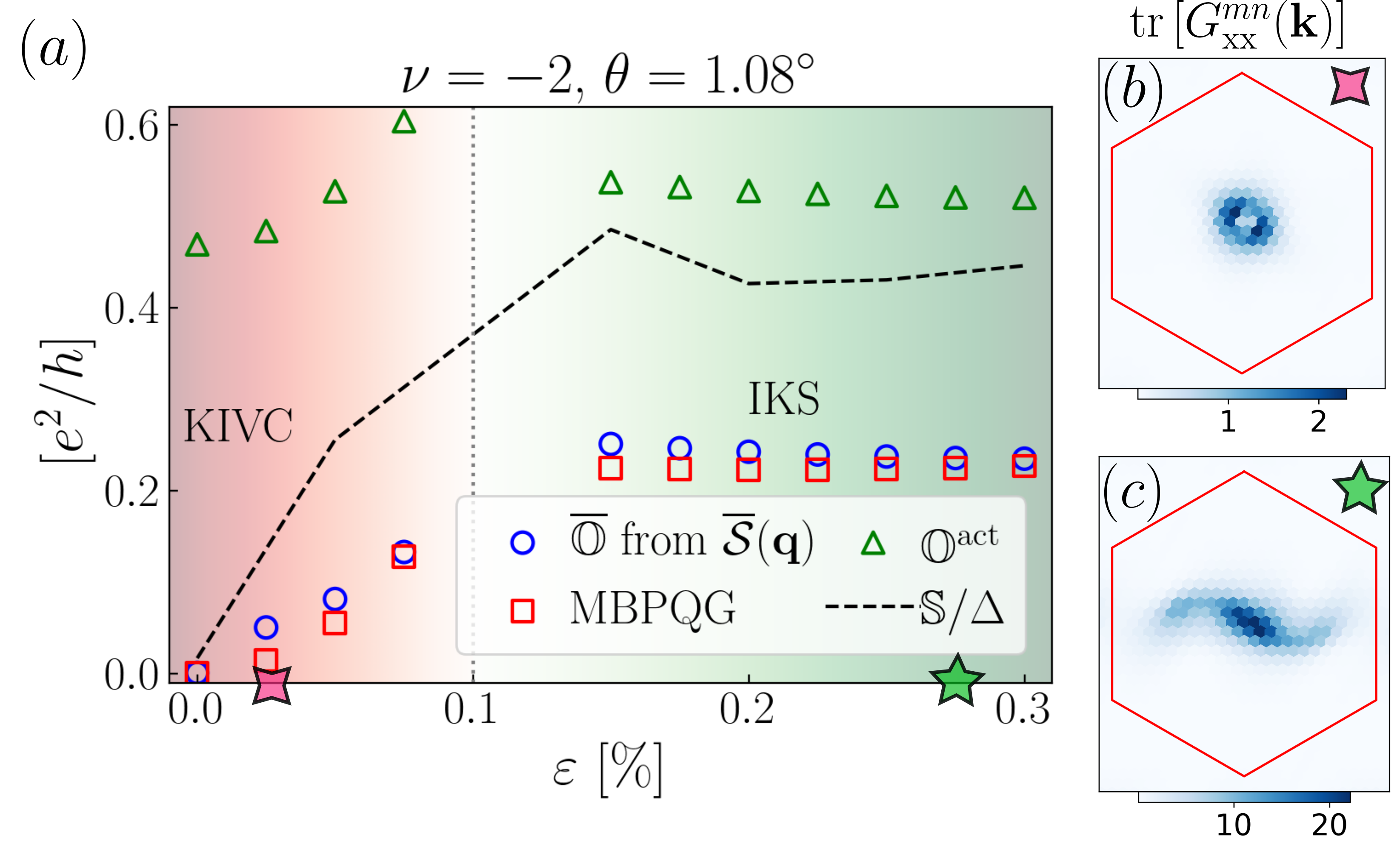} 
\caption{(a) Optical absorption spectral weight as a function of heterostrain ($\varepsilon$) at $T=0$ obtained with respect to the insulating Hartree-Fock ground-state for TBG at $\nu=-2$ and $\theta=1.08^0$. The blue circles correspond to $\overline{\mathbb{O}}$ extracted by fitting the small-$q$ behavior of $\overline{\mathcal{S}}(\q)$ using Eqn.\ref{eq:Struct}, and the black dashed line corresponds to the upper bound, $\overline{\mathbb{S}}/\Delta_{\rm gap}$. The red squares correspond to $\overline{\mathbb{O}}$ evaluated using the many-body projected quantum geometry (MBPQG) in Eq.~\ref{MBPQG}.   { The $\overline{\mathbb{O}}$ and $\overline{\mathbb{S}}$ denote the projected optical absorption with the frequency cutoff $\Lambda$ in between active bands and remove conduction bands, as indicated in Fig.\ref{fig:model}. The green triangles denote the contribution to the full sum-rule $\mathbb{O}$ from the active bands.}  
MBPQG for (b) KIVC order at $\varepsilon=0.025\%$, and (c) IKS order at $\varepsilon=0.275\%$ in the $\q_{\mathrm{IKS}}$ boosted Brillouin zone.} 
\label{fig:Obar}
\end{figure}

In the chiral-flat band limit, $H_{\rm{TBG}}$ has an exact $U(4)\times U(4)$ symmetry \cite{tarnopolsky_origin_2019}, leading to a large manifold of degenerate ground states at all integer fillings.    {Interestingly, the emergent symmetry transformation leaves $\hat{\overline{X}}$ invariant, yielding an emergent ``dipole” conservation law and a {\it vanishing} $\overline{\mathbb{O}}=0$, in spite of a non-vanishing quantum metric associated with the renormalized (gapped) electronic Bloch wavefunctions in the insulating state (see the green triangle at $\epsilon = 0$ in Fig.~\ref{fig:Obar}).  }To see this, note that in the sub-lattice basis of TBG, we have 
$\hat{\overline{X}} = \sum_{\k} c^\dag_{\k,\alpha,\mu}  (i \delta_{\alpha,\alpha'}  \partial_{\k_x}  + \mathcal{A}_{\k, \alpha \alpha',\mu}) c^{\phantom\dagger}_{\k, \alpha',\mu}$, $\mathcal{A}_{\k,\alpha \alpha',\mu}^\nu = i \langle u_{\k,\alpha,\mu} | \partial_{k_\nu} u_{\k,\alpha',\mu}\rangle$ the multi-orbital Berry connection for valley $\mu$. Using a gauge-fixing scheme \cite{bultinck_ground_2020} with no non-abelian Berry connection between the two sub-lattices, and the combination of $C_{2z}$ and particle-hole symmetry (with $\mathcal{A}_{\k, 1,1} = \mathcal{A}_{\k}$),
\beq
\hat{\overline{X}} = \sum_{\k} \hat{c}_{\k}^\dag (i \partial_{\k_x} \mathbf{1} + \mathcal{A}_{\k}^x \sigma_z\tau_z)\hat{c}^{\phantom\dagger}_{\k},
\label{eqn:PXP}
\eeq
where $\sigma$ and $\tau$ act on the sub-lattice and valley basis, respectively. The ground-state projector commutes with $\hat{\overline{X}}$, leading to the absence of any low-energy optical absorption spectral weight. Note that unlike the LL case, where the vanishing $\hat{\overline{X}}$ is tied to the low-energy effective Hamiltonian, in MATBG it is a property only of the ground-state manifold in the chiral flat-band limit.   {There is a conceptually simple reason for the vanishing absorption in this chiral flat-band limit, where the integer filling insulators correspond to a subset of fully filled quantum numbers (combining spin and valley). Since the dipole transition is within the same quantum number sector, for each sector, it is either completely filled or empty, and no optical transition is allowed.} Staying within this manifold, a perturbative correction due to the bare bandwidth $\hk~\sim O(t)$,  leads to $\overline{\mathbb{O}}\sim O[(t/V)^2]$. 

To quantify these corrections to $\overline{\mathbb{O}}$, we perform self-consistent Hartree-Fock (HF) calculations on the continuum model including realistic perturbations away from the chiral-flat band limit to approximate the robust insulating ground-states at filling $\nu=-2$ (Appendix \ref{app:HFTBG}); the computations can be straightforwardly generalized to other integer fillings. Deviations from the chiral flat-band limit tend to pick intervalley coherent states, with the precise state being largely determined in experiments by the extrinsic effects of heterostrain ($\varepsilon$). The introduction of small amounts of strain at the graphene scale is enhanced by the moir\'e pattern, and significantly increase the single-particle bandwidth in addition to breaking various point-group symmetries. To investigate the optical absorption, we calculate $\overline{\mathbb{O}}$ by exploiting its connection to the low-$q$ behavior of $\overline{\mathcal{S}}(\q)$ as a function of  $\varepsilon$; this is shown in Fig.~\ref{fig:Obar}(a). At small $\varepsilon$, the ground-state is a spin-polarized Kramers-intervalley coherent state (KIVC), which is approximately annihilated by the density operator $\delta\rho_{\q}$ \cite{bultinck_ground_2020}, leading to a large suppression of $\overline{\mathcal{S}}(\q)$ and a small optical absorption spectral weight. Thus, in the limit of a small $\varepsilon$, the insulating states remain largely optically ``dark". As $\varepsilon$ is increased, the ground state projector develops a small $O(t/V)$ component parallel to the order parameter of a $C_2\mathcal{T}$ and $U(1)_V$ preserving semi-metal \cite{bultinck_ground_2020}; the state remains gapped overall. This component is off-diagonal in the sublattice basis and thus no longer commutes with $\overline{X}$ in Eqn.~\ref{eqn:PXP}, and is then responsible for the enhancement of $\overline{\mathbb{O}}$. Above a critical value of strain, the combined effect of Coulomb renormalization of the bands and strain-induced enhancement of the kinetic energy triggers a translation symmetry breaking order; at $\nu=-2$ this leads to the incommensurate Kekule spiral (IKS) both within Hartree-Fock mean-field and DMRG studies \cite{kwan_kekule_2021,SoejimaDMRG}. In our calculation, this occurs for $\varepsilon\gtrsim0.1\%$, where the state is no longer described as part of the manifold of degenerate ground-states of the chiral model, and instead features a non-trivial winding texture of the inter-Chern coherent component of the order parameter \cite{kwan2024texturedexcitoninsulators, wang2024cherntexturedexcitoninsulatorsvalley}. This strong momentum dependence of the order parameter is inherited from the underlying topology of the TBG bands and due to its Chern-coherent component, it anticommutes with $\overline{X}$ leading to a further enhancement of $\overline{\mathbb{O}}$.   

It is also possible to bound the optical absorption spectral weight: $\overline{\mathbb{O}}\leq \overline{\mathbb{S}}/\Delta_{\tn{gap}}$, where $\overline{\mathbb{S}}$ represents the low-energy ``partial" $f-$sum-rule associated with the same interacting theory defined in $\mathbb{H}$ (and has been evaluated previously in Ref.~\cite{opticalsumrule2023}),
\begin{subequations}
\beq
\overline{\mathbb{S}} &\equiv& \int_0^\Lambda \tn{Re}[\sigma_{\rm{xx}}^{\tn{eff}}(\omega)]~d\omega = \frac{\pi}{2}\langle K_{\rm{xx}}^{\rm{eff}}\rangle,\\
\langle K_{\rm{xx}}^{\rm{eff}}\rangle &=& - \frac{e^2}{h}\frac{1}{A}\tn{Tr}\left( [\hat{\overline{X}},[\hat{\overline{X}},\hat\rho_G]]\heff\right),
\eeq
\end{subequations}
and $\Delta_{\rm{gap}}$ represents the many-body gap in the insulating state.  
Note that in using the bound on $\overline{\mathbb{O}}$ in terms of $\overline{\mathbb{S}}$, we have assumed that the insulator hosts no sub-gap optical response, i.e. $\tn{Re}[\sigma_{\rm{xx}}^{\tn{eff}}(\omega)]=0$ for $\omega\leq\Delta_{\rm{gap}}$. The comparison between $\overline{\mathbb{O}}$ and this bound is shown in Fig.~\ref{fig:Obar}(a). Here, we see that the bound closely tracks the behavior of the projected quantum weight. However, near the transition point at $\varepsilon\approx 0.1\%$, the gap on the KIVC side is reduced by the tendency towards the $C_{2}\mathcal{T}$preserving semi-metal, which makes the bound less accurate as the transition is approached from the low $\varepsilon$ side. In general this bound is expected to be inaccurate near a quantum phase transition where $\Delta_{\mathrm{gap}}\rightarrow 0$ but $\overline{\mathbb{S}}$ remains finite; at $\nu=-2$ the system remains gapped for the range of $\varepsilon$ studied in Fig.~\ref{fig:Obar}. 

Let us now turn to the possible connections between $\overline{\mathbb{O}}$ and a ``geometric" quantity constructed out of the Bloch wavefunctions associated with the active bands. We note at the outset that the quantum metric associated with the bare non-interacting bands is clearly irrelevant for the full response, $\overline{\mathbb{O}}$. For any insulating Slater determinant state obtained within the Hartree-Fock approximation, we find that $\overline{\mathbb{O}}$ can simply be obtained by replacing the quantum metric ($g_{\mu \nu}$) in the sum-rule for $\mathbb{O}$ in Eq.~\ref{eq:SWM} by a ``projected" multi-band quantum metric $G_{\mu \nu}$ (Appendix \ref{sec:SF}),
\begin{subequations}
\beq
\overline{\mathbb{O}} &=& \frac{\pi e^2}{\hbar} \delta_{\mu\nu} \sum_\k \tn{Tr}[G_{\mu\nu}],~\tn{where}\label{MBPQG}\\
G_{\mu\nu}^{mn}(\k) &=& \langle \partial_\mu \tilde{u}_{m,\k} |[\mathcal{P}_{\text{IR}}(\k)- \mathcal{P}_{\text{occ}}(\k)]|\partial_\nu \tilde{u}_{n,\k} \rangle.\label{MBPQGb}
\eeq
\end{subequations}
Here, the trace is taken over the occupied Hartree-Fock bands, and $|\tilde{u}_{n,\k} \rangle$ denotes the Hartree-Fock renormalized Bloch wavefunction. The projectors, 
\begin{subequations}
\beq
\mathcal{P}_{\text{IR}}(\k) = \sum_{m \in \tn{IR}} |\tilde{u}_{m,\k} \rangle \langle \tilde{u}_{m,\k}|,\\
\mathcal{P}_{\text{occ}}(\k) = \sum_{m \in \tn{occ}} |\tilde{u}_{m,\k} \rangle \langle \tilde{u}_{m,\k}|,
\eeq
\end{subequations}
 are associated with the (occupied) Hartree-Fock bands in the IR theory.   {We dub $G_{\mu\nu}^{mn}(\k)$ ``many-body projected quantum geometry" (MBPQG) to highlight its importance in accounting for the ``partial" sum-rule associated with the interaction-renormalized bands. MBPQG is different from the usual multi-band quantum geometric tensor \cite{OnishiFu24a, bouhon2023quantum} in two ways. First, MBPQG invokes the renormalized Bloch wavefunction, reflecting its many-body nature. Secondly, MBPQG only takes into account transitions within the restricted IR-Hilbert space, as manifest in the $\mathcal{P}_{\text{IR}}(\k)$ term, and hence is ``projected". }Hence, the integral of $G_{\mu\nu}^{mm}(\k)$ is not related to the minimal spread of any many-body Wannier wavefunction associated with the renormalized band. In our example of TBG, the Slater-insulators in the chiral flat-band limit have a vanishing $\sum_\k \tn{Tr}[G_{\mu\nu}]$ even though they are not atomic insulators. It is easy to verify that MBPQG is positive semi-definite and bounded from above by the band quantum geometry (Appendix \ref{sec:SF}).  { Hence, we can view MBPQG as accounting for the optically active part of the spreading of the Wannier wavefunction at low-frequencies}. In the {\it non-interacting} limit, the MBPQG is reduced to the partial-sum rule in Ref.\cite{OnishiFu24a}.
 We have evaluated $\overline{\mathbb{O}}$ using Eq.~\ref{MBPQG} in Fig.~\ref{fig:Obar}(a). To help visualize the MBPQG associated with the interaction-renormalized (filled) bands, we plot $\delta_{\mu\nu} \tn{Tr}[G_{\mu\nu}](\k)$ in the continuum model for two different values of $\varepsilon$ in Fig.~\ref{fig:Obar}(b) in the KIVC and IKS phases, respectively.   {For comparison, we also show $\mathbb{O}^{\rm{act}}$, which denotes the contribution in the full $\mathbb{O}$ from the partially filled flat bands. We note that 
 \begin{subequations}
\beq
\mathbb{O}^{\rm{act}} &=& \frac{\pi e^2}{\hbar} \delta_{\mu\nu} \sum_\k \tn{Tr}[G_{\mu\nu}^{\rm{act}}],~\tn{with}\label{MBQG}\\
G_{\mu\nu}^{mn,\rm{act}}(\k) &=& \langle \partial_\mu \tilde{u}_{m,\k} |[\mathbb{I}- \mathcal{P}_{\text{occ}}(\k)]|\partial_\nu \tilde{u}_{n,\k} \rangle,\label{MBQGb}
\eeq
\end{subequations}
where the $G_{\mu\nu}^{\rm{act}}$ can be viewed as the many-body quantum geometry (MBQG) of the renormalized Bloch wavefunction. We note that the completely filled {\it remote} bands also contribute to the optical sum-rule, $\mathbb{O}$. Since the remote-band contribution and $\mathbb{O}^{\rm{act}}$ fall into the same frequency range in the optical absorption, they are not separable in general, which makes it hard to infer the property of the active bands solely from $\mathbb{O}$. Moreover, the contribution due to the remote bands can be large. We have numerically calculated the optical absorption from the eight fully-filled remote bands that are closest to the active bands by doing Hartree-Fock including these remote bands. Deep in the IKS phase at $\epsilon=0.3\%$, the contribution to the optical sum-rule is given by $\mathbb{O}|_{\rm{8\,bands}}\approx 5.1e^2/h$, which is much larger than the contribution from the active bands $\mathbb{O}^{\rm{act}}$ plotted in Fig. \ref{fig:Obar}. }


\subsection{Fractional Chern insulators}
\label{sec:FCI}
A number of recent experimental breakthroughs have led to the observation of FCI \cite{XGW11,TN11,BAB11,Sheng2011}, or fractional quantum anomalous Hall (FQAH) insulators \cite{FQAHrev}, in moir\'e transition metal dichalcogenide (TMD) \cite{xiaodong,FaiFCI,tingxin} and multilayer graphene materials \cite{LongJu}. We address here the fraction of the total spectral weight, $\mathbb{O}$, that is contained in $\overline{\mathbb{O}}$ for these insulating states. Regardless of the microscopic mechanism (which may involve various forms of symmetry-breaking \cite{cecile2020chern,Xie2021fractional,Li2021spontaneous,wilhelm2021, liu2024fci}), the low-energy physics for many of the FCI/FQAH insulators is described via Coulomb interactions projected to a set of nearly flat and isolated topological (Chern) bands having near-ideal quantum geometry (``vortexability") \cite{vortexability}; see also effects of interaction-induced band-mixing across the single-particle gap on the stability of FCI \cite{Sheng2011, Kourtis2014fractional, grushin15,sharma2024, yu2024fractional}. 
We will ignore any band-mixing effects and focus only on the ``projected" limit in this manuscript, where many of the recently discovered FCI states can be approximated by a generalized Laughlin-like character \cite{ledwith2020fractional,sheffer2021, wangcano, dong23manybody, shi2024prb, nicolas}. The low-energy theory is then described by placing $\Lambda$ inside the bandgap, $E_{\rm{band}}$, to the remote bands. A recent work \cite{WolfAMD} has addressed the question of the low-energy optical absorption in FCI's numerically for vortexable bands starting from the LLL limit, and argued that the low-frequency response is ``weak". In what follows, we will be able to shed some complementary analytical insights into what controls the weak response and make connections to the moir\'e TMD materials; we return to comparisons with this work in Sec.~\ref{sec:perB}.

We begin by summarizing the results for a low-energy theory that describes the classic problem of interactions projected to the lowest (spinless) LL; $\Lambda$ is now a scale that resides in the cyclotron-gap to the first LL. Recall that in the absence of disorder, $\hat{\overline{X}}$ is an emergent conserved operator, and thus the many-body energy eigenstates $|n\rangle\in\mathbb{H}$ can be simultaneously diagonalized in the $\hat{\overline{X}}-$basis. Thus, we immediately infer that $\overline{\mathbb{O}}=0$ in this theory. In fact, $\overline{\mathbb{O}}$ vanishes at {\it any} filling of the lowest Landau level (LLL), even when the spectrum does {\it not} have a gap while satisfying Eq.~\ref{ins:def} \cite{ASLF24}.   {The latter is a consequence of Kohn's theorem, stating that the $f-$sum-rule is saturated at the cyclotron frequency with $\tn{Re}[\sigma_{\tn{xx}}(\omega)]=0$ for $\omega<\omega_c$ since the dipole transition is purely inter-LL \cite{kohn1961,yip89}.} The vanishing $\overline{\mathbb{O}}$ in the LLL theory for both quantum Hall insulators and for metallic phases, such as the composite Fermi liquid \cite{HLR}, is consistent with the known results for the projected structure factor in Eq.~\ref{PQW}, which vanishes as $\overline{\mathcal{S}}(\q)\sim q^4$ in the former \cite{GMP,HaldaneSF}, and as $q^3$ (with additional $\log q$ corrections) in the latter \cite{read,prashant,XCW}, respectively. Interestingly, within this projected theory, $\overline{\mathbb{O}}(\beta\rightarrow0)$ in Eq.~\ref{highT} also vanishes. Therefore, given the structure of the many-body matrix elements of $\hat{\overline{X}}$, in the theory for a partially filled LLL, the entire absorption spectral weight associated with $\mathbb{O}$ is saturated by $(\mathbb{O}-\overline{\mathbb{O}})$ at all temperatures.   {Clearly, the LLL has a non-vanishing (uniform) quantum geometry, which turns out to be {\it irrelevant} for determining the {\it low-energy} optical absorption.  However, as we show below, the {\it fluctuations} in quantum geometry can control the low-energy spectral weight for perturbed LLLs, which captures a broad class of fractional Chern insulators.}

The vanishing low-energy spectral weight in LLL is due to the conservation of center-of-mass (or dipole moment) at low-energy. In general, any term that breaks dipole symmetry contributes to the low-energy spectral weight. Let us first give a generic argument for the strength of this response, and then we will discuss individual cases. By definition, $\overline{\mathbb{O}}$ is positive semi-definite and it has a lower bound, namely, 
\beq
{\overline{\mathbb{O}}}\bigg|_{T=0} &\geq& \pi e^2 |\langle \Psi_e|\hat{\overline{X}}|\Psi_g\rangle|^2= \pi e^2 \frac{\left|\langle \Psi_e|\left[\hat{\overline{X}}, \heff\right]|\Psi_g\rangle\right|^2}{\Delta^2}, \nn\\
\eeq
where $|\Psi_e\rangle$ is the first excited state and $\Delta = E_e - E_g$ is the energy gap. If we consider a gapped fractional quantum Hall state obtained by perturbing away from the dipole-conserving limit, this automatically implies that ${\overline{\mathbb{O}}}\bigg|_{T=0}$ is in general finite. Note that for systems with open boundary condition, $\Delta$ and $|\Psi_e\rangle$ should be viewed as bulk gap and bulk excitation, respectively. For systems on a torus, $\hat{\overline{X}}$ should be viewed as taking the $\q \rightarrow 0$ limit in $\bar{\rho}_\q$. In order to vary $\q$ continuously, one then needs to perform flux-threading in the torus. Numerical exact-diagonalization studies that target only the low-lying states and the associated structure-factor for relatively large system-sizes can in principle be used to estimate the strength of this low-energy response.

For LLL, two types of dipole symmetry breaking perturbation have been pointed out (not involving disorder \cite{AMdis}): (i) a spatially periodic potential, and (ii) a spatially periodic magnetic field. We consider them individually in the next two subsections.

\subsubsection{Lowest Landau-level in a periodic potential}
For a weak periodic potential, the low-energy Hilbert space is still described in terms of the original LLL, and the projected periodic potential can be written as $H_{\rm{pot}} = \sum_{\vec{G}} V_{\vec{G}} \bar{\rho}_{\vec{G}}$, where $\vec{G}$ is the reciprocal lattice vector associated with the periodic potential and $V_{\vec{G}}$ represents its strength. Using the single mode approximation \cite{GMP}, and a Laughlin-like wavefunction for the ground-state \cite{wu2016moire}, we can approximate
\begin{subequations}
\beq
{\overline{\mathbb{O}}} &\approx& \frac{e^2}{4 \hbar} \sum_{\vec{G}} l_B^2 |\vec{G}|^2 \frac{|V_{\vec{G}}|^2}{\Delta_{\vec{G}}^2} \overline{{\cal{S}}}_0(\vec{G}),\\
{\overline{\mathbb{S}}} &\approx& \frac{e^2}{4 \hbar} \sum_{\vec{G}} l_B^2 |\vec{G}|^2 \frac{|V_{\vec{G}}|^2}{\Delta_{\vec{G}}} \overline{{\cal{S}}}_0(\vec{G}),
\eeq
\end{subequations} 
where $l_B$ is the magnetic length, $\Delta_{\vec{G}}$ denotes the gap for the collective excitation at $\vec{G}$, and $\overline{{\cal{S}}}_0(\vec{G})$ is the structure factor evaluated with respect to the Laughlin state. 

\subsubsection{Lowest Landau-level in a periodic magnetic field}
\label{sec:perB}

Periodic magnetic field in the LLL has been proposed as a theoretical tool to study (fractional) Chern insulating states across moir\'e systems \cite{ledwith2020fractional, sheffer2021,dong2022diracelectronperiodicmagnetic, vortexability, wangcano, WolfAMD, nicolas, estienne2023, li2024variationalmappingchernbands}. 
A uniform magnetic field gives rise to uniform Berry curvature in the magnetic Brillouin zone and the fluctuation of the magnetic field in real space controls the fluctuation of the Berry curvature in momentum space. The LLL in the periodic magnetic field is given by the zero modes of the operator $(-i \bar{\partial} - \bar{A}) $, where $\bar{\partial} \equiv \partial_x + i \partial_y$ and $\bar{A} \equiv A_x + i A_y$, $A_\mu$ being the vector potential \cite{aharonov1979ground, dubrovin1980ground}. Therefore, the wavefunction can be written as $\psi_{LLL}(\vec{r}) = f(z) e^{-\varphi(\vec{r})}$, where $\partial_\mu^2 \varphi(\vec{r}) = B(\vec{r})$. Hence, the projected Hilbert space has a different basis compared to the LLL under uniform magnetic field and the dipole conservation is broken from the projection in $\mathcal{H}_{\text{eff}}$. For $B(\vec{r})$ that is periodic in space, there is discrete magnetic translation symmetry, which allows us to define a magnetic Brillouin zone. The relationship between $B(\vec{r})$ and the Berry curvature $\mathcal{B}_\k$ can be obtained from the Bloch wavefunction, but the generic analytical form is cumbersome so we only give an approximate expression. If $B(\vec{r}) \equiv  B_0 + B_1(\vec{r}) = 1 + \phi_1 \sum_j e^{i \vec{k}_j \cdot \vec{r}}$, where $B_0 =1$ denotes the uniform part and $\vec{k}_j$ denotes the reciprocal lattice vectors, to the leading order in $\phi_1$, $\mathcal{B}_\k \approx  1- \phi_1 \sum_j e^{i \vec{k}_j \wedge \vec{k} - \frac{|\vec{k}_j|^2}{4} }$\cite{mao2024upper}.   {Therefore, $\phi_1$ controls both the fluctuation the magnetic field in real-space and the fluctuation of the Berry curvature in momentum space.}
 
\begin{figure}
    \centering
    \includegraphics[width=1.0\linewidth]{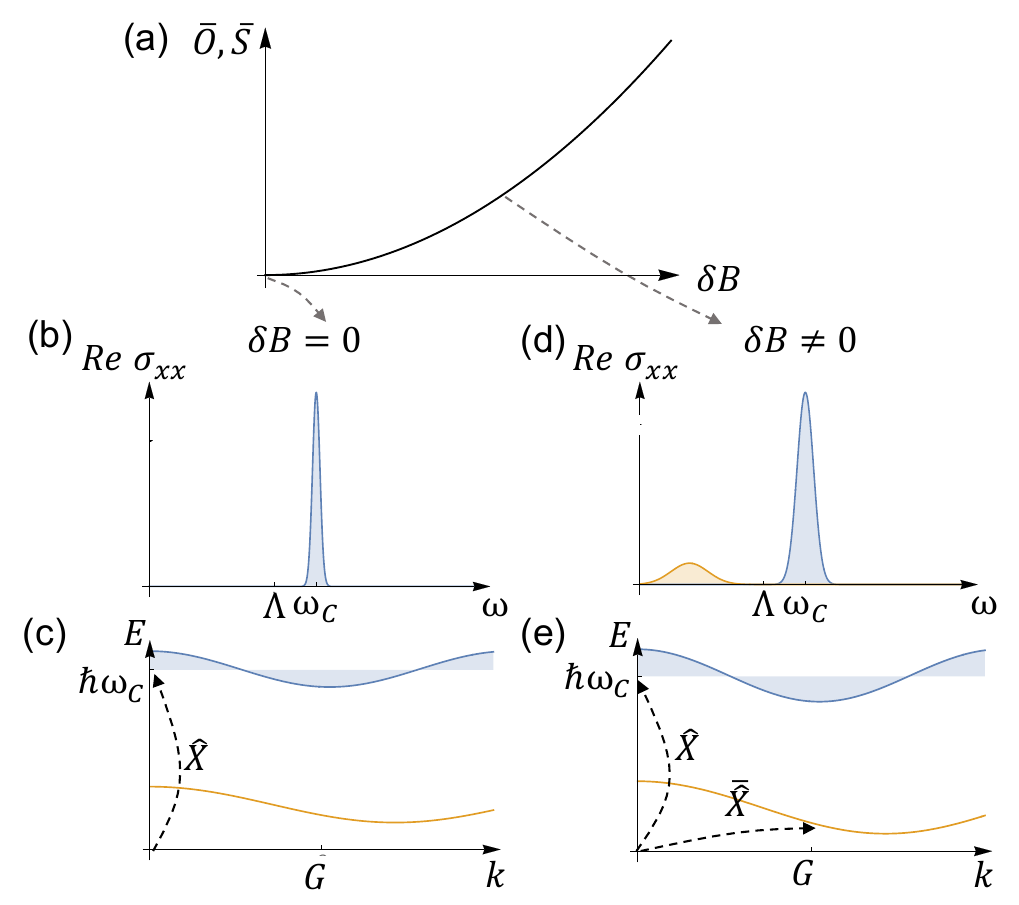}
    \caption{Schematic summary of the optical sum-rules and allowed transitions in the LLL. (a) For a small fluctuation of the magnetic field, $\delta B$, the strength of $\overline{\mathbb{O}}$ and $\overline{\mathbb{S}}$ is quadratic in $\delta B$. For a uniform [periodic] $B$, the optical absorption and the dispersion of the collective modes are sketched in panels (b) [(d)] and (c) [(e)], respectively. The periodic $B$ can change the energetic of the states of higher LL, resulting in broader optical absorption near $\omega_c$ [panel (d)]. The breaking of continuous magnetic translation symmetry by the spatial fluctuation $\delta B$ enables intra-LL optical transition, which excites collective modes at momentum of the reciprocal lattice vector [panel (e)].}
    \label{fig:gLL}
\end{figure}

We summarize our results of the low-energy spectral weight schematically in Fig.~\ref{fig:gLL}a. For the LLL under a uniform magnetic field, Kohn's theorem dictates that the dipole transition is purely inter-LL (Fig.~\ref{fig:gLL}b). However, under a periodic magnetic field, the fluctuation of the magnetic field $\delta B \sim \phi_1$ enables intra-LL dipole transition (Fig.~\ref{fig:gLL}d), resulting in a finite $\overline{\mathbb{S}}$
 and $\overline{\mathbb{O}}$, proportional to $(\delta B)^2$ to the leading order. 

For a small $\phi_1$, the dipole matrix element can be written as,
\beq
\hat{\overline{X}} = \hat{\overline{X}}_0 + \sum_{n=1}^\infty (\phi_1)^n \hat{\overline{X}}_n,
\eeq
where the subscript `$0$' denotes the matrix of the projected position operator under $B_0$, and the subscript `$n$' denotes the $n$-th order correction. Recall that $[\hat{\overline{X}}_0, \hat{\overline{Y}}_0] = -i/B_0$, which is the so-called ``emergent" guiding center \cite{si, wang2023origin}.   {To the leading order in $\phi_1$, we will see that only $\hat{\overline{X}}_1$ is important. To proceed, we will first derive the expression for $\overline{\mathbb{O}}$ to the leading order in $\phi_1$ and then derive the algebraic expression for $\hat{\overline{X}}_1$.}

Since the center-of-mass coordinate no longer commutes with the projected Hamiltonian, the ground state can be written as $|\widetilde{\Psi}\rangle  = |\Psi^0\rangle + \sum_{n=1}^\infty (\phi_1)^n |\Psi^n\rangle$, where $|\Psi^0\rangle$ is the eigenstate of $\hat{\overline{X}}_0$.

The leading order contribution in ${\overline{\mathbb{O}}}\bigg|_{T=0}$ is of order $\phi_1^2$, and the coefficient is positive semi-definite by definition. To be explicit, ${\overline{\mathbb{O}}}\bigg|_{T=0} = \pi e^2 \phi_1^2 \langle \eta |(1-|\Psi^0\rangle \langle \Psi^0|) |\eta\rangle + O(\phi^4)$, where $|\eta\rangle = \hat{\overline{X}}_0 |\Psi^1\rangle + \hat{\overline{X}}_1 |\Psi^0\rangle$. Although we cannot solve for $|\Psi^0\rangle$ and $|\Psi^1\rangle$ exactly, we can make progress in estimating the projected spectral weight by approximating the ground state $|\widetilde{\Psi}\rangle$ at fractional fillings by generalized Laughlin states \cite{ledwith2020fractional}, which amounts to ignoring the contribution from $|\Psi^n\rangle$ for $n \geq 1$.

On the plane, the generalized Laughlin state can be written as $\Psi_{gL}(\vec{x}_1,...,\vec{x}_N) = R^{-1} \prod_{i<j} (z_i - z_j)^m e^{-\sum_i \varphi(\vec{x}_i)}$, recalling that $\partial_\mu^2 \varphi(\vec{r}) = B(\vec{r})$ and therefore $\varphi(\vec{r}) = |z|^2/4 + \tilde{\varphi}(\vec{r})$ for $B(\vec{r}) = 1 + \tilde{B}(\vec{r})$. Without the $\tilde{\varphi}(\vec{x}_i)$ term, $\Psi_{gL}$ is the original Laughlin wavefunction with filling fraction $\nu = 1/m$, and $R$ being the normalization constant. We note that $|\Psi_{gL}\rangle$ can be viewed as a special $|\Psi^0\rangle$ that is annihilated by $\hat{\overline{X}}_0$ \cite{si}. However, since the generic projected density-density interaction does not conserve the ``emergent" guiding center,  $|\Psi_{gL}\rangle$ is not the exact ground state except for some special pseudo-potential Hamiltonian \cite{wangcano}. Nonetheless, we assume $|\Psi_{gL}\rangle$ is a good approximation to the true ground state, and a direct evaluation of $\overline{\mathbb{O}}$ with respect to this state yields,
\beq
{\overline{\mathbb{O}}}_{gL} &\equiv& \pi e^2 \langle \Psi_{gL}| \hat{\overline{X}} (1- |\Psi_{gL}\rangle \langle \Psi_{gL} |) \hat{\overline{X}} |\Psi_{gL} \rangle \nn\\
&=&\pi e^2\phi_1^2 \langle \Psi^0_{gL}| \hat{\overline{X}}_1 (1- |\Psi^0_{gL}\rangle \langle \Psi^0_{gL} |) \hat{\overline{X}}_1 |\Psi^0_{gL} \rangle + O(\phi_1^4),\nn\\
\eeq
where $|\Psi^0_{gL} \rangle$ denotes the Laughlin wavefunction under uniform magnetic field. Note that the correction from the periodic part of the magnetic field to the wavefunction only appears at a higher order in $O(\phi_1^2)$.

In order to obtain $\hat{\overline{X}}_1$ explicitly, we consider the projected density operator under a periodic magnetic field. From the LLL wavefunction, we have \cite{dubrovin1980ground,mao2024upper},
\beq
\hat{\overline{\rho}}_{\vec{q}} = \hat{\overline{\rho}}^0_{\vec{q}} + \frac{\phi_1}{|G|^2} \sum_j \hat{\overline{\rho}}^0_{\vec{q}+\vec{k}_j}\bigg[2 - \bigg(e^{\frac{\bar{k}_j q}{2}} +e^{\frac{k_j \bar{q}}{2}}\bigg)\bigg],
\eeq
where $q = q_x + iq_y$ and analogously for $k_j$, and the superscript `$0$' denotes the density operators that satisfy Girvin-MacDonald-Platzman (GMP) algebra \cite{GMP}, 
\begin{subequations}
\beq
[\tilde{\rho}_{\vec{q}_1}^0 , \tilde{\rho}_{\vec{q}_2}^0  ] &=& 2i \sin{\frac{\vec{q}_1 \wedge \vec{q}_2}{2}}\tilde{\rho}^0_{\vec{q}_1+ \vec{q}_2},~\tn{where}\\
\tilde{\rho}_{\vec{q}}^0 &=& \hat{\overline{\rho}}^0 e^{\frac{|\vec{q}|^2}{4}}, \tn{and}~~\vec{q}_1 \wedge \vec{q}_2 = \epsilon_{\mu\nu} q_{1,\mu} q_{2,\nu}.
\eeq 
\end{subequations}
In the small $\vec{q}$ limit of $\hat{\overline{\rho}}_{\vec{q}}$, we have $\hat{\overline{\rho}}_{\vec{q}} = i \vec{q} \cdot \hat{\overline{X}}$ (see Appendix \ref{app:perB}) and
\beq
\hat{\overline{X}}_1 &=& -\frac{i}{ |G|^2}\sum_j k_{j,x} \hat{\overline{\rho}}_{\vec{k}_j}^0,
\eeq
which gives 
\beq
{\overline{\mathbb{O}}}_{gL} = \frac{\pi e^2 \phi_1^2}{|G|^4 l_B^2} \sum_j k_{j,x}^2 \overline{{\cal{S}}}_0(\vec{k}_j).
\label{eq:O_gl}
\eeq

Similarly, we can obtain the partial $f-$sum-rule, $\overline{\mathbb{S}}_{gL}$. Again, the leading order contribution is of order $\phi_1^2$ and therefore we only need to consider the order $\phi_1$ term in $\mathcal{H}_{\text{eff}}$,
\begin{subequations}
\beq
\overline{\mathbb{S}}_{gL} &\equiv& -\frac{e^2}{4} \langle \Psi_{gL}| \left[\hat{\overline{X}} ,\left[\hat{\overline{X}} , \mathcal{H}_{\text{eff}}\right]\right]|\Psi_{gL} \rangle\\
&=&-\frac{e^2}{4} \phi_1^2 \langle \Psi_{gL}^0| \left[\hat{\overline{X}}_1 ,\left[\hat{\overline{X}}_1 , \mathcal{H}_{\text{eff}}^0\right]\right]|\Psi_{gL}^0 \rangle\nn\\
&~&-\frac{e^2}{4} \phi_1^2 \langle \Psi_{gL}^0| \left[\hat{\overline{X}}_1 ,\left[\hat{\overline{X}}_0 , \mathcal{H}_{\text{eff}}^1\right]\right]|\Psi_{gL}^0 \rangle + O(\phi_1^4),\nn\\
\eeq
\end{subequations}
where $\mathcal{H}_{\text{eff}} \equiv \sum_{\vec{q}} V_{\vec{q}} \hat{\overline{\rho}}_{\vec{q}} \hat{\overline{\rho}}_{-\vec{q}} = \mathcal{H}_{\text{eff}}^0 + \phi_1 \mathcal{H}_{\text{eff}}^1 + O(\phi_1^2)$. From discrete rotation symmetry on a lattice, one finds that the second term in the last line of the above equation vanishes and therefore,
\beq
\overline{\mathbb{S}}_{gL} &=& \frac{e^2 \phi_1^2 }{4|G|^4 l_B^2}   \sum_j k_{j,x}^2  \bar{f}_0(\vec{k}_j),
\label{eq:S_gl}
\eeq
where $\bar{f}_0(\vec{k}_j) = \langle \left[\hat{\overline{\rho}}_{\vec{k}_j}^0,\left[\hat{\overline{\rho}}_{-\vec{k}_j}^0,\mathcal{H}^0_{\text{eff}}\right]\right]\rangle$ is the projected oscillator strength \cite{GMP}.

\begin{figure}
    \centering
    \includegraphics[width=\linewidth]{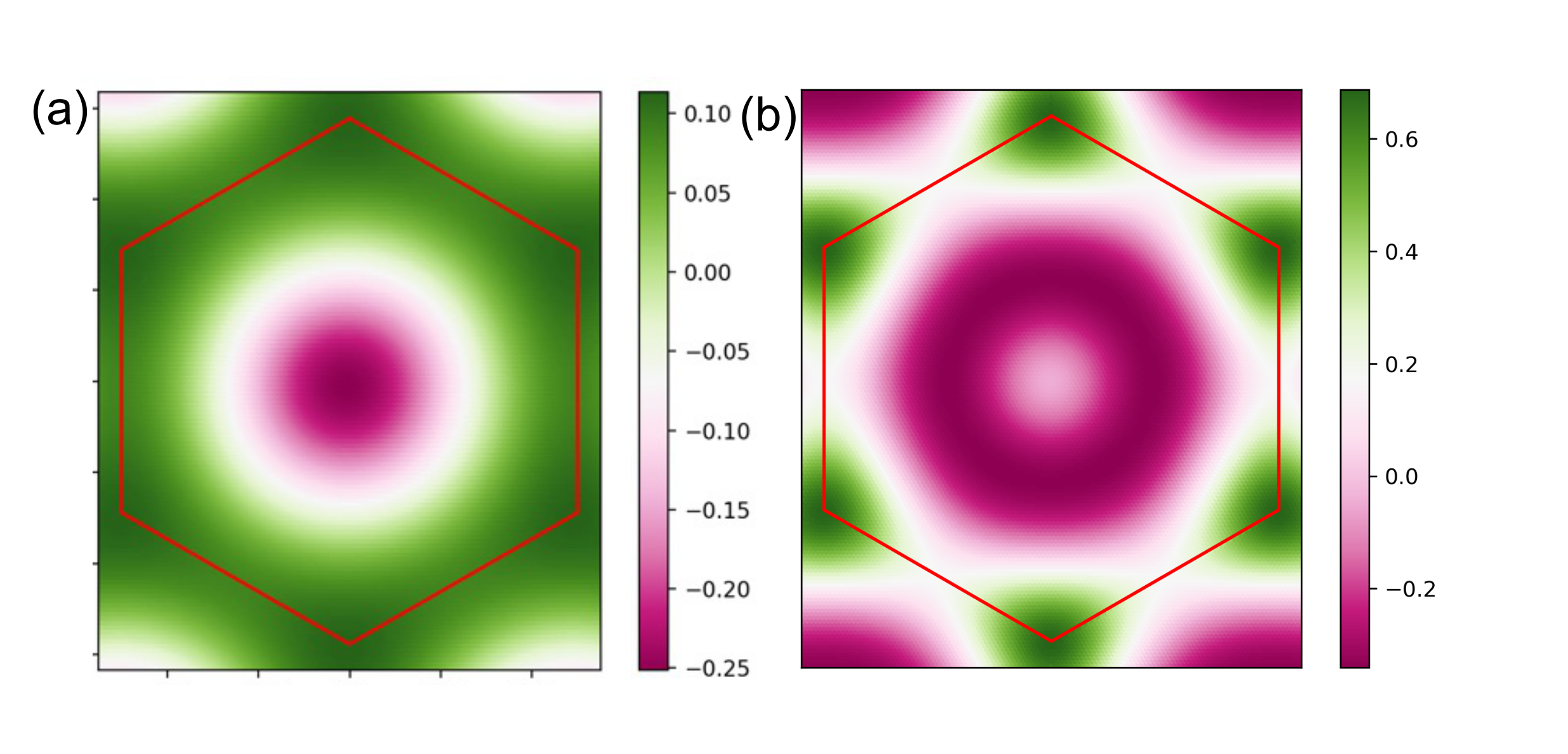}
    \caption{Berry curvature fluctuation calculated using the  (a) effective magnetic field in Eq.~\ref{Beffr}, and (b) continuum model for twisted homobilayer MoTe$_2$. The red hexagon marks the Brillouin zone, where the background uniform Berry curvature is normalized to $1$. The effective magnetic field is obtained from the unitary transformation formalism in Ref.~\cite{nicolas} using the same parameters as the continuum model for unstrained MoTe$_2$ at twist angle $1.5^\circ$  \cite{wu2019topological}. For the effective magnetic field, we retain the first five Fourier components (excluding the zero momenta) to be -0.251, -0.316, 0.478, -0.138, 0.0804. }
    \label{fig:berry_com}
\end{figure}

Let us now apply our formalism to the correlated insulators in twisted homobilayer TMD, and address the question of how large is $\overline{\mathbb{O}}$   {(and $\overline{\mathbb{S}}$)}. First, we would like to comment on the validity of the leading order expansion in $\phi_1$. Although $\phi_1$ quantifies the fluctuation of the magnetic field in space, since the wavefunction involves $\tilde{\varphi}(\vec{x})$, the expansion parameter is actually $\phi_1/(|G|^2 l_B^2)$, which can be parametrically smaller than $\phi_1$. For example, for a periodic modulation with $2\pi-$flux per unit cell, $|G|^2 l_B^2 = 2\pi$ for square lattice and $4\pi/\sqrt{3}$ for triangular lattice. In twisted homobilayer TMD systems, the moir\'e lattice potential generates an effective magnetic field $B_{\text{eff}}(\vec{r})$ \cite{nicolas}. We keep the first five shells of the reciprocal lattice vector and write,
\beq
B_{\text{eff}}(\vec{r}) = 1 + \sum_{n=1}^5 \sum_{j} \phi_{n,j} e^{i \vec{k}_{n,j} \cdot \vec{r}},
\label{Beffr}
\eeq 
where $n$ labels the shell and $j$ labels the different reciprocal lattice vectors within one shell of the same $|\vec{k}_{n,j}|$. From the first-order expansion, the Berry curvature can be written as $B_{\text{eff}}(\vec{k}) = 1 - \sum_{n=1}^5 \sum_{j} \phi_{n,j} e^{i \vec{k}_{n,j} \wedge \vec{k} - \frac{|\vec{k}_{n,j}|^2}{4}}$. We compare $B_{\text{eff}}(\vec{k})$ with the Berry curvature obtained from the continuum model of the same system in Fig.~\ref{fig:berry_com} and observe that the first-order approximation already provides qualitative agreement with the continuum model. 

In order to compute $\overline{\mathbb{O}}_{gL},~\overline{\mathbb{S}}_{gL}$ in Eq.~\ref{eq:O_gl} and \ref{eq:S_gl}, we need the numerical values of $\overline{\mathcal{S}}_0(\vec{k}_j)$ and $\bar{f}_0(\vec{k}_j)$. These can be obtained from the known Monte-Carlo results for $\bar{\mathcal{S}}_0(\vec{q})$, and the relationship between $\bar{\mathcal{S}}_0(\vec{q})$ and $\bar{f}_0(\vec{q})$ for the Laughlin state at $\nu = 1/3$   {(see Table I and related equations in Ref.~\cite{GMP})}. Here we use the $1/3$ generalized Laughlin state to estimate the intra-band transition of the fractional Chern insulator at filling fraction $-2/3$ of the twisted bilayer TMD systems \cite{FaiFCI,xiaodong}. Counting the electron filling, $\nu=-2/3$ corresponds to a fully filled valence band for one valley, which is inert given that valley degrees of freedom (d.o.f.) is a good quantum number, and a $1/3$ filled valence band for the other valley. 
 
To take into account multiple momentum shells, we replace $\frac{k_{j,x}^2}{|G|^4}$ by $\frac{k_{n,j,x}^2}{|\vec{k}_{n,j}|^4}$, and obtain ${\overline{\mathbb{O}}}_{gL} \approx 2.73\times 10^{-4}  e^2/\hbar$ and ${\overline{\mathbb{S}}}_{gL} \approx 2.76 \times 10^{-5} (e^2/ \epsilon l_B)$, where the interaction is taken to be Coulomb, $V(r) = e^2/\epsilon r$, with $\epsilon$ being the dielectric constant. Here we ignore the contribution from the gate screening since $|\vec{k}_1| d \gg 1$ for the first momentum shell and a typical gate distance 300 $\mathring{A}$ \cite{xiaodong}.
Interestingly, both of these optical spectral weights are numerically ``small" in the relevant units for a weak $(\delta B)$, suggesting that most of the absorption would be associated with excitations to the remote degrees of freedom, contained in $\mathbb{O}-\overline{\mathbb{O}}$. This is in line with previous complementary analysis that highlighted the ``dark" nature of the FCI states in the moir\'e setting \cite{WolfAMD}. 

The strength of the effective magnetic field does not depend on the twist angle except for an overall scaling factor that does not affect ${\overline{\mathbb{O}}}_{gL}$ and ${\overline{\mathbb{S}}}_{gL}$. However, the twist angle controls the deviation of the periodic LLL limit \cite{WolfAMD}. As long as the FCI state still persists, the leading order perturbation can be viewed as a periodic potential applied to LLL and the discussion in the previous section follows. To give an estimate of the magnitude, the strength of the periodic potential $V_{\vec{G}}$ is of the same order as the band-width, which depends on the deviation of the twist angle $\theta$ from the angle where the bare bandwidth is minimized. The gap $\Delta_{\vec{G}} \approx 0.2 (e^2/ \epsilon l_B)$ from SMA. For $\theta = 3.7^\circ$ and dielectric constant $\epsilon = 8$ \cite{xiaodong}, $\Delta_{\vec{G}} \approx 6.6$ meV. Hence for a bandwidth $\sim 10~$meV, $V_{\vec{G}}/\Delta_{\vec{G}} \sim O(1)$. Therefore, the dispersion induced by twist angle contributes to $\overline{\mathbb{O}}\sim \overline{\mathcal{S}}_0(\vec{G}) e^2/\hbar = 0.021 e^2/\hbar$ and $\overline{\mathbb{S}} \sim V_{\vec{G}} \overline{\mathcal{S}}_0(\vec{G}) \sim 0.21~$meV. The reduction of the spectral weight from the ``bare" bandwidth comes from the smallness of the projected structure factor $\overline{\mathcal{S}}_0(\vec{G})$.

  {
\subsubsection{Comparison of $\overline{\mathbb{O}}$ and $\mathbb{O}$}
In the LLL, due to Kohn's theorem, $\overline{\mathbb{O}} = 0$ for optical absorption below the cyclotron frequency. The quantity $\mathbb{O}$, due to inter LL transitions, can be related to the (unprojected) structure factor $\mathcal{S}(\q)$. Since the quantum metric is uniform, $\mathbb{O} = \frac{e^2}{2\hbar} \nu$, where $\nu$ is the filling fraction \cite{GMP,can2014prl}.

In the case of FCI where the active bands mimic the LLL, the $\overline{\mathbb{O}}$ that we have calculated in the previous two subsections should be understood as the intraband absorption of the partially filled active band. If we ignore the contribution to the optical absorption from completed-filled remote bands, the contribution from the FCI would be the same as the LLL case, $\mathbb{O}^{\text{act}} = \frac{e^2}{2\hbar} \nu$. This is because the active band satisfies the ideal quantum geometry \cite{wangcano} and the topological bound for $\mathbb{O}^{\text{act}}$ is therefore saturated \cite{OnishiFu24c}. A recent numerical study of $\mathbb{O}^{\text{act}}$ for twisted TMD verifies the closeness to the bound around magic angle \cite{zaklama2025structurefactortopologicalbound}. Again, isolating the $\mathbb{O}^{\text{act}}$ part from the total $\mathbb{O}$, that includes contribution from remote filled bands is not practical once the optical frequency starts to allow for remote band transitions.
}

\subsection{Correlated charge-density wave insulator in trivial flat-bands}
\label{sec:chiral}
Let us now turn our attention to topologically trivial flat-bands which display  intertwined phases driven by interaction effects near commensurate fillings. In this regard, one of us has recently analyzed a model of spinful electrons ($s=\uparrow,\downarrow$) with two orbitals ($\l=1,2$) on a square lattice (with spacing, $a$, set to unity) using non-perturbative quantum Monte-Carlo (QMC) computations \cite{chiralPRL}. The Hamiltonian is given by, $H=H_K + H_I$, where
\begin{subequations}
\beq\label{eq:Ham0}
H_K &=& -t \sum_{\k} \hat{\mathbf{c}}^{\dagger}_{\k} \left(\tau_x \sin{\alpha_{\k}} + \sigma_z \tau_y \cos{\alpha_{\k}} + \mu \tau_0 \right) \hat{\mathbf{c}}^{\phantom{\dagger}}_{\k}, \nn\\ \\
\alpha_{\k}&=&\zeta (\cos{k_x }+\cos{k_y }),\nn\\
H_I &=& - \frac{U}{2} \sum_{\r,\l}  \delta \hat{n}_{\r,\l}^2
    + V \sum_{\langle \r,\r'\rangle,\l} \delta \hat{n}_{\r,\l} ~\delta \hat{n}_{\r',\l} \,. \label{chiral}
\eeq
\end{subequations}
We have combined the electronic operators into the vector, $\hat{\mathbf{c}}^{\dagger}_{\k}$, with the Pauli-matrices $\hat\sigma$ and $\hat\tau$ acting on the spin and orbital indices, respectively. Note that $H_K$ hosts two completely flat bands with energies, $\ve_\k=\pm t$, regardless of the parameter, $\zeta$. The latter controls the minimal spatial extent of the localized Wannier functions with an integrated quantum metric, $\zeta^2 /2$. The local interaction terms acting on the shifted density operator, $\delta \hat{n}_{\r,l} = \sum_s \hat{c}^{\dagger}_{\r,l,s}\hat{c}^{\phantom{\dagger}}_{\r,l,s} - 1$, include an attraction, $U$, and nearest-neighbor repulsion, $V$, which controls the competition between distinct many-body ground-states. In the non-interacting limit at a finite filling ($\nu$), the only insulating phases are band-insulators at $\nu=2$ associated with a fully filled $\ve_\k=-t$ band with the chemical potential inside the bandgap.  Including all of the inter-band transitions between the filled and empty states ($\ve_\k=t$ band) leads to,
\beq
\mathbb{O} = 2 \frac{\pi  e^2}{\hbar} \sum_k (\partial_x \alpha_k)^2 = \frac{\pi e^2 \zeta^2}{\hbar},
\label{chiralSWM}
\eeq
in line with the expectations of the standard SWM result in this simplified setting of just two bands (each of which is two-fold spin degenerate). Clearly in the same band-insulating state a weak $U,~V$ (compared to the bandgap) does not affect the result.

\begin{figure}[pth!]
\centering
\includegraphics[width=\linewidth]{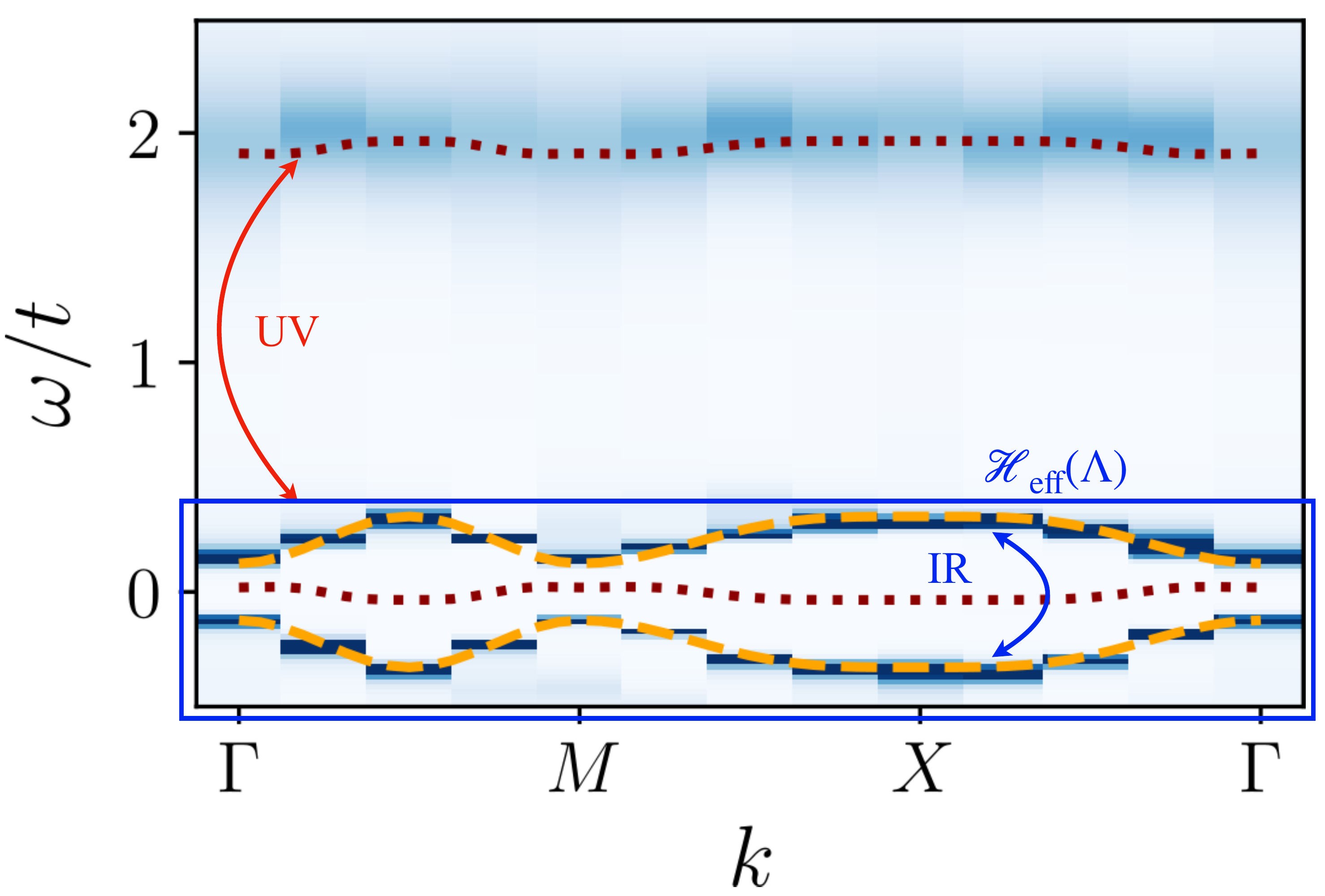} 
\caption{The gapped single-electron spectral functions for the model in Eq.~\ref{chiral} obtained via (analytically continued) QMC computations in the CDW phase at $\nu=1$ (see Ref.~\cite{chiralPRL} for details). The microscopic parameters are $U/t=1,~V/t=0.08$. Red dotted lines denote the original flat bands while the dashed orange lines represent the interaction-renormalized bands. In the band-insulator at $\nu=2$, the UV interband transitions determine $\mathbb{O}$ as in Eq.~\ref{chiralSWM}. In the effective theory, ${\cal{H}}_{\rm{eff}}$, the inter-band transitions within the low-energy subspace controls $\overline{\mathbb{O}}$ (Eq.~\ref{projSWMchiral}).} 
\label{fig:chiral}
\end{figure}

Using QMC, we have previously obtained the numerically exact many-body phase-diagram in the ``projected" limit where $\{V,U\}\lesssim E_{\tn{gap}}(=2t)$ for the entire range of filling of the lower flat-band ($\ve_\k=-t$), as a function of $\zeta$ and different ratios of $V/U$. Notably, for a finite $\zeta$ and $V=0$, and for $U/t=1$ the ground state is a superconductor at all fillings $0<\nu<2$. However, for $V\ll U$ and half-filling of the lower band ($\nu=1$), the ground-state is a charge-density wave (CDW) insulator of pre-formed Cooper-pairs. We are thus interested in $\overline{\mathbb{O}}$ for the insulating state when $\Lambda$ lies inside the bandgap to the remote bands, i.e. the remote bands are projected out (Fig.~\ref{fig:chiral}). Note that for a finite doping, $\delta\nu$, away from this commensurate filling the ground-state is a ``supersolid" (SS), which implies that $\overline{\mathbb{O}}$ diverges with $\delta\nu=|\nu-1|$. 

Firstly, in the limit of $\zeta\rightarrow0$, which is effectively a decoupled ``classical" limit, $\overline{O}\rightarrow0$ vanishes trivially; a chiral symmetry forbids an $O(\zeta)$ contribution. Second, since the CDW-insulator at $\nu=1$ melts and yields a superconductor for $V\rightarrow0$, we expect $\overline{\mathbb{O}}$ to diverge in this limit. Finally, the QMC also found a continuous quantum phase transition from a CDW insulator to a SS across a critical $\zeta_c$ at a fixed $V$ and filling $\nu=1$. This transition must also be accompanied by a diverging $\overline{\mathbb{O}}$. Combining all of these constraints at a fixed $\nu=1$, we expect,
\beq
\overline{\mathbb{O}} = \zeta^2|\zeta_c-\zeta|^{-|\alpha|} {\cal{G}}_\zeta\bigg(\frac{V}{U}\bigg),
\label{projSWMchiral}
\eeq
where ${\cal{G}}_\zeta(x\rightarrow0)\rightarrow\infty$ exhibits a smooth dependence on $\zeta$, and $|\alpha|$ is a universal critical exponent. Notably, once again, we observe a complex low-energy absorption spectral weight that is not simply related to the bare quantum metric, $\zeta$. Moreover, given that the CDW to SS transition does not exhibit any single-particle gap-closing (Fig.~\ref{fig:chiral}), or any singular features in the single-particle response, the above quantity can not be interpreted simply as the quantum geometry of the interaction-renormalized bands either (Fig.~\ref{fig:chiral}).   {Again, the full $\mathbb{O}$ includes the contribution from $\overline{\mathbb{O}}$ and the optical transitions to the remote bands $\mathbb{O}_{\text{int}}$ (the red arrow in Fig. \ref{fig:chiral}). The specific value of $\mathbb{O}_{\text{int}}$ depends on the details of the insulating state, but we expect it to be on the order of the value of the band insulator given by Eq.~\ref{chiralSWM}. }

\section{Outlook}
\label{sec:outlook}
Electronic insulators that arise due to strong interaction-induced effects at a partial band-filling represent one of the most fascinating quantum phases of matter. Their low-energy optical absorption encodes useful information about the many-body spectrum and the nature of dipole transitions between the filled and empty states, respectively. In this manuscript, we have analyzed an optical sum-rule for insulators associated with a low-energy effective Hamiltonian in a variety of interacting ``flat-band" systems hosting non-trivial band-geometry. By focusing on solvable examples in the non-perturbative limit where the interaction scale overwhelms the bare kinetic energy, we have systematically disentangled the effects of emergent symmetries tied to the ground-state manifold and their band quantum geometry on the low-energy sum-rule. An interesting future direction is to extend these ideas to the realm of non-linear response functions and associated sum-rules \cite{oshikawaNL,jankowski2024quantized, avdoshkin2024multi}.

Investigating the fate of these low-energy sum-rules in correlated Mott insulators, including those exhibiting fractionalization \cite{QSL_rev}, remains an exciting frontier. For a variety of such quantum spin-liquid insulators with a finite charge-gap, the electrically neutral sub-gap excitations contribute to a power-law conductivity that vanishes in the limit of $\omega\rightarrow0$ \cite{Ng,potter,huh}. The exact exponents depend on the dominant dissipation mechanism, and in the absence of any controlled analytical methods can be computed using the Ioffe-Larkin composition rule \cite{IL}. For example, there exist specific spin-liquid ground-states \cite{potter} where $\tn{Re}[\sigma_{\rm{xx}}(\omega)]\sim (e^2t^2\omega^2/hU^4)$, where $t,~U$ represent a characteristic hoping and interaction energy, respectively. Clearly, the low-energy absorption spectral weight restricted to frequencies comparable to the $O(U)$ charge-gap will be finite, and will generically depend on the ratios of the microscopic energy-scales. The connections to any underlying quantum-geometry, which is often ignored in standard discussions of Mott insulators, remains an interesting future direction. It will be especially interesting to extend some of the ideas inspired by the study of quantum Hall-like wavefunctions to pursue this direction, without invoking standard parton-based Ioffe-Larkin based methods,   {and extend the current framework to partially filled higher Landau-levels}. On the experimental front, moir\'e materials present a unique platform to investigate these many-body phases \cite{FaiMIT}, though measurements of their optical response in the energy window of interest remains an outstanding challenge.

We have also revealed an intriguing connection between a specific thermally regulated partial sum-rule and the quantum Fisher information.   {The QFI has been argued to be a sensitive witness to multipartite entanglement both in spin systems and in fermionic systems \cite{hauke2016measuring,hauke21}. In our explicit computations in the context of the partial sum-rules at low-temperatures, the response can either vanish or be small even for ground-states with non-trivial entanglement (e.g. Laughlin-like states). This is consistent with the fact that multipartite entanglement bounds the QFI from below, as a vanishing sum-rule would indicate the multipartite entanglement is at least minimal \cite{QFI12a}.} Sharpening the underlying physical principles that describe the connections between measurable response functions in a restricted energy window, their underlying quantum geometry, and various metrics of quantum entanglement remains an exciting future research direction.

{\it Acknowledgements.-} DC thanks E. Berg, J. Dong, S. Kim, P. Ledwith, O. Lesser, Z. Papic, and A. Reddy for a number of fruitful discussions. We thank J. Hofmann for his help in preparing Fig.~\ref{fig:chiral}. JFMV thanks Y. Kwan and J. Herzog-Arbeitman for fruitful discussions. We also thank Y. Onishi for helpful discussions on connections to some of the results in Ref.~\cite{OnishiFu24a}. The numerical computations in this work are supported in part by a NSF CAREER grant (DMR-2237522), and a Sloan Research Fellowship to DC. This research was supported in part by grant NSF PHY-2309135 to the Kavli Institute for Theoretical Physics (KITP).

{\it Note added:} A recent manuscript has also proposed a connection between the UV sum-rule and quantum Fisher information \cite{BB24}; the two manuscripts have minimal overlap.

\bibliographystyle{apsrev4-2-prx}
\bibliography{refs}

\begin{thebibliography}{157}%
\makeatletter
\providecommand \@ifxundefined [1]{%
 \@ifx{#1\undefined}
}%
\providecommand \@ifnum [1]{%
 \ifnum #1\expandafter \@firstoftwo
 \else \expandafter \@secondoftwo
 \fi
}%
\providecommand \@ifx [1]{%
 \ifx #1\expandafter \@firstoftwo
 \else \expandafter \@secondoftwo
 \fi
}%
\providecommand \natexlab [1]{#1}%
\providecommand \enquote  [1]{``#1''}%
\providecommand \bibnamefont  [1]{#1}%
\providecommand \bibfnamefont [1]{#1}%
\providecommand \citenamefont [1]{#1}%
\providecommand \href@noop [0]{\@secondoftwo}%
\providecommand \href [0]{\begingroup \@sanitize@url \@href}%
\providecommand \@href[1]{\@@startlink{#1}\@@href}%
\providecommand \@@href[1]{\endgroup#1\@@endlink}%
\providecommand \@sanitize@url [0]{\catcode `\\12\catcode `\$12\catcode
  `\&12\catcode `\#12\catcode `\^12\catcode `\_12\catcode `\%12\relax}%
\providecommand \@@startlink[1]{}%
\providecommand \@@endlink[0]{}%
\providecommand \url  [0]{\begingroup\@sanitize@url \@url }%
\providecommand \@url [1]{\endgroup\@href {#1}{\urlprefix }}%
\providecommand \urlprefix  [0]{URL }%
\providecommand \Eprint [0]{\href }%
\providecommand \doibase [0]{https://doi.org/}%
\providecommand \selectlanguage [0]{\@gobble}%
\providecommand \bibinfo  [0]{\@secondoftwo}%
\providecommand \bibfield  [0]{\@secondoftwo}%
\providecommand \translation [1]{[#1]}%
\providecommand \BibitemOpen [0]{}%
\providecommand \bibitemStop [0]{}%
\providecommand \bibitemNoStop [0]{.\EOS\space}%
\providecommand \EOS [0]{\spacefactor3000\relax}%
\providecommand \BibitemShut  [1]{\csname bibitem#1\endcsname}%
\let\auto@bib@innerbib\@empty
\bibitem [{\citenamefont {Martin}(2020)}]{martin2020electronic}%
  \BibitemOpen
  \bibfield  {author} {\bibinfo {author} {\bibfnamefont {R.~M.}\ \bibnamefont
  {Martin}},\ }\href@noop {} {\emph {\bibinfo {title} {Electronic structure:
  basic theory and practical methods}}}\ (\bibinfo  {publisher} {Cambridge
  university press},\ \bibinfo {address} {Cambridge, UK},\ \bibinfo {year}
  {2020})\BibitemShut {NoStop}%
\bibitem [{\citenamefont {Imada}\ \emph {et~al.}(1998)\citenamefont {Imada},
  \citenamefont {Fujimori},\ and\ \citenamefont {Tokura}}]{revMIT}%
  \BibitemOpen
  \bibfield  {author} {\bibinfo {author} {\bibfnamefont {M.}~\bibnamefont
  {Imada}}, \bibinfo {author} {\bibfnamefont {A.}~\bibnamefont {Fujimori}},\
  and\ \bibinfo {author} {\bibfnamefont {Y.}~\bibnamefont {Tokura}},\
  }\bibfield  {title} {\emph {\bibinfo {title} {Metal-insulator transitions}},\
  }\href {https://doi.org/10.1103/RevModPhys.70.1039} {\bibfield  {journal}
  {\bibinfo  {journal} {Rev. Mod. Phys.}\ }\textbf {\bibinfo {volume} {70}},\
  \bibinfo {pages} {1039} (\bibinfo {year} {1998})}\BibitemShut {NoStop}%
\bibitem [{\citenamefont {Lee}\ and\ \citenamefont
  {Ramakrishnan}(1985)}]{LeeRama}%
  \BibitemOpen
  \bibfield  {author} {\bibinfo {author} {\bibfnamefont {P.~A.}\ \bibnamefont
  {Lee}}\ and\ \bibinfo {author} {\bibfnamefont {T.~V.}\ \bibnamefont
  {Ramakrishnan}},\ }\bibfield  {title} {\emph {\bibinfo {title} {Disordered
  electronic systems}},\ }\href {https://doi.org/10.1103/RevModPhys.57.287}
  {\bibfield  {journal} {\bibinfo  {journal} {Rev. Mod. Phys.}\ }\textbf
  {\bibinfo {volume} {57}},\ \bibinfo {pages} {287} (\bibinfo {year}
  {1985})}\BibitemShut {NoStop}%
\bibitem [{\citenamefont {Hasan}\ and\ \citenamefont {Kane}(2010)}]{revTI}%
  \BibitemOpen
  \bibfield  {author} {\bibinfo {author} {\bibfnamefont {M.~Z.}\ \bibnamefont
  {Hasan}}\ and\ \bibinfo {author} {\bibfnamefont {C.~L.}\ \bibnamefont
  {Kane}},\ }\bibfield  {title} {\emph {\bibinfo {title} {Colloquium:
  Topological insulators}},\ }\href
  {https://doi.org/10.1103/RevModPhys.82.3045} {\bibfield  {journal} {\bibinfo
  {journal} {Rev. Mod. Phys.}\ }\textbf {\bibinfo {volume} {82}},\ \bibinfo
  {pages} {3045} (\bibinfo {year} {2010})}\BibitemShut {NoStop}%
\bibitem [{\citenamefont {Kohn}(1964)}]{kohn}%
  \BibitemOpen
  \bibfield  {author} {\bibinfo {author} {\bibfnamefont {W.}~\bibnamefont
  {Kohn}},\ }\bibfield  {title} {\emph {\bibinfo {title} {Theory of the
  insulating state}},\ }\href {https://doi.org/10.1103/PhysRev.133.A171}
  {\bibfield  {journal} {\bibinfo  {journal} {Phys. Rev.}\ }\textbf {\bibinfo
  {volume} {133}},\ \bibinfo {pages} {A171} (\bibinfo {year}
  {1964})}\BibitemShut {NoStop}%
\bibitem [{\citenamefont {Scalapino}\ \emph {et~al.}(1993)\citenamefont
  {Scalapino}, \citenamefont {White},\ and\ \citenamefont {Zhang}}]{scalapino}%
  \BibitemOpen
  \bibfield  {author} {\bibinfo {author} {\bibfnamefont {D.~J.}\ \bibnamefont
  {Scalapino}}, \bibinfo {author} {\bibfnamefont {S.~R.}\ \bibnamefont
  {White}},\ and\ \bibinfo {author} {\bibfnamefont {S.}~\bibnamefont {Zhang}},\
  }\bibfield  {title} {\emph {\bibinfo {title} {Insulator, metal, or
  superconductor: The criteria}},\ }\href
  {https://doi.org/10.1103/PhysRevB.47.7995} {\bibfield  {journal} {\bibinfo
  {journal} {Phys. Rev. B}\ }\textbf {\bibinfo {volume} {47}},\ \bibinfo
  {pages} {7995} (\bibinfo {year} {1993})}\BibitemShut {NoStop}%
\bibitem [{\citenamefont {Souza}\ \emph {et~al.}(2000)\citenamefont {Souza},
  \citenamefont {Wilkens},\ and\ \citenamefont {Martin}}]{SWM}%
  \BibitemOpen
  \bibfield  {author} {\bibinfo {author} {\bibfnamefont {I.}~\bibnamefont
  {Souza}}, \bibinfo {author} {\bibfnamefont {T.}~\bibnamefont {Wilkens}},\
  and\ \bibinfo {author} {\bibfnamefont {R.~M.}\ \bibnamefont {Martin}},\
  }\bibfield  {title} {\emph {\bibinfo {title} {Polarization and localization
  in insulators: Generating function approach}},\ }\href
  {https://doi.org/10.1103/PhysRevB.62.1666} {\bibfield  {journal} {\bibinfo
  {journal} {Phys. Rev. B}\ }\textbf {\bibinfo {volume} {62}},\ \bibinfo
  {pages} {1666} (\bibinfo {year} {2000})}\BibitemShut {NoStop}%
\bibitem [{\citenamefont {T\"orm\"a}(2023)}]{torma}%
  \BibitemOpen
  \bibfield  {author} {\bibinfo {author} {\bibfnamefont {P.}~\bibnamefont
  {T\"orm\"a}},\ }\bibfield  {title} {\emph {\bibinfo {title} {Essay: Where can
  quantum geometry lead us?}},\ }\href
  {https://doi.org/10.1103/PhysRevLett.131.240001} {\bibfield  {journal}
  {\bibinfo  {journal} {Phys. Rev. Lett.}\ }\textbf {\bibinfo {volume} {131}},\
  \bibinfo {pages} {240001} (\bibinfo {year} {2023})}\BibitemShut {NoStop}%
\bibitem [{\citenamefont {Provost}\ and\ \citenamefont
  {Vallee}(1980)}]{provost}%
  \BibitemOpen
  \bibfield  {author} {\bibinfo {author} {\bibfnamefont {J.~P.}\ \bibnamefont
  {Provost}}\ and\ \bibinfo {author} {\bibfnamefont {G.}~\bibnamefont
  {Vallee}},\ }\bibfield  {title} {\emph {\bibinfo {title} {{Riemannian
  structure on manifolds of quantum states}}},\ }\href@noop {} {\bibfield
  {journal} {\bibinfo  {journal} {Communications in Mathematical Physics}\
  }\textbf {\bibinfo {volume} {76}},\ \bibinfo {pages} {289 } (\bibinfo {year}
  {1980})}\BibitemShut {NoStop}%
\bibitem [{\citenamefont {T{\"o}rm{\"a}}\ \emph {et~al.}(2022)\citenamefont
  {T{\"o}rm{\"a}}, \citenamefont {Peotta},\ and\ \citenamefont
  {Bernevig}}]{QGreview}%
  \BibitemOpen
  \bibfield  {author} {\bibinfo {author} {\bibfnamefont {P.}~\bibnamefont
  {T{\"o}rm{\"a}}}, \bibinfo {author} {\bibfnamefont {S.}~\bibnamefont
  {Peotta}},\ and\ \bibinfo {author} {\bibfnamefont {B.~A.}\ \bibnamefont
  {Bernevig}},\ }\bibfield  {title} {\emph {\bibinfo {title}
  {Superconductivity, superfluidity and quantum geometry in twisted multilayer
  systems}},\ }\href {https://doi.org/10.1038/s42254-022-00466-y} {\bibfield
  {journal} {\bibinfo  {journal} {Nature Reviews Physics}\ }\textbf {\bibinfo
  {volume} {4}},\ \bibinfo {pages} {528} (\bibinfo {year} {2022})}\BibitemShut
  {NoStop}%
\bibitem [{\citenamefont {Roy}(2014)}]{Roy14}%
  \BibitemOpen
  \bibfield  {author} {\bibinfo {author} {\bibfnamefont {R.}~\bibnamefont
  {Roy}},\ }\bibfield  {title} {\emph {\bibinfo {title} {Band geometry of
  fractional topological insulators}},\ }\href
  {https://doi.org/10.1103/PhysRevB.90.165139} {\bibfield  {journal} {\bibinfo
  {journal} {Phys. Rev. B}\ }\textbf {\bibinfo {volume} {90}},\ \bibinfo
  {pages} {165139} (\bibinfo {year} {2014})}\BibitemShut {NoStop}%
\bibitem [{\citenamefont {Parameswaran}\ \emph {et~al.}(2013)\citenamefont
  {Parameswaran}, \citenamefont {Roy},\ and\ \citenamefont {Sondhi}}]{crphys}%
  \BibitemOpen
  \bibfield  {author} {\bibinfo {author} {\bibfnamefont {S.~A.}\ \bibnamefont
  {Parameswaran}}, \bibinfo {author} {\bibfnamefont {R.}~\bibnamefont {Roy}},\
  and\ \bibinfo {author} {\bibfnamefont {S.~L.}\ \bibnamefont {Sondhi}},\
  }\bibfield  {title} {\emph {\bibinfo {title} {Fractional quantum {Hall}
  physics in topological flat bands}},\ }\href
  {https://doi.org/10.1016/j.crhy.2013.04.003} {\bibfield  {journal} {\bibinfo
  {journal} {Comptes Rendus. Physique}\ }\textbf {\bibinfo {volume} {14}},\
  \bibinfo {pages} {816} (\bibinfo {year} {2013})}\BibitemShut {NoStop}%
\bibitem [{\citenamefont {Ledwith}\ \emph {et~al.}(2020)\citenamefont
  {Ledwith}, \citenamefont {Tarnopolsky}, \citenamefont {Khalaf},\ and\
  \citenamefont {Vishwanath}}]{ledwith2020fractional}%
  \BibitemOpen
  \bibfield  {author} {\bibinfo {author} {\bibfnamefont {P.~J.}\ \bibnamefont
  {Ledwith}}, \bibinfo {author} {\bibfnamefont {G.}~\bibnamefont
  {Tarnopolsky}}, \bibinfo {author} {\bibfnamefont {E.}~\bibnamefont
  {Khalaf}},\ and\ \bibinfo {author} {\bibfnamefont {A.}~\bibnamefont
  {Vishwanath}},\ }\bibfield  {title} {\emph {\bibinfo {title} {Fractional
  chern insulator states in twisted bilayer graphene: An analytical
  approach}},\ }\href {https://doi.org/10.1103/PhysRevResearch.2.023237}
  {\bibfield  {journal} {\bibinfo  {journal} {Phys. Rev. Res.}\ }\textbf
  {\bibinfo {volume} {2}},\ \bibinfo {pages} {023237} (\bibinfo {year}
  {2020})}\BibitemShut {NoStop}%
\bibitem [{\citenamefont {Andrews}\ \emph {et~al.}(2024)\citenamefont
  {Andrews}, \citenamefont {Raja}, \citenamefont {Mishra}, \citenamefont
  {Zaletel},\ and\ \citenamefont {Roy}}]{andrews24}%
  \BibitemOpen
  \bibfield  {author} {\bibinfo {author} {\bibfnamefont {B.}~\bibnamefont
  {Andrews}}, \bibinfo {author} {\bibfnamefont {M.}~\bibnamefont {Raja}},
  \bibinfo {author} {\bibfnamefont {N.}~\bibnamefont {Mishra}}, \bibinfo
  {author} {\bibfnamefont {M.~P.}\ \bibnamefont {Zaletel}},\ and\ \bibinfo
  {author} {\bibfnamefont {R.}~\bibnamefont {Roy}},\ }\bibfield  {title} {\emph
  {\bibinfo {title} {Stability of fractional chern insulators with a non-landau
  level continuum limit}},\ }\href
  {https://doi.org/10.1103/PhysRevB.109.245111} {\bibfield  {journal} {\bibinfo
   {journal} {Phys. Rev. B}\ }\textbf {\bibinfo {volume} {109}},\ \bibinfo
  {pages} {245111} (\bibinfo {year} {2024})}\BibitemShut {NoStop}%
\bibitem [{\citenamefont {Ahn}\ \emph {et~al.}(2020)\citenamefont {Ahn},
  \citenamefont {Guo},\ and\ \citenamefont {Nagaosa}}]{ahn20}%
  \BibitemOpen
  \bibfield  {author} {\bibinfo {author} {\bibfnamefont {J.}~\bibnamefont
  {Ahn}}, \bibinfo {author} {\bibfnamefont {G.-Y.}\ \bibnamefont {Guo}},\ and\
  \bibinfo {author} {\bibfnamefont {N.}~\bibnamefont {Nagaosa}},\ }\bibfield
  {title} {\emph {\bibinfo {title} {Low-frequency divergence and quantum
  geometry of the bulk photovoltaic effect in topological semimetals}},\ }\href
  {https://doi.org/10.1103/PhysRevX.10.041041} {\bibfield  {journal} {\bibinfo
  {journal} {Phys. Rev. X}\ }\textbf {\bibinfo {volume} {10}},\ \bibinfo
  {pages} {041041} (\bibinfo {year} {2020})}\BibitemShut {NoStop}%
\bibitem [{\citenamefont {Holder}\ \emph {et~al.}(2020)\citenamefont {Holder},
  \citenamefont {Kaplan},\ and\ \citenamefont {Yan}}]{holder20}%
  \BibitemOpen
  \bibfield  {author} {\bibinfo {author} {\bibfnamefont {T.}~\bibnamefont
  {Holder}}, \bibinfo {author} {\bibfnamefont {D.}~\bibnamefont {Kaplan}},\
  and\ \bibinfo {author} {\bibfnamefont {B.}~\bibnamefont {Yan}},\ }\bibfield
  {title} {\emph {\bibinfo {title} {Consequences of time-reversal-symmetry
  breaking in the light-matter interaction: Berry curvature, quantum metric,
  and diabatic motion}},\ }\href
  {https://doi.org/10.1103/PhysRevResearch.2.033100} {\bibfield  {journal}
  {\bibinfo  {journal} {Phys. Rev. Res.}\ }\textbf {\bibinfo {volume} {2}},\
  \bibinfo {pages} {033100} (\bibinfo {year} {2020})}\BibitemShut {NoStop}%
\bibitem [{\citenamefont {Ma}\ \emph {et~al.}(2021)\citenamefont {Ma},
  \citenamefont {Grushin},\ and\ \citenamefont {Burch}}]{Ma2021}%
  \BibitemOpen
  \bibfield  {author} {\bibinfo {author} {\bibfnamefont {Q.}~\bibnamefont
  {Ma}}, \bibinfo {author} {\bibfnamefont {A.~G.}\ \bibnamefont {Grushin}},\
  and\ \bibinfo {author} {\bibfnamefont {K.~S.}\ \bibnamefont {Burch}},\
  }\bibfield  {title} {\emph {\bibinfo {title} {Topology and geometry under the
  nonlinear electromagnetic spotlight}},\ }\href
  {https://doi.org/10.1038/s41563-021-00992-7} {\bibfield  {journal} {\bibinfo
  {journal} {Nature Materials}\ }\textbf {\bibinfo {volume} {20}},\ \bibinfo
  {pages} {1601} (\bibinfo {year} {2021})}\BibitemShut {NoStop}%
\bibitem [{\citenamefont {Topp}\ \emph {et~al.}(2021)\citenamefont {Topp},
  \citenamefont {Eckhardt}, \citenamefont {Kennes}, \citenamefont {Sentef},\
  and\ \citenamefont {T\"orm\"a}}]{sentef21}%
  \BibitemOpen
  \bibfield  {author} {\bibinfo {author} {\bibfnamefont {G.~E.}\ \bibnamefont
  {Topp}}, \bibinfo {author} {\bibfnamefont {C.~J.}\ \bibnamefont {Eckhardt}},
  \bibinfo {author} {\bibfnamefont {D.~M.}\ \bibnamefont {Kennes}}, \bibinfo
  {author} {\bibfnamefont {M.~A.}\ \bibnamefont {Sentef}},\ and\ \bibinfo
  {author} {\bibfnamefont {P.}~\bibnamefont {T\"orm\"a}},\ }\bibfield  {title}
  {\emph {\bibinfo {title} {Light-matter coupling and quantum geometry in
  moir\'e materials}},\ }\href {https://doi.org/10.1103/PhysRevB.104.064306}
  {\bibfield  {journal} {\bibinfo  {journal} {Phys. Rev. B}\ }\textbf {\bibinfo
  {volume} {104}},\ \bibinfo {pages} {064306} (\bibinfo {year}
  {2021})}\BibitemShut {NoStop}%
\bibitem [{\citenamefont {Orenstein}\ \emph {et~al.}(2021)\citenamefont
  {Orenstein}, \citenamefont {Moore}, \citenamefont {Morimoto}, \citenamefont
  {Torchinsky}, \citenamefont {Harter},\ and\ \citenamefont
  {Hsieh}}]{annrev21}%
  \BibitemOpen
  \bibfield  {author} {\bibinfo {author} {\bibfnamefont {J.}~\bibnamefont
  {Orenstein}}, \bibinfo {author} {\bibfnamefont {J.}~\bibnamefont {Moore}},
  \bibinfo {author} {\bibfnamefont {T.}~\bibnamefont {Morimoto}}, \bibinfo
  {author} {\bibfnamefont {D.}~\bibnamefont {Torchinsky}}, \bibinfo {author}
  {\bibfnamefont {J.}~\bibnamefont {Harter}},\ and\ \bibinfo {author}
  {\bibfnamefont {D.}~\bibnamefont {Hsieh}},\ }\bibfield  {title} {\emph
  {\bibinfo {title} {Topology and symmetry of quantum materials via nonlinear
  optical responses}},\ }\href
  {https://doi.org/https://doi.org/10.1146/annurev-conmatphys-031218-013712}
  {\bibfield  {journal} {\bibinfo  {journal} {Annual Review of Condensed Matter
  Physics}\ }\textbf {\bibinfo {volume} {12}},\ \bibinfo {pages} {247}
  (\bibinfo {year} {2021})}\BibitemShut {NoStop}%
\bibitem [{\citenamefont {Kozii}\ \emph {et~al.}(2021)\citenamefont {Kozii},
  \citenamefont {Avdoshkin}, \citenamefont {Zhong},\ and\ \citenamefont
  {Moore}}]{Moore21}%
  \BibitemOpen
  \bibfield  {author} {\bibinfo {author} {\bibfnamefont {V.}~\bibnamefont
  {Kozii}}, \bibinfo {author} {\bibfnamefont {A.}~\bibnamefont {Avdoshkin}},
  \bibinfo {author} {\bibfnamefont {S.}~\bibnamefont {Zhong}},\ and\ \bibinfo
  {author} {\bibfnamefont {J.~E.}\ \bibnamefont {Moore}},\ }\bibfield  {title}
  {\emph {\bibinfo {title} {Intrinsic anomalous hall conductivity in a
  nonuniform electric field}},\ }\href
  {https://doi.org/10.1103/PhysRevLett.126.156602} {\bibfield  {journal}
  {\bibinfo  {journal} {Phys. Rev. Lett.}\ }\textbf {\bibinfo {volume} {126}},\
  \bibinfo {pages} {156602} (\bibinfo {year} {2021})}\BibitemShut {NoStop}%
\bibitem [{\citenamefont {Ahn}\ \emph {et~al.}(2022)\citenamefont {Ahn},
  \citenamefont {Guo}, \citenamefont {Nagaosa},\ and\ \citenamefont
  {Vishwanath}}]{Ahn2022}%
  \BibitemOpen
  \bibfield  {author} {\bibinfo {author} {\bibfnamefont {J.}~\bibnamefont
  {Ahn}}, \bibinfo {author} {\bibfnamefont {G.-Y.}\ \bibnamefont {Guo}},
  \bibinfo {author} {\bibfnamefont {N.}~\bibnamefont {Nagaosa}},\ and\ \bibinfo
  {author} {\bibfnamefont {A.}~\bibnamefont {Vishwanath}},\ }\bibfield  {title}
  {\emph {\bibinfo {title} {Riemannian geometry of resonant optical
  responses}},\ }\href {https://doi.org/10.1038/s41567-021-01465-z} {\bibfield
  {journal} {\bibinfo  {journal} {Nature Physics}\ }\textbf {\bibinfo {volume}
  {18}},\ \bibinfo {pages} {290} (\bibinfo {year} {2022})}\BibitemShut
  {NoStop}%
\bibitem [{\citenamefont {Chaudhary}\ \emph {et~al.}(2022)\citenamefont
  {Chaudhary}, \citenamefont {Lewandowski},\ and\ \citenamefont
  {Refael}}]{CL22}%
  \BibitemOpen
  \bibfield  {author} {\bibinfo {author} {\bibfnamefont {S.}~\bibnamefont
  {Chaudhary}}, \bibinfo {author} {\bibfnamefont {C.}~\bibnamefont
  {Lewandowski}},\ and\ \bibinfo {author} {\bibfnamefont {G.}~\bibnamefont
  {Refael}},\ }\bibfield  {title} {\emph {\bibinfo {title} {Shift-current
  response as a probe of quantum geometry and electron-electron interactions in
  twisted bilayer graphene}},\ }\href
  {https://doi.org/10.1103/PhysRevResearch.4.013164} {\bibfield  {journal}
  {\bibinfo  {journal} {Phys. Rev. Res.}\ }\textbf {\bibinfo {volume} {4}},\
  \bibinfo {pages} {013164} (\bibinfo {year} {2022})}\BibitemShut {NoStop}%
\bibitem [{\citenamefont {Mitscherling}\ and\ \citenamefont
  {Holder}(2022)}]{holder22}%
  \BibitemOpen
  \bibfield  {author} {\bibinfo {author} {\bibfnamefont {J.}~\bibnamefont
  {Mitscherling}}\ and\ \bibinfo {author} {\bibfnamefont {T.}~\bibnamefont
  {Holder}},\ }\bibfield  {title} {\emph {\bibinfo {title} {Bound on
  resistivity in flat-band materials due to the quantum metric}},\ }\href
  {https://doi.org/10.1103/PhysRevB.105.085154} {\bibfield  {journal} {\bibinfo
   {journal} {Phys. Rev. B}\ }\textbf {\bibinfo {volume} {105}},\ \bibinfo
  {pages} {085154} (\bibinfo {year} {2022})}\BibitemShut {NoStop}%
\bibitem [{\citenamefont {Tai}\ and\ \citenamefont {Claassen}(2023)}]{MC23}%
  \BibitemOpen
  \bibfield  {author} {\bibinfo {author} {\bibfnamefont {W.~T.}\ \bibnamefont
  {Tai}}\ and\ \bibinfo {author} {\bibfnamefont {M.}~\bibnamefont {Claassen}},\
  }\href {https://arxiv.org/abs/2303.01597} {\bibinfo {title}
  {Quantum-geometric light-matter coupling in correlated quantum materials}}
  (\bibinfo {year} {2023}),\ \Eprint {https://arxiv.org/abs/2303.01597}
  {arXiv:2303.01597 [cond-mat.str-el]} \BibitemShut {NoStop}%
\bibitem [{\citenamefont {Adak}\ \emph {et~al.}(2024)\citenamefont {Adak},
  \citenamefont {Sinha}, \citenamefont {Agarwal},\ and\ \citenamefont
  {Deshmukh}}]{deshmukh}%
  \BibitemOpen
  \bibfield  {author} {\bibinfo {author} {\bibfnamefont {P.~C.}\ \bibnamefont
  {Adak}}, \bibinfo {author} {\bibfnamefont {S.}~\bibnamefont {Sinha}},
  \bibinfo {author} {\bibfnamefont {A.}~\bibnamefont {Agarwal}},\ and\ \bibinfo
  {author} {\bibfnamefont {M.~M.}\ \bibnamefont {Deshmukh}},\ }\bibfield
  {title} {\emph {\bibinfo {title} {Tunable moir{\'e} materials for probing
  berry physics and topology}},\ }\href
  {https://doi.org/10.1038/s41578-024-00671-4} {\bibfield  {journal} {\bibinfo
  {journal} {Nature Reviews Materials}\ }\textbf {\bibinfo {volume} {9}},\
  \bibinfo {pages} {481} (\bibinfo {year} {2024})}\BibitemShut {NoStop}%
\bibitem [{\citenamefont {Bouhon}\ \emph {et~al.}(2023)\citenamefont {Bouhon},
  \citenamefont {Timmel},\ and\ \citenamefont {Slager}}]{bouhon2023quantum}%
  \BibitemOpen
  \bibfield  {author} {\bibinfo {author} {\bibfnamefont {A.}~\bibnamefont
  {Bouhon}}, \bibinfo {author} {\bibfnamefont {A.}~\bibnamefont {Timmel}},\
  and\ \bibinfo {author} {\bibfnamefont {R.-J.}\ \bibnamefont {Slager}},\
  }\bibfield  {title} {\emph {\bibinfo {title} {Quantum geometry beyond
  projective single bands}},\ }\href@noop {} {\bibfield  {journal} {\bibinfo
  {journal} {arXiv preprint arXiv:2303.02180}\ } (\bibinfo {year}
  {2023})}\BibitemShut {NoStop}%
\bibitem [{\citenamefont {Jankowski}\ \emph {et~al.}(2023)\citenamefont
  {Jankowski}, \citenamefont {Morris}, \citenamefont {Bouhon}, \citenamefont
  {{\"U}nal},\ and\ \citenamefont {Slager}}]{jankowski2023optical}%
  \BibitemOpen
  \bibfield  {author} {\bibinfo {author} {\bibfnamefont {W.~J.}\ \bibnamefont
  {Jankowski}}, \bibinfo {author} {\bibfnamefont {A.~S.}\ \bibnamefont
  {Morris}}, \bibinfo {author} {\bibfnamefont {A.}~\bibnamefont {Bouhon}},
  \bibinfo {author} {\bibfnamefont {F.~N.}\ \bibnamefont {{\"U}nal}},\ and\
  \bibinfo {author} {\bibfnamefont {R.-J.}\ \bibnamefont {Slager}},\ }\bibfield
   {title} {\emph {\bibinfo {title} {Optical manifestations of topological
  euler class in electronic materials}},\ }\href@noop {} {\bibfield  {journal}
  {\bibinfo  {journal} {arXiv preprint arXiv:2311.07545}\ } (\bibinfo {year}
  {2023})}\BibitemShut {NoStop}%
\bibitem [{\citenamefont {McKay}\ \emph {et~al.}(2024)\citenamefont {McKay},
  \citenamefont {Mahmood},\ and\ \citenamefont {Bradlyn}}]{Bradlyn24a}%
  \BibitemOpen
  \bibfield  {author} {\bibinfo {author} {\bibfnamefont {R.~C.}\ \bibnamefont
  {McKay}}, \bibinfo {author} {\bibfnamefont {F.}~\bibnamefont {Mahmood}},\
  and\ \bibinfo {author} {\bibfnamefont {B.}~\bibnamefont {Bradlyn}},\
  }\bibfield  {title} {\emph {\bibinfo {title} {Charge conservation beyond
  uniformity: Spatially inhomogeneous electromagnetic response in periodic
  solids}},\ }\href {https://doi.org/10.1103/PhysRevX.14.011058} {\bibfield
  {journal} {\bibinfo  {journal} {Phys. Rev. X}\ }\textbf {\bibinfo {volume}
  {14}},\ \bibinfo {pages} {011058} (\bibinfo {year} {2024})}\BibitemShut
  {NoStop}%
\bibitem [{\citenamefont {Bradlyn}\ and\ \citenamefont
  {Abbamonte}(2024)}]{bradlyn24b}%
  \BibitemOpen
  \bibfield  {author} {\bibinfo {author} {\bibfnamefont {B.}~\bibnamefont
  {Bradlyn}}\ and\ \bibinfo {author} {\bibfnamefont {P.}~\bibnamefont
  {Abbamonte}},\ }\href {https://arxiv.org/abs/2404.16144} {\bibinfo {title}
  {Spectral density and sum rules for second-order response functions}}
  (\bibinfo {year} {2024}),\ \Eprint {https://arxiv.org/abs/2404.16144}
  {arXiv:2404.16144 [cond-mat.mes-hall]} \BibitemShut {NoStop}%
\bibitem [{\citenamefont {Jankowski}\ and\ \citenamefont
  {Slager}(2024)}]{jankowski2024quantized}%
  \BibitemOpen
  \bibfield  {author} {\bibinfo {author} {\bibfnamefont {W.~J.}\ \bibnamefont
  {Jankowski}}\ and\ \bibinfo {author} {\bibfnamefont {R.-J.}\ \bibnamefont
  {Slager}},\ }\bibfield  {title} {\emph {\bibinfo {title} {Quantized shift
  response in multi-gap topological phases}},\ }\href@noop {} {\bibfield
  {journal} {\bibinfo  {journal} {arXiv preprint arXiv:2402.13245}\ } (\bibinfo
  {year} {2024})}\BibitemShut {NoStop}%
\bibitem [{\citenamefont {Antebi}\ \emph {et~al.}(2024)\citenamefont {Antebi},
  \citenamefont {Mitscherling},\ and\ \citenamefont
  {Holder}}]{antebi2024drude}%
  \BibitemOpen
  \bibfield  {author} {\bibinfo {author} {\bibfnamefont {O.}~\bibnamefont
  {Antebi}}, \bibinfo {author} {\bibfnamefont {J.}~\bibnamefont
  {Mitscherling}},\ and\ \bibinfo {author} {\bibfnamefont {T.}~\bibnamefont
  {Holder}},\ }\bibfield  {title} {\emph {\bibinfo {title} {The drude weight of
  a flatband metal}},\ }\href@noop {} {\bibfield  {journal} {\bibinfo
  {journal} {arXiv preprint arXiv:2407.09599}\ } (\bibinfo {year}
  {2024})}\BibitemShut {NoStop}%
\bibitem [{\citenamefont {Avdoshkin}\ \emph {et~al.}(2024)\citenamefont
  {Avdoshkin}, \citenamefont {Mitscherling},\ and\ \citenamefont
  {Moore}}]{avdoshkin2024multi}%
  \BibitemOpen
  \bibfield  {author} {\bibinfo {author} {\bibfnamefont {A.}~\bibnamefont
  {Avdoshkin}}, \bibinfo {author} {\bibfnamefont {J.}~\bibnamefont
  {Mitscherling}},\ and\ \bibinfo {author} {\bibfnamefont {J.~E.}\ \bibnamefont
  {Moore}},\ }\bibfield  {title} {\emph {\bibinfo {title} {The multi-state
  geometry of shift current and polarization}},\ }\href@noop {} {\bibfield
  {journal} {\bibinfo  {journal} {arXiv preprint arXiv:2409.16358}\ } (\bibinfo
  {year} {2024})}\BibitemShut {NoStop}%
\bibitem [{\citenamefont {Paul}(2024)}]{NP24}%
  \BibitemOpen
  \bibfield  {author} {\bibinfo {author} {\bibfnamefont {N.}~\bibnamefont
  {Paul}},\ }\bibfield  {title} {\emph {\bibinfo {title} {Area-law entanglement
  from quantum geometry}},\ }\href
  {https://doi.org/10.1103/PhysRevB.109.085146} {\bibfield  {journal} {\bibinfo
   {journal} {Phys. Rev. B}\ }\textbf {\bibinfo {volume} {109}},\ \bibinfo
  {pages} {085146} (\bibinfo {year} {2024})}\BibitemShut {NoStop}%
\bibitem [{\citenamefont {{Tam}}\ \emph {et~al.}(2024)\citenamefont {{Tam}},
  \citenamefont {{Herzog-Arbeitman}},\ and\ \citenamefont {{Yu}}}]{PMT}%
  \BibitemOpen
  \bibfield  {author} {\bibinfo {author} {\bibfnamefont {P.~M.}\ \bibnamefont
  {{Tam}}}, \bibinfo {author} {\bibfnamefont {J.}~\bibnamefont
  {{Herzog-Arbeitman}}},\ and\ \bibinfo {author} {\bibfnamefont
  {J.}~\bibnamefont {{Yu}}},\ }\bibfield  {title} {\emph {\bibinfo {title}
  {{Corner Charge Fluctuation as an Observable for Quantum Geometry and
  Entanglement in Two-dimensional Insulators}}},\ }\href
  {https://doi.org/10.48550/arXiv.2406.17023} {\bibfield  {journal} {\bibinfo
  {journal} {arXiv e-prints}\ ,\ \bibinfo {eid} {arXiv:2406.17023}} (\bibinfo
  {year} {2024})},\ \Eprint {https://arxiv.org/abs/2406.17023}
  {arXiv:2406.17023 [cond-mat.mes-hall]} \BibitemShut {NoStop}%
\bibitem [{\citenamefont {{Wu}}\ \emph {et~al.}(2024)\citenamefont {{Wu}},
  \citenamefont {{Cai}}, \citenamefont {{Cheng}},\ and\ \citenamefont
  {{Kumar}}}]{XCW}%
  \BibitemOpen
  \bibfield  {author} {\bibinfo {author} {\bibfnamefont {X.-C.}\ \bibnamefont
  {{Wu}}}, \bibinfo {author} {\bibfnamefont {K.-L.}\ \bibnamefont {{Cai}}},
  \bibinfo {author} {\bibfnamefont {M.}~\bibnamefont {{Cheng}}},\ and\ \bibinfo
  {author} {\bibfnamefont {P.}~\bibnamefont {{Kumar}}},\ }\bibfield  {title}
  {\emph {\bibinfo {title} {{Corner Charge Fluctuations and Many-Body Quantum
  Geometry}}},\ }\href {https://doi.org/10.48550/arXiv.2408.16057} {\bibfield
  {journal} {\bibinfo  {journal} {arXiv e-prints}\ ,\ \bibinfo {eid}
  {arXiv:2408.16057}} (\bibinfo {year} {2024})},\ \Eprint
  {https://arxiv.org/abs/2408.16057} {arXiv:2408.16057 [cond-mat.str-el]}
  \BibitemShut {NoStop}%
\bibitem [{\citenamefont {{Kruchkov}}\ and\ \citenamefont
  {{Ryu}}(2024)}]{SR24}%
  \BibitemOpen
  \bibfield  {author} {\bibinfo {author} {\bibfnamefont {A.}~\bibnamefont
  {{Kruchkov}}}\ and\ \bibinfo {author} {\bibfnamefont {S.}~\bibnamefont
  {{Ryu}}},\ }\bibfield  {title} {\emph {\bibinfo {title} {{Entanglement
  entropy in lattice models with quantum metric}}},\ }\href
  {https://doi.org/10.48550/arXiv.2408.10314} {\bibfield  {journal} {\bibinfo
  {journal} {arXiv e-prints}\ ,\ \bibinfo {eid} {arXiv:2408.10314}} (\bibinfo
  {year} {2024})},\ \Eprint {https://arxiv.org/abs/2408.10314}
  {arXiv:2408.10314 [cond-mat.str-el]} \BibitemShut {NoStop}%
\bibitem [{\citenamefont {Resta}\ and\ \citenamefont
  {Sorella}(1999)}]{RestaSorella}%
  \BibitemOpen
  \bibfield  {author} {\bibinfo {author} {\bibfnamefont {R.}~\bibnamefont
  {Resta}}\ and\ \bibinfo {author} {\bibfnamefont {S.}~\bibnamefont
  {Sorella}},\ }\bibfield  {title} {\emph {\bibinfo {title} {Electron
  localization in the insulating state}},\ }\href
  {https://doi.org/10.1103/PhysRevLett.82.370} {\bibfield  {journal} {\bibinfo
  {journal} {Phys. Rev. Lett.}\ }\textbf {\bibinfo {volume} {82}},\ \bibinfo
  {pages} {370} (\bibinfo {year} {1999})}\BibitemShut {NoStop}%
\bibitem [{\citenamefont {Resta}(2006)}]{Resta}%
  \BibitemOpen
  \bibfield  {author} {\bibinfo {author} {\bibfnamefont {R.}~\bibnamefont
  {Resta}},\ }\bibfield  {title} {\emph {\bibinfo {title} {Polarization
  fluctuations in insulators and metals: New and old theories merge}},\ }\href
  {https://doi.org/10.1103/PhysRevLett.96.137601} {\bibfield  {journal}
  {\bibinfo  {journal} {Phys. Rev. Lett.}\ }\textbf {\bibinfo {volume} {96}},\
  \bibinfo {pages} {137601} (\bibinfo {year} {2006})}\BibitemShut {NoStop}%
\bibitem [{\citenamefont {{Kudinov}}(1991)}]{Kudinov}%
  \BibitemOpen
  \bibfield  {author} {\bibinfo {author} {\bibfnamefont {E.~K.}\ \bibnamefont
  {{Kudinov}}},\ }\bibfield  {title} {\emph {\bibinfo {title} {{Difference
  between insulating and conducting states}}},\ }\href
  {https://doi.org/10.48550/arXiv.cond-mat/9902361} {\bibfield  {journal}
  {\bibinfo  {journal} {Sov. Phys. Solid State}\ }\textbf {\bibinfo {volume}
  {33}},\ \bibinfo {eid} {cond-mat/9902361} (\bibinfo {year} {1991})},\ \Eprint
  {https://arxiv.org/abs/cond-mat/9902361} {arXiv:cond-mat/9902361
  [cond-mat.supr-con]} \BibitemShut {NoStop}%
\bibitem [{\citenamefont {Onishi}\ and\ \citenamefont
  {Fu}(2024)}]{OnishiFu24a}%
  \BibitemOpen
  \bibfield  {author} {\bibinfo {author} {\bibfnamefont {Y.}~\bibnamefont
  {Onishi}}\ and\ \bibinfo {author} {\bibfnamefont {L.}~\bibnamefont {Fu}},\
  }\bibfield  {title} {\emph {\bibinfo {title} {Fundamental bound on
  topological gap}},\ }\href {https://doi.org/10.1103/PhysRevX.14.011052}
  {\bibfield  {journal} {\bibinfo  {journal} {Phys. Rev. X}\ }\textbf {\bibinfo
  {volume} {14}},\ \bibinfo {pages} {011052} (\bibinfo {year}
  {2024})}\BibitemShut {NoStop}%
\bibitem [{\citenamefont {{Souza}}\ \emph {et~al.}(2024)\citenamefont
  {{Souza}}, \citenamefont {{Martin}},\ and\ \citenamefont
  {{Stengel}}}]{souza24}%
  \BibitemOpen
  \bibfield  {author} {\bibinfo {author} {\bibfnamefont {I.}~\bibnamefont
  {{Souza}}}, \bibinfo {author} {\bibfnamefont {R.~M.}\ \bibnamefont
  {{Martin}}},\ and\ \bibinfo {author} {\bibfnamefont {M.}~\bibnamefont
  {{Stengel}}},\ }\bibfield  {title} {\emph {\bibinfo {title} {{Optical bounds
  on many-electron localization}}},\ }\href
  {https://doi.org/10.48550/arXiv.2407.17908} {\bibfield  {journal} {\bibinfo
  {journal} {arXiv e-prints}\ ,\ \bibinfo {eid} {arXiv:2407.17908}} (\bibinfo
  {year} {2024})},\ \Eprint {https://arxiv.org/abs/2407.17908}
  {arXiv:2407.17908 [cond-mat.mtrl-sci]} \BibitemShut {NoStop}%
\bibitem [{\citenamefont {{Batra}}\ and\ \citenamefont
  {{Feldman}}(2024)}]{feldman}%
  \BibitemOpen
  \bibfield  {author} {\bibinfo {author} {\bibfnamefont {N.}~\bibnamefont
  {{Batra}}}\ and\ \bibinfo {author} {\bibfnamefont {D.~E.}\ \bibnamefont
  {{Feldman}}},\ }\bibfield  {title} {\emph {\bibinfo {title} {{A Bound on
  Topological Gap from Newton's Laws}}},\ }\href
  {https://doi.org/10.48550/arXiv.2407.17603} {\bibfield  {journal} {\bibinfo
  {journal} {arXiv e-prints}\ ,\ \bibinfo {eid} {arXiv:2407.17603}} (\bibinfo
  {year} {2024})},\ \Eprint {https://arxiv.org/abs/2407.17603}
  {arXiv:2407.17603 [cond-mat.mes-hall]} \BibitemShut {NoStop}%
\bibitem [{\citenamefont {{Verma}}\ and\ \citenamefont
  {{Queiroz}}(2024)}]{VQ24a}%
  \BibitemOpen
  \bibfield  {author} {\bibinfo {author} {\bibfnamefont {N.}~\bibnamefont
  {{Verma}}}\ and\ \bibinfo {author} {\bibfnamefont {R.}~\bibnamefont
  {{Queiroz}}},\ }\bibfield  {title} {\emph {\bibinfo {title} {{Instantaneous
  Response and Quantum Geometry of Insulators}}},\ }\href
  {https://doi.org/10.48550/arXiv.2403.07052} {\bibfield  {journal} {\bibinfo
  {journal} {arXiv e-prints}\ ,\ \bibinfo {eid} {arXiv:2403.07052}} (\bibinfo
  {year} {2024})},\ \Eprint {https://arxiv.org/abs/2403.07052}
  {arXiv:2403.07052 [cond-mat.mes-hall]} \BibitemShut {NoStop}%
\bibitem [{\citenamefont {Verma}\ and\ \citenamefont {Queiroz}(2025)}]{VQ24b}%
  \BibitemOpen
  \bibfield  {author} {\bibinfo {author} {\bibfnamefont {N.}~\bibnamefont
  {Verma}}\ and\ \bibinfo {author} {\bibfnamefont {R.}~\bibnamefont
  {Queiroz}},\ }\bibfield  {title} {\emph {\bibinfo {title} {Framework to
  measure quantum metric from step response}},\ }\href
  {https://doi.org/10.1103/PhysRevLett.134.106403} {\bibfield  {journal}
  {\bibinfo  {journal} {Phys. Rev. Lett.}\ }\textbf {\bibinfo {volume} {134}},\
  \bibinfo {pages} {106403} (\bibinfo {year} {2025})}\BibitemShut {NoStop}%
\bibitem [{\citenamefont {{Onishi}}\ and\ \citenamefont
  {{Fu}}(2024{\natexlab{a}})}]{OnishiFu24b}%
  \BibitemOpen
  \bibfield  {author} {\bibinfo {author} {\bibfnamefont {Y.}~\bibnamefont
  {{Onishi}}}\ and\ \bibinfo {author} {\bibfnamefont {L.}~\bibnamefont
  {{Fu}}},\ }\bibfield  {title} {\emph {\bibinfo {title} {{Quantum weight}}},\
  }\href {https://doi.org/10.48550/arXiv.2401.13847} {\bibfield  {journal}
  {\bibinfo  {journal} {arXiv e-prints}\ ,\ \bibinfo {eid} {arXiv:2401.13847}}
  (\bibinfo {year} {2024}{\natexlab{a}})},\ \Eprint
  {https://arxiv.org/abs/2401.13847} {arXiv:2401.13847 [cond-mat.str-el]}
  \BibitemShut {NoStop}%
\bibitem [{\citenamefont {Bistritzer}\ and\ \citenamefont
  {MacDonald}(2011)}]{Bistritzer2011}%
  \BibitemOpen
  \bibfield  {author} {\bibinfo {author} {\bibfnamefont {R.}~\bibnamefont
  {Bistritzer}}\ and\ \bibinfo {author} {\bibfnamefont {A.~H.}\ \bibnamefont
  {MacDonald}},\ }\bibfield  {title} {\emph {\bibinfo {title} {Moir{\'e} bands
  in twisted double-layer graphene}},\ }\href
  {https://doi.org/10.1073/pnas.1108174108} {\bibfield  {journal} {\bibinfo
  {journal} {Proceedings of the National Academy of Sciences}\ }\textbf
  {\bibinfo {volume} {108}},\ \bibinfo {pages} {12233} (\bibinfo {year}
  {2011})},\ \Eprint
  {https://arxiv.org/abs/https://www.pnas.org/content/108/30/12233.full.pdf}
  {https://www.pnas.org/content/108/30/12233.full.pdf} \BibitemShut {NoStop}%
\bibitem [{\citenamefont {Marzari}\ and\ \citenamefont
  {Vanderbilt}(1997)}]{MarzariVanderbilt}%
  \BibitemOpen
  \bibfield  {author} {\bibinfo {author} {\bibfnamefont {N.}~\bibnamefont
  {Marzari}}\ and\ \bibinfo {author} {\bibfnamefont {D.}~\bibnamefont
  {Vanderbilt}},\ }\bibfield  {title} {\emph {\bibinfo {title} {Maximally
  localized generalized wannier functions for composite energy bands}},\ }\href
  {https://doi.org/10.1103/PhysRevB.56.12847} {\bibfield  {journal} {\bibinfo
  {journal} {Phys. Rev. B}\ }\textbf {\bibinfo {volume} {56}},\ \bibinfo
  {pages} {12847} (\bibinfo {year} {1997})}\BibitemShut {NoStop}%
\bibitem [{\citenamefont {Ghosh}\ \emph {et~al.}(2024)\citenamefont {Ghosh},
  \citenamefont {Onishi}, \citenamefont {Xu}, \citenamefont {Lin},
  \citenamefont {Fu},\ and\ \citenamefont {Bansil}}]{Barun}%
  \BibitemOpen
  \bibfield  {author} {\bibinfo {author} {\bibfnamefont {B.}~\bibnamefont
  {Ghosh}}, \bibinfo {author} {\bibfnamefont {Y.}~\bibnamefont {Onishi}},
  \bibinfo {author} {\bibfnamefont {S.-Y.}\ \bibnamefont {Xu}}, \bibinfo
  {author} {\bibfnamefont {H.}~\bibnamefont {Lin}}, \bibinfo {author}
  {\bibfnamefont {L.}~\bibnamefont {Fu}},\ and\ \bibinfo {author}
  {\bibfnamefont {A.}~\bibnamefont {Bansil}},\ }\bibfield  {title} {\emph
  {\bibinfo {title} {Probing quantum geometry through optical conductivity and
  magnetic circular dichroism}},\ }\href
  {https://doi.org/10.1126/sciadv.ado1761} {\bibfield  {journal} {\bibinfo
  {journal} {Science Advances}\ }\textbf {\bibinfo {volume} {10}},\ \bibinfo
  {pages} {eado1761} (\bibinfo {year} {2024})},\ \Eprint
  {https://arxiv.org/abs/https://www.science.org/doi/pdf/10.1126/sciadv.ado1761}
  {https://www.science.org/doi/pdf/10.1126/sciadv.ado1761} \BibitemShut
  {NoStop}%
\bibitem [{\citenamefont {Hyllus}\ \emph {et~al.}(2012)\citenamefont {Hyllus},
  \citenamefont {Laskowski}, \citenamefont {Krischek}, \citenamefont
  {Schwemmer}, \citenamefont {Wieczorek}, \citenamefont {Weinfurter},
  \citenamefont {Pezz\'e},\ and\ \citenamefont {Smerzi}}]{QFI12a}%
  \BibitemOpen
  \bibfield  {author} {\bibinfo {author} {\bibfnamefont {P.}~\bibnamefont
  {Hyllus}}, \bibinfo {author} {\bibfnamefont {W.}~\bibnamefont {Laskowski}},
  \bibinfo {author} {\bibfnamefont {R.}~\bibnamefont {Krischek}}, \bibinfo
  {author} {\bibfnamefont {C.}~\bibnamefont {Schwemmer}}, \bibinfo {author}
  {\bibfnamefont {W.}~\bibnamefont {Wieczorek}}, \bibinfo {author}
  {\bibfnamefont {H.}~\bibnamefont {Weinfurter}}, \bibinfo {author}
  {\bibfnamefont {L.}~\bibnamefont {Pezz\'e}},\ and\ \bibinfo {author}
  {\bibfnamefont {A.}~\bibnamefont {Smerzi}},\ }\bibfield  {title} {\emph
  {\bibinfo {title} {Fisher information and multiparticle entanglement}},\
  }\href {https://doi.org/10.1103/PhysRevA.85.022321} {\bibfield  {journal}
  {\bibinfo  {journal} {Phys. Rev. A}\ }\textbf {\bibinfo {volume} {85}},\
  \bibinfo {pages} {022321} (\bibinfo {year} {2012})}\BibitemShut {NoStop}%
\bibitem [{\citenamefont {T\'oth}(2012)}]{QFI12b}%
  \BibitemOpen
  \bibfield  {author} {\bibinfo {author} {\bibfnamefont {G.}~\bibnamefont
  {T\'oth}},\ }\bibfield  {title} {\emph {\bibinfo {title} {Multipartite
  entanglement and high-precision metrology}},\ }\href
  {https://doi.org/10.1103/PhysRevA.85.022322} {\bibfield  {journal} {\bibinfo
  {journal} {Phys. Rev. A}\ }\textbf {\bibinfo {volume} {85}},\ \bibinfo
  {pages} {022322} (\bibinfo {year} {2012})}\BibitemShut {NoStop}%
\bibitem [{\citenamefont {Strobel}\ \emph {et~al.}(2014)\citenamefont
  {Strobel}, \citenamefont {Muessel}, \citenamefont {Linnemann}, \citenamefont
  {Zibold}, \citenamefont {Hume}, \citenamefont {Pezzè}, \citenamefont
  {Smerzi},\ and\ \citenamefont {Oberthaler}}]{QFI14}%
  \BibitemOpen
  \bibfield  {author} {\bibinfo {author} {\bibfnamefont {H.}~\bibnamefont
  {Strobel}}, \bibinfo {author} {\bibfnamefont {W.}~\bibnamefont {Muessel}},
  \bibinfo {author} {\bibfnamefont {D.}~\bibnamefont {Linnemann}}, \bibinfo
  {author} {\bibfnamefont {T.}~\bibnamefont {Zibold}}, \bibinfo {author}
  {\bibfnamefont {D.~B.}\ \bibnamefont {Hume}}, \bibinfo {author}
  {\bibfnamefont {L.}~\bibnamefont {Pezzè}}, \bibinfo {author} {\bibfnamefont
  {A.}~\bibnamefont {Smerzi}},\ and\ \bibinfo {author} {\bibfnamefont {M.~K.}\
  \bibnamefont {Oberthaler}},\ }\bibfield  {title} {\emph {\bibinfo {title}
  {Fisher information and entanglement of non-gaussian spin states}},\ }\href
  {https://doi.org/10.1126/science.1250147} {\bibfield  {journal} {\bibinfo
  {journal} {Science}\ }\textbf {\bibinfo {volume} {345}},\ \bibinfo {pages}
  {424} (\bibinfo {year} {2014})},\ \Eprint
  {https://arxiv.org/abs/https://www.science.org/doi/pdf/10.1126/science.1250147}
  {https://www.science.org/doi/pdf/10.1126/science.1250147} \BibitemShut
  {NoStop}%
\bibitem [{\citenamefont {Hauke}\ \emph {et~al.}(2016)\citenamefont {Hauke},
  \citenamefont {Heyl}, \citenamefont {Tagliacozzo},\ and\ \citenamefont
  {Zoller}}]{hauke2016measuring}%
  \BibitemOpen
  \bibfield  {author} {\bibinfo {author} {\bibfnamefont {P.}~\bibnamefont
  {Hauke}}, \bibinfo {author} {\bibfnamefont {M.}~\bibnamefont {Heyl}},
  \bibinfo {author} {\bibfnamefont {L.}~\bibnamefont {Tagliacozzo}},\ and\
  \bibinfo {author} {\bibfnamefont {P.}~\bibnamefont {Zoller}},\ }\bibfield
  {title} {\emph {\bibinfo {title} {Measuring multipartite entanglement through
  dynamic susceptibilities}},\ }\href@noop {} {\bibfield  {journal} {\bibinfo
  {journal} {Nature Physics}\ }\textbf {\bibinfo {volume} {12}},\ \bibinfo
  {pages} {778} (\bibinfo {year} {2016})}\BibitemShut {NoStop}%
\bibitem [{\citenamefont {Cao}\ \emph {et~al.}(2018{\natexlab{a}})\citenamefont
  {Cao}, \citenamefont {Fatemi}, \citenamefont {Demir}, \citenamefont {Fang},
  \citenamefont {Tomarken}, \citenamefont {Luo}, \citenamefont
  {Sanchez-Yamagishi}, \citenamefont {Watanabe}, \citenamefont {Taniguchi},
  \citenamefont {Kaxiras}, \citenamefont {Ashoori},\ and\ \citenamefont
  {Jarillo-Herrero}}]{Cao2018}%
  \BibitemOpen
  \bibfield  {author} {\bibinfo {author} {\bibfnamefont {Y.}~\bibnamefont
  {Cao}}, \bibinfo {author} {\bibfnamefont {V.}~\bibnamefont {Fatemi}},
  \bibinfo {author} {\bibfnamefont {A.}~\bibnamefont {Demir}}, \bibinfo
  {author} {\bibfnamefont {S.}~\bibnamefont {Fang}}, \bibinfo {author}
  {\bibfnamefont {S.~L.}\ \bibnamefont {Tomarken}}, \bibinfo {author}
  {\bibfnamefont {J.~Y.}\ \bibnamefont {Luo}}, \bibinfo {author} {\bibfnamefont
  {J.~D.}\ \bibnamefont {Sanchez-Yamagishi}}, \bibinfo {author} {\bibfnamefont
  {K.}~\bibnamefont {Watanabe}}, \bibinfo {author} {\bibfnamefont
  {T.}~\bibnamefont {Taniguchi}}, \bibinfo {author} {\bibfnamefont
  {E.}~\bibnamefont {Kaxiras}}, \bibinfo {author} {\bibfnamefont {R.~C.}\
  \bibnamefont {Ashoori}},\ and\ \bibinfo {author} {\bibfnamefont
  {P.}~\bibnamefont {Jarillo-Herrero}},\ }\bibfield  {title} {\emph {\bibinfo
  {title} {Correlated insulator behaviour at half-filling in magic-angle
  graphene superlattices}},\ }\href {https://doi.org/10.1038/nature26154}
  {\bibfield  {journal} {\bibinfo  {journal} {Nature}\ }\textbf {\bibinfo
  {volume} {556}},\ \bibinfo {pages} {80} (\bibinfo {year}
  {2018}{\natexlab{a}})}\BibitemShut {NoStop}%
\bibitem [{\citenamefont {Cao}\ \emph {et~al.}(2018{\natexlab{b}})\citenamefont
  {Cao}, \citenamefont {Fatemi}, \citenamefont {Fang}, \citenamefont
  {Watanabe}, \citenamefont {Taniguchi}, \citenamefont {Kaxiras},\ and\
  \citenamefont {Jarillo-Herrero}}]{Cao2018b}%
  \BibitemOpen
  \bibfield  {author} {\bibinfo {author} {\bibfnamefont {Y.}~\bibnamefont
  {Cao}}, \bibinfo {author} {\bibfnamefont {V.}~\bibnamefont {Fatemi}},
  \bibinfo {author} {\bibfnamefont {S.}~\bibnamefont {Fang}}, \bibinfo {author}
  {\bibfnamefont {K.}~\bibnamefont {Watanabe}}, \bibinfo {author}
  {\bibfnamefont {T.}~\bibnamefont {Taniguchi}}, \bibinfo {author}
  {\bibfnamefont {E.}~\bibnamefont {Kaxiras}},\ and\ \bibinfo {author}
  {\bibfnamefont {P.}~\bibnamefont {Jarillo-Herrero}},\ }\bibfield  {title}
  {\emph {\bibinfo {title} {Unconventional superconductivity in magic-angle
  graphene superlattices}},\ }\href {https://doi.org/10.1038/nature26160}
  {\bibfield  {journal} {\bibinfo  {journal} {Nature}\ }\textbf {\bibinfo
  {volume} {556}},\ \bibinfo {pages} {43} (\bibinfo {year}
  {2018}{\natexlab{b}})}\BibitemShut {NoStop}%
\bibitem [{\citenamefont {Yankowitz}\ \emph {et~al.}(2019)\citenamefont
  {Yankowitz}, \citenamefont {Chen}, \citenamefont {Polshyn}, \citenamefont
  {Zhang}, \citenamefont {Watanabe}, \citenamefont {Taniguchi}, \citenamefont
  {Graf}, \citenamefont {Young},\ and\ \citenamefont {Dean}}]{Yankowitz_2019}%
  \BibitemOpen
  \bibfield  {author} {\bibinfo {author} {\bibfnamefont {M.}~\bibnamefont
  {Yankowitz}}, \bibinfo {author} {\bibfnamefont {S.}~\bibnamefont {Chen}},
  \bibinfo {author} {\bibfnamefont {H.}~\bibnamefont {Polshyn}}, \bibinfo
  {author} {\bibfnamefont {Y.}~\bibnamefont {Zhang}}, \bibinfo {author}
  {\bibfnamefont {K.}~\bibnamefont {Watanabe}}, \bibinfo {author}
  {\bibfnamefont {T.}~\bibnamefont {Taniguchi}}, \bibinfo {author}
  {\bibfnamefont {D.}~\bibnamefont {Graf}}, \bibinfo {author} {\bibfnamefont
  {A.~F.}\ \bibnamefont {Young}},\ and\ \bibinfo {author} {\bibfnamefont
  {C.~R.}\ \bibnamefont {Dean}},\ }\bibfield  {title} {\emph {\bibinfo {title}
  {Tuning superconductivity in twisted bilayer graphene}},\ }\href
  {https://doi.org/10.1126/science.aav1910} {\bibfield  {journal} {\bibinfo
  {journal} {Science}\ }\textbf {\bibinfo {volume} {363}},\ \bibinfo {pages}
  {1059} (\bibinfo {year} {2019})}\BibitemShut {NoStop}%
\bibitem [{\citenamefont {Lu}\ \emph {et~al.}(2019)\citenamefont {Lu},
  \citenamefont {Stepanov}, \citenamefont {Yang}, \citenamefont {Xie},
  \citenamefont {Aamir}, \citenamefont {Das}, \citenamefont {Urgell},
  \citenamefont {Watanabe}, \citenamefont {Taniguchi}, \citenamefont {Zhang},
  \citenamefont {Bachtold}, \citenamefont {MacDonald},\ and\ \citenamefont
  {Efetov}}]{Lu2019}%
  \BibitemOpen
  \bibfield  {author} {\bibinfo {author} {\bibfnamefont {X.}~\bibnamefont
  {Lu}}, \bibinfo {author} {\bibfnamefont {P.}~\bibnamefont {Stepanov}},
  \bibinfo {author} {\bibfnamefont {W.}~\bibnamefont {Yang}}, \bibinfo {author}
  {\bibfnamefont {M.}~\bibnamefont {Xie}}, \bibinfo {author} {\bibfnamefont
  {M.~A.}\ \bibnamefont {Aamir}}, \bibinfo {author} {\bibfnamefont
  {I.}~\bibnamefont {Das}}, \bibinfo {author} {\bibfnamefont {C.}~\bibnamefont
  {Urgell}}, \bibinfo {author} {\bibfnamefont {K.}~\bibnamefont {Watanabe}},
  \bibinfo {author} {\bibfnamefont {T.}~\bibnamefont {Taniguchi}}, \bibinfo
  {author} {\bibfnamefont {G.}~\bibnamefont {Zhang}}, \bibinfo {author}
  {\bibfnamefont {A.}~\bibnamefont {Bachtold}}, \bibinfo {author}
  {\bibfnamefont {A.~H.}\ \bibnamefont {MacDonald}},\ and\ \bibinfo {author}
  {\bibfnamefont {D.~K.}\ \bibnamefont {Efetov}},\ }\bibfield  {title} {\emph
  {\bibinfo {title} {Superconductors, orbital magnets and correlated states in
  magic-angle bilayer graphene}},\ }\href
  {https://doi.org/10.1038/s41586-019-1695-0} {\bibfield  {journal} {\bibinfo
  {journal} {Nature}\ }\textbf {\bibinfo {volume} {574}},\ \bibinfo {pages}
  {653} (\bibinfo {year} {2019})}\BibitemShut {NoStop}%
\bibitem [{\citenamefont {Park}\ \emph {et~al.}(2021)\citenamefont {Park},
  \citenamefont {Cao}, \citenamefont {Watanabe}, \citenamefont {Taniguchi},\
  and\ \citenamefont {Jarillo-Herrero}}]{Park_2021}%
  \BibitemOpen
  \bibfield  {author} {\bibinfo {author} {\bibfnamefont {J.~M.}\ \bibnamefont
  {Park}}, \bibinfo {author} {\bibfnamefont {Y.}~\bibnamefont {Cao}}, \bibinfo
  {author} {\bibfnamefont {K.}~\bibnamefont {Watanabe}}, \bibinfo {author}
  {\bibfnamefont {T.}~\bibnamefont {Taniguchi}},\ and\ \bibinfo {author}
  {\bibfnamefont {P.}~\bibnamefont {Jarillo-Herrero}},\ }\bibfield  {title}
  {\emph {\bibinfo {title} {Flavour hund's coupling, chern gaps and charge
  diffusivity in moir{\'e} graphene}},\ }\href
  {https://doi.org/10.1038/s41586-021-03366-w} {\bibfield  {journal} {\bibinfo
  {journal} {Nature}\ }\textbf {\bibinfo {volume} {592}},\ \bibinfo {pages}
  {43} (\bibinfo {year} {2021})}\BibitemShut {NoStop}%
\bibitem [{\citenamefont {Zondiner}\ \emph {et~al.}(2020)\citenamefont
  {Zondiner}, \citenamefont {Rozen}, \citenamefont {Rodan-Legrain},
  \citenamefont {Cao}, \citenamefont {Queiroz}, \citenamefont {Taniguchi},
  \citenamefont {Watanabe}, \citenamefont {Oreg}, \citenamefont {von Oppen},
  \citenamefont {Stern},\ and\ \citenamefont {et~al.}}]{Zondiner_2020}%
  \BibitemOpen
  \bibfield  {author} {\bibinfo {author} {\bibfnamefont {U.}~\bibnamefont
  {Zondiner}}, \bibinfo {author} {\bibfnamefont {A.}~\bibnamefont {Rozen}},
  \bibinfo {author} {\bibfnamefont {D.}~\bibnamefont {Rodan-Legrain}}, \bibinfo
  {author} {\bibfnamefont {Y.}~\bibnamefont {Cao}}, \bibinfo {author}
  {\bibfnamefont {R.}~\bibnamefont {Queiroz}}, \bibinfo {author} {\bibfnamefont
  {T.}~\bibnamefont {Taniguchi}}, \bibinfo {author} {\bibfnamefont
  {K.}~\bibnamefont {Watanabe}}, \bibinfo {author} {\bibfnamefont
  {Y.}~\bibnamefont {Oreg}}, \bibinfo {author} {\bibfnamefont {F.}~\bibnamefont
  {von Oppen}}, \bibinfo {author} {\bibfnamefont {A.}~\bibnamefont {Stern}},\
  and\ \bibinfo {author} {\bibnamefont {et~al.}},\ }\bibfield  {title} {\emph
  {\bibinfo {title} {Cascade of phase transitions and dirac revivals in
  magic-angle graphene}},\ }\href {https://doi.org/10.1038/s41586-020-2373-y}
  {\bibfield  {journal} {\bibinfo  {journal} {Nature}\ }\textbf {\bibinfo
  {volume} {582}},\ \bibinfo {pages} {203} (\bibinfo {year}
  {2020})}\BibitemShut {NoStop}%
\bibitem [{\citenamefont {Uri}\ \emph {et~al.}(2020)\citenamefont {Uri},
  \citenamefont {Grover}, \citenamefont {Cao}, \citenamefont {Crosse},
  \citenamefont {Bagani}, \citenamefont {Rodan-Legrain}, \citenamefont
  {Myasoedov}, \citenamefont {Watanabe}, \citenamefont {Taniguchi},
  \citenamefont {Moon} \emph {et~al.}}]{uri2020mapping}%
  \BibitemOpen
  \bibfield  {author} {\bibinfo {author} {\bibfnamefont {A.}~\bibnamefont
  {Uri}}, \bibinfo {author} {\bibfnamefont {S.}~\bibnamefont {Grover}},
  \bibinfo {author} {\bibfnamefont {Y.}~\bibnamefont {Cao}}, \bibinfo {author}
  {\bibfnamefont {J.~A.}\ \bibnamefont {Crosse}}, \bibinfo {author}
  {\bibfnamefont {K.}~\bibnamefont {Bagani}}, \bibinfo {author} {\bibfnamefont
  {D.}~\bibnamefont {Rodan-Legrain}}, \bibinfo {author} {\bibfnamefont
  {Y.}~\bibnamefont {Myasoedov}}, \bibinfo {author} {\bibfnamefont
  {K.}~\bibnamefont {Watanabe}}, \bibinfo {author} {\bibfnamefont
  {T.}~\bibnamefont {Taniguchi}}, \bibinfo {author} {\bibfnamefont
  {P.}~\bibnamefont {Moon}}, \emph {et~al.},\ }\bibfield  {title} {\emph
  {\bibinfo {title} {Mapping the twist-angle disorder and landau levels in
  magic-angle graphene}},\ }\href@noop {} {\bibfield  {journal} {\bibinfo
  {journal} {Nature}\ }\textbf {\bibinfo {volume} {581}},\ \bibinfo {pages}
  {47} (\bibinfo {year} {2020})}\BibitemShut {NoStop}%
\bibitem [{\citenamefont {Saito}\ \emph {et~al.}(2020)\citenamefont {Saito},
  \citenamefont {Ge}, \citenamefont {Watanabe}, \citenamefont {Taniguchi},\
  and\ \citenamefont {Young}}]{saito2020independent}%
  \BibitemOpen
  \bibfield  {author} {\bibinfo {author} {\bibfnamefont {Y.}~\bibnamefont
  {Saito}}, \bibinfo {author} {\bibfnamefont {J.}~\bibnamefont {Ge}}, \bibinfo
  {author} {\bibfnamefont {K.}~\bibnamefont {Watanabe}}, \bibinfo {author}
  {\bibfnamefont {T.}~\bibnamefont {Taniguchi}},\ and\ \bibinfo {author}
  {\bibfnamefont {A.~F.}\ \bibnamefont {Young}},\ }\bibfield  {title} {\emph
  {\bibinfo {title} {Independent superconductors and correlated insulators in
  twisted bilayer graphene}},\ }\href@noop {} {\bibfield  {journal} {\bibinfo
  {journal} {Nature Physics}\ }\textbf {\bibinfo {volume} {16}},\ \bibinfo
  {pages} {926} (\bibinfo {year} {2020})}\BibitemShut {NoStop}%
\bibitem [{\citenamefont {Saito}\ \emph
  {et~al.}(2021{\natexlab{a}})\citenamefont {Saito}, \citenamefont {Yang},
  \citenamefont {Ge}, \citenamefont {Liu}, \citenamefont {Taniguchi},
  \citenamefont {Watanabe}, \citenamefont {Li}, \citenamefont {Berg},\ and\
  \citenamefont {Young}}]{saito2021isospin}%
  \BibitemOpen
  \bibfield  {author} {\bibinfo {author} {\bibfnamefont {Y.}~\bibnamefont
  {Saito}}, \bibinfo {author} {\bibfnamefont {F.}~\bibnamefont {Yang}},
  \bibinfo {author} {\bibfnamefont {J.}~\bibnamefont {Ge}}, \bibinfo {author}
  {\bibfnamefont {X.}~\bibnamefont {Liu}}, \bibinfo {author} {\bibfnamefont
  {T.}~\bibnamefont {Taniguchi}}, \bibinfo {author} {\bibfnamefont
  {K.}~\bibnamefont {Watanabe}}, \bibinfo {author} {\bibfnamefont
  {J.}~\bibnamefont {Li}}, \bibinfo {author} {\bibfnamefont {E.}~\bibnamefont
  {Berg}},\ and\ \bibinfo {author} {\bibfnamefont {A.~F.}\ \bibnamefont
  {Young}},\ }\bibfield  {title} {\emph {\bibinfo {title} {Isospin pomeranchuk
  effect in twisted bilayer graphene}},\ }\href@noop {} {\bibfield  {journal}
  {\bibinfo  {journal} {Nature}\ }\textbf {\bibinfo {volume} {592}},\ \bibinfo
  {pages} {220} (\bibinfo {year} {2021}{\natexlab{a}})}\BibitemShut {NoStop}%
\bibitem [{\citenamefont {Cao}\ \emph {et~al.}(2021)\citenamefont {Cao},
  \citenamefont {Rodan-Legrain}, \citenamefont {Park}, \citenamefont {Yuan},
  \citenamefont {Watanabe}, \citenamefont {Taniguchi}, \citenamefont
  {Fernandes}, \citenamefont {Fu},\ and\ \citenamefont
  {Jarillo-Herrero}}]{Cao_2021}%
  \BibitemOpen
  \bibfield  {author} {\bibinfo {author} {\bibfnamefont {Y.}~\bibnamefont
  {Cao}}, \bibinfo {author} {\bibfnamefont {D.}~\bibnamefont {Rodan-Legrain}},
  \bibinfo {author} {\bibfnamefont {J.~M.}\ \bibnamefont {Park}}, \bibinfo
  {author} {\bibfnamefont {N.~F.~Q.}\ \bibnamefont {Yuan}}, \bibinfo {author}
  {\bibfnamefont {K.}~\bibnamefont {Watanabe}}, \bibinfo {author}
  {\bibfnamefont {T.}~\bibnamefont {Taniguchi}}, \bibinfo {author}
  {\bibfnamefont {R.~M.}\ \bibnamefont {Fernandes}}, \bibinfo {author}
  {\bibfnamefont {L.}~\bibnamefont {Fu}},\ and\ \bibinfo {author}
  {\bibfnamefont {P.}~\bibnamefont {Jarillo-Herrero}},\ }\bibfield  {title}
  {\emph {\bibinfo {title} {Nematicity and competing orders in superconducting
  magic-angle graphene}},\ }\href {https://doi.org/10.1126/science.abc2836}
  {\bibfield  {journal} {\bibinfo  {journal} {Science}\ }\textbf {\bibinfo
  {volume} {372}},\ \bibinfo {pages} {264} (\bibinfo {year}
  {2021})}\BibitemShut {NoStop}%
\bibitem [{\citenamefont {Liu}\ \emph {et~al.}(2021)\citenamefont {Liu},
  \citenamefont {Wang}, \citenamefont {Watanabe}, \citenamefont {Taniguchi},
  \citenamefont {Vafek},\ and\ \citenamefont {Li}}]{liu2021tuning}%
  \BibitemOpen
  \bibfield  {author} {\bibinfo {author} {\bibfnamefont {X.}~\bibnamefont
  {Liu}}, \bibinfo {author} {\bibfnamefont {Z.}~\bibnamefont {Wang}}, \bibinfo
  {author} {\bibfnamefont {K.}~\bibnamefont {Watanabe}}, \bibinfo {author}
  {\bibfnamefont {T.}~\bibnamefont {Taniguchi}}, \bibinfo {author}
  {\bibfnamefont {O.}~\bibnamefont {Vafek}},\ and\ \bibinfo {author}
  {\bibfnamefont {J.}~\bibnamefont {Li}},\ }\bibfield  {title} {\emph {\bibinfo
  {title} {Tuning electron correlation in magic-angle twisted bilayer graphene
  using coulomb screening}},\ }\href@noop {} {\bibfield  {journal} {\bibinfo
  {journal} {Science}\ }\textbf {\bibinfo {volume} {371}},\ \bibinfo {pages}
  {1261} (\bibinfo {year} {2021})}\BibitemShut {NoStop}%
\bibitem [{\citenamefont {Das}\ \emph {et~al.}(2021)\citenamefont {Das},
  \citenamefont {Lu}, \citenamefont {Herzog-Arbeitman}, \citenamefont {Song},
  \citenamefont {Watanabe}, \citenamefont {Taniguchi}, \citenamefont
  {Bernevig},\ and\ \citenamefont {Efetov}}]{Das_2021}%
  \BibitemOpen
  \bibfield  {author} {\bibinfo {author} {\bibfnamefont {I.}~\bibnamefont
  {Das}}, \bibinfo {author} {\bibfnamefont {X.}~\bibnamefont {Lu}}, \bibinfo
  {author} {\bibfnamefont {J.}~\bibnamefont {Herzog-Arbeitman}}, \bibinfo
  {author} {\bibfnamefont {Z.-D.}\ \bibnamefont {Song}}, \bibinfo {author}
  {\bibfnamefont {K.}~\bibnamefont {Watanabe}}, \bibinfo {author}
  {\bibfnamefont {T.}~\bibnamefont {Taniguchi}}, \bibinfo {author}
  {\bibfnamefont {B.~A.}\ \bibnamefont {Bernevig}},\ and\ \bibinfo {author}
  {\bibfnamefont {D.~K.}\ \bibnamefont {Efetov}},\ }\bibfield  {title} {\emph
  {\bibinfo {title} {Symmetry-broken chern insulators and rashba-like
  landau-level crossings in magic-angle bilayer graphene}},\ }\href@noop {}
  {\bibfield  {journal} {\bibinfo  {journal} {Nature Physics}\ }\textbf
  {\bibinfo {volume} {17}},\ \bibinfo {pages} {710} (\bibinfo {year}
  {2021})}\BibitemShut {NoStop}%
\bibitem [{\citenamefont {Rozen}\ \emph {et~al.}(2021)\citenamefont {Rozen},
  \citenamefont {Park}, \citenamefont {Zondiner}, \citenamefont {Cao},
  \citenamefont {Rodan-Legrain}, \citenamefont {Taniguchi}, \citenamefont
  {Watanabe}, \citenamefont {Oreg}, \citenamefont {Stern}, \citenamefont {Berg}
  \emph {et~al.}}]{rozen2021entropic}%
  \BibitemOpen
  \bibfield  {author} {\bibinfo {author} {\bibfnamefont {A.}~\bibnamefont
  {Rozen}}, \bibinfo {author} {\bibfnamefont {J.~M.}\ \bibnamefont {Park}},
  \bibinfo {author} {\bibfnamefont {U.}~\bibnamefont {Zondiner}}, \bibinfo
  {author} {\bibfnamefont {Y.}~\bibnamefont {Cao}}, \bibinfo {author}
  {\bibfnamefont {D.}~\bibnamefont {Rodan-Legrain}}, \bibinfo {author}
  {\bibfnamefont {T.}~\bibnamefont {Taniguchi}}, \bibinfo {author}
  {\bibfnamefont {K.}~\bibnamefont {Watanabe}}, \bibinfo {author}
  {\bibfnamefont {Y.}~\bibnamefont {Oreg}}, \bibinfo {author} {\bibfnamefont
  {A.}~\bibnamefont {Stern}}, \bibinfo {author} {\bibfnamefont
  {E.}~\bibnamefont {Berg}}, \emph {et~al.},\ }\bibfield  {title} {\emph
  {\bibinfo {title} {Entropic evidence for a pomeranchuk effect in magic-angle
  graphene}},\ }\href@noop {} {\bibfield  {journal} {\bibinfo  {journal}
  {Nature}\ }\textbf {\bibinfo {volume} {592}},\ \bibinfo {pages} {214}
  (\bibinfo {year} {2021})}\BibitemShut {NoStop}%
\bibitem [{\citenamefont {Serlin}\ \emph {et~al.}(2020)\citenamefont {Serlin},
  \citenamefont {Tschirhart}, \citenamefont {Polshyn}, \citenamefont {Zhang},
  \citenamefont {Zhu}, \citenamefont {Watanabe}, \citenamefont {Taniguchi},
  \citenamefont {Balents},\ and\ \citenamefont {Young}}]{Serlin900}%
  \BibitemOpen
  \bibfield  {author} {\bibinfo {author} {\bibfnamefont {M.}~\bibnamefont
  {Serlin}}, \bibinfo {author} {\bibfnamefont {C.~L.}\ \bibnamefont
  {Tschirhart}}, \bibinfo {author} {\bibfnamefont {H.}~\bibnamefont {Polshyn}},
  \bibinfo {author} {\bibfnamefont {Y.}~\bibnamefont {Zhang}}, \bibinfo
  {author} {\bibfnamefont {J.}~\bibnamefont {Zhu}}, \bibinfo {author}
  {\bibfnamefont {K.}~\bibnamefont {Watanabe}}, \bibinfo {author}
  {\bibfnamefont {T.}~\bibnamefont {Taniguchi}}, \bibinfo {author}
  {\bibfnamefont {L.}~\bibnamefont {Balents}},\ and\ \bibinfo {author}
  {\bibfnamefont {A.~F.}\ \bibnamefont {Young}},\ }\bibfield  {title} {\emph
  {\bibinfo {title} {Intrinsic quantized anomalous hall effect in a moir{\'e}
  heterostructure}},\ }\href {https://doi.org/10.1126/science.aay5533}
  {\bibfield  {journal} {\bibinfo  {journal} {Science}\ }\textbf {\bibinfo
  {volume} {367}},\ \bibinfo {pages} {900} (\bibinfo {year}
  {2020})}\BibitemShut {NoStop}%
\bibitem [{\citenamefont {Sharpe}\ \emph {et~al.}(2019)\citenamefont {Sharpe},
  \citenamefont {Fox}, \citenamefont {Barnard}, \citenamefont {Finney},
  \citenamefont {Watanabe}, \citenamefont {Taniguchi}, \citenamefont
  {Kastner},\ and\ \citenamefont {Goldhaber-Gordon}}]{Sharpe_2019}%
  \BibitemOpen
  \bibfield  {author} {\bibinfo {author} {\bibfnamefont {A.~L.}\ \bibnamefont
  {Sharpe}}, \bibinfo {author} {\bibfnamefont {E.~J.}\ \bibnamefont {Fox}},
  \bibinfo {author} {\bibfnamefont {A.~W.}\ \bibnamefont {Barnard}}, \bibinfo
  {author} {\bibfnamefont {J.}~\bibnamefont {Finney}}, \bibinfo {author}
  {\bibfnamefont {K.}~\bibnamefont {Watanabe}}, \bibinfo {author}
  {\bibfnamefont {T.}~\bibnamefont {Taniguchi}}, \bibinfo {author}
  {\bibfnamefont {M.~A.}\ \bibnamefont {Kastner}},\ and\ \bibinfo {author}
  {\bibfnamefont {D.}~\bibnamefont {Goldhaber-Gordon}},\ }\bibfield  {title}
  {\emph {\bibinfo {title} {Emergent ferromagnetism near three-quarters filling
  in twisted bilayer graphene}},\ }\href
  {https://doi.org/10.1126/science.aaw3780} {\bibfield  {journal} {\bibinfo
  {journal} {Science}\ }\textbf {\bibinfo {volume} {365}},\ \bibinfo {pages}
  {605} (\bibinfo {year} {2019})}\BibitemShut {NoStop}%
\bibitem [{\citenamefont {Stepanov}\ \emph {et~al.}(2020)\citenamefont
  {Stepanov}, \citenamefont {Das}, \citenamefont {Lu}, \citenamefont
  {Fahimniya}, \citenamefont {Watanabe}, \citenamefont {Taniguchi},
  \citenamefont {Koppens}, \citenamefont {Lischner}, \citenamefont {Levitov},\
  and\ \citenamefont {Efetov}}]{Stepanov_2020}%
  \BibitemOpen
  \bibfield  {author} {\bibinfo {author} {\bibfnamefont {P.}~\bibnamefont
  {Stepanov}}, \bibinfo {author} {\bibfnamefont {I.}~\bibnamefont {Das}},
  \bibinfo {author} {\bibfnamefont {X.}~\bibnamefont {Lu}}, \bibinfo {author}
  {\bibfnamefont {A.}~\bibnamefont {Fahimniya}}, \bibinfo {author}
  {\bibfnamefont {K.}~\bibnamefont {Watanabe}}, \bibinfo {author}
  {\bibfnamefont {T.}~\bibnamefont {Taniguchi}}, \bibinfo {author}
  {\bibfnamefont {F.~H.~L.}\ \bibnamefont {Koppens}}, \bibinfo {author}
  {\bibfnamefont {J.}~\bibnamefont {Lischner}}, \bibinfo {author}
  {\bibfnamefont {L.}~\bibnamefont {Levitov}},\ and\ \bibinfo {author}
  {\bibfnamefont {D.~K.}\ \bibnamefont {Efetov}},\ }\bibfield  {title} {\emph
  {\bibinfo {title} {Untying the insulating and superconducting orders in
  magic-angle graphene}},\ }\href {https://doi.org/10.1038/s41586-020-2459-6}
  {\bibfield  {journal} {\bibinfo  {journal} {Nature}\ }\textbf {\bibinfo
  {volume} {583}},\ \bibinfo {pages} {375} (\bibinfo {year}
  {2020})}\BibitemShut {NoStop}%
\bibitem [{\citenamefont {Wu}\ \emph {et~al.}(2021)\citenamefont {Wu},
  \citenamefont {Zhang}, \citenamefont {Watanabe}, \citenamefont {Taniguchi},\
  and\ \citenamefont {Andrei}}]{Wu_2021}%
  \BibitemOpen
  \bibfield  {author} {\bibinfo {author} {\bibfnamefont {S.}~\bibnamefont
  {Wu}}, \bibinfo {author} {\bibfnamefont {Z.}~\bibnamefont {Zhang}}, \bibinfo
  {author} {\bibfnamefont {K.}~\bibnamefont {Watanabe}}, \bibinfo {author}
  {\bibfnamefont {T.}~\bibnamefont {Taniguchi}},\ and\ \bibinfo {author}
  {\bibfnamefont {E.~Y.}\ \bibnamefont {Andrei}},\ }\bibfield  {title} {\emph
  {\bibinfo {title} {Chern insulators, van hove singularities and topological
  flat bands in magic-angle twisted bilayer graphene}},\ }\href
  {https://doi.org/10.1038/s41563-020-00911-2} {\bibfield  {journal} {\bibinfo
  {journal} {Nature Materials}\ }\textbf {\bibinfo {volume} {20}},\ \bibinfo
  {pages} {488} (\bibinfo {year} {2021})}\BibitemShut {NoStop}%
\bibitem [{\citenamefont {Saito}\ \emph
  {et~al.}(2021{\natexlab{b}})\citenamefont {Saito}, \citenamefont {Ge},
  \citenamefont {Rademaker}, \citenamefont {Watanabe}, \citenamefont
  {Taniguchi}, \citenamefont {Abanin},\ and\ \citenamefont
  {Young}}]{Saito_2021Hofstadter}%
  \BibitemOpen
  \bibfield  {author} {\bibinfo {author} {\bibfnamefont {Y.}~\bibnamefont
  {Saito}}, \bibinfo {author} {\bibfnamefont {J.}~\bibnamefont {Ge}}, \bibinfo
  {author} {\bibfnamefont {L.}~\bibnamefont {Rademaker}}, \bibinfo {author}
  {\bibfnamefont {K.}~\bibnamefont {Watanabe}}, \bibinfo {author}
  {\bibfnamefont {T.}~\bibnamefont {Taniguchi}}, \bibinfo {author}
  {\bibfnamefont {D.~A.}\ \bibnamefont {Abanin}},\ and\ \bibinfo {author}
  {\bibfnamefont {A.~F.}\ \bibnamefont {Young}},\ }\bibfield  {title} {\emph
  {\bibinfo {title} {Hofstadter subband ferromagnetism and symmetry-broken
  chern insulators in twisted bilayer graphene}},\ }\href
  {https://doi.org/10.1038/s41567-020-01129-4} {\bibfield  {journal} {\bibinfo
  {journal} {Nature Physics}\ }\textbf {\bibinfo {volume} {17}},\ \bibinfo
  {pages} {478} (\bibinfo {year} {2021}{\natexlab{b}})}\BibitemShut {NoStop}%
\bibitem [{\citenamefont {Nuckolls}\ \emph {et~al.}(2023)\citenamefont
  {Nuckolls}, \citenamefont {Lee}, \citenamefont {Oh}, \citenamefont {Wong},
  \citenamefont {Soejima}, \citenamefont {Hong}, \citenamefont {Călugăru},
  \citenamefont {Herzog-Arbeitman}, \citenamefont {Bernevig}, \citenamefont
  {Watanabe}, \citenamefont {Taniguchi}, \citenamefont {Regnault},
  \citenamefont {Zaletel},\ and\ \citenamefont
  {Yazdani}}]{nuckolls2023quantum}%
  \BibitemOpen
  \bibfield  {author} {\bibinfo {author} {\bibfnamefont {K.~P.}\ \bibnamefont
  {Nuckolls}}, \bibinfo {author} {\bibfnamefont {R.~L.}\ \bibnamefont {Lee}},
  \bibinfo {author} {\bibfnamefont {M.}~\bibnamefont {Oh}}, \bibinfo {author}
  {\bibfnamefont {D.}~\bibnamefont {Wong}}, \bibinfo {author} {\bibfnamefont
  {T.}~\bibnamefont {Soejima}}, \bibinfo {author} {\bibfnamefont {J.~P.}\
  \bibnamefont {Hong}}, \bibinfo {author} {\bibfnamefont {D.}~\bibnamefont
  {Călugăru}}, \bibinfo {author} {\bibfnamefont {J.}~\bibnamefont
  {Herzog-Arbeitman}}, \bibinfo {author} {\bibfnamefont {B.~A.}\ \bibnamefont
  {Bernevig}}, \bibinfo {author} {\bibfnamefont {K.}~\bibnamefont {Watanabe}},
  \bibinfo {author} {\bibfnamefont {T.}~\bibnamefont {Taniguchi}}, \bibinfo
  {author} {\bibfnamefont {N.}~\bibnamefont {Regnault}}, \bibinfo {author}
  {\bibfnamefont {M.~P.}\ \bibnamefont {Zaletel}},\ and\ \bibinfo {author}
  {\bibfnamefont {A.}~\bibnamefont {Yazdani}},\ }\href@noop {} {\bibinfo
  {title} {Quantum textures of the many-body wavefunctions in magic-angle
  graphene}} (\bibinfo {year} {2023}),\ \Eprint
  {https://arxiv.org/abs/2303.00024} {arXiv:2303.00024 [cond-mat.mes-hall]}
  \BibitemShut {NoStop}%
\bibitem [{\citenamefont {Grover}\ \emph {et~al.}(2022)\citenamefont {Grover},
  \citenamefont {Bocarsly}, \citenamefont {Uri}, \citenamefont {Stepanov},
  \citenamefont {Battista}, \citenamefont {Roy}, \citenamefont {Xiao},
  \citenamefont {Meltzer}, \citenamefont {Myasoedov}, \citenamefont {Pareek},
  \citenamefont {Watanabe}, \citenamefont {Taniguchi}, \citenamefont {Yan},
  \citenamefont {Stern}, \citenamefont {Berg}, \citenamefont {Efetov},\ and\
  \citenamefont {Zeldov}}]{Grover2022mosaic}%
  \BibitemOpen
  \bibfield  {author} {\bibinfo {author} {\bibfnamefont {S.}~\bibnamefont
  {Grover}}, \bibinfo {author} {\bibfnamefont {M.}~\bibnamefont {Bocarsly}},
  \bibinfo {author} {\bibfnamefont {A.}~\bibnamefont {Uri}}, \bibinfo {author}
  {\bibfnamefont {P.}~\bibnamefont {Stepanov}}, \bibinfo {author}
  {\bibfnamefont {G.~D.}\ \bibnamefont {Battista}}, \bibinfo {author}
  {\bibfnamefont {I.}~\bibnamefont {Roy}}, \bibinfo {author} {\bibfnamefont
  {J.}~\bibnamefont {Xiao}}, \bibinfo {author} {\bibfnamefont {A.~Y.}\
  \bibnamefont {Meltzer}}, \bibinfo {author} {\bibfnamefont {Y.}~\bibnamefont
  {Myasoedov}}, \bibinfo {author} {\bibfnamefont {K.}~\bibnamefont {Pareek}},
  \bibinfo {author} {\bibfnamefont {K.}~\bibnamefont {Watanabe}}, \bibinfo
  {author} {\bibfnamefont {T.}~\bibnamefont {Taniguchi}}, \bibinfo {author}
  {\bibfnamefont {B.}~\bibnamefont {Yan}}, \bibinfo {author} {\bibfnamefont
  {A.}~\bibnamefont {Stern}}, \bibinfo {author} {\bibfnamefont
  {E.}~\bibnamefont {Berg}}, \bibinfo {author} {\bibfnamefont {D.~K.}\
  \bibnamefont {Efetov}},\ and\ \bibinfo {author} {\bibfnamefont
  {E.}~\bibnamefont {Zeldov}},\ }\bibfield  {title} {\emph {\bibinfo {title}
  {Chern mosaic and berry-curvature magnetism in magic-angle graphene}},\
  }\href {https://doi.org/10.1038/s41567-022-01635-7} {\bibfield  {journal}
  {\bibinfo  {journal} {Nature Physics}\ }\textbf {\bibinfo {volume} {18}},\
  \bibinfo {pages} {885} (\bibinfo {year} {2022})}\BibitemShut {NoStop}%
\bibitem [{\citenamefont {Yu}\ \emph {et~al.}(2022)\citenamefont {Yu},
  \citenamefont {Foutty}, \citenamefont {Han}, \citenamefont {Barber},
  \citenamefont {Schattner}, \citenamefont {Watanabe}, \citenamefont
  {Taniguchi}, \citenamefont {Phillips}, \citenamefont {Shen}, \citenamefont
  {Kivelson},\ and\ \citenamefont {Feldman}}]{Yu2022hofstadter}%
  \BibitemOpen
  \bibfield  {author} {\bibinfo {author} {\bibfnamefont {J.}~\bibnamefont
  {Yu}}, \bibinfo {author} {\bibfnamefont {B.~A.}\ \bibnamefont {Foutty}},
  \bibinfo {author} {\bibfnamefont {Z.}~\bibnamefont {Han}}, \bibinfo {author}
  {\bibfnamefont {M.~E.}\ \bibnamefont {Barber}}, \bibinfo {author}
  {\bibfnamefont {Y.}~\bibnamefont {Schattner}}, \bibinfo {author}
  {\bibfnamefont {K.}~\bibnamefont {Watanabe}}, \bibinfo {author}
  {\bibfnamefont {T.}~\bibnamefont {Taniguchi}}, \bibinfo {author}
  {\bibfnamefont {P.}~\bibnamefont {Phillips}}, \bibinfo {author}
  {\bibfnamefont {Z.-X.}\ \bibnamefont {Shen}}, \bibinfo {author}
  {\bibfnamefont {S.~A.}\ \bibnamefont {Kivelson}},\ and\ \bibinfo {author}
  {\bibfnamefont {B.~E.}\ \bibnamefont {Feldman}},\ }\bibfield  {title} {\emph
  {\bibinfo {title} {Correlated hofstadter spectrum and flavour phase diagram
  in magic-angle twisted bilayer graphene}},\ }\href
  {https://doi.org/10.1038/s41567-022-01589-w} {\bibfield  {journal} {\bibinfo
  {journal} {Nature Physics}\ }\textbf {\bibinfo {volume} {18}},\ \bibinfo
  {pages} {825} (\bibinfo {year} {2022})}\BibitemShut {NoStop}%
\bibitem [{\citenamefont {Yu}\ \emph {et~al.}(2023)\citenamefont {Yu},
  \citenamefont {Foutty}, \citenamefont {Kwan}, \citenamefont {Barber},
  \citenamefont {Watanabe}, \citenamefont {Taniguchi}, \citenamefont {Shen},
  \citenamefont {Parameswaran},\ and\ \citenamefont
  {Feldman}}]{Yu2022skyrmion}%
  \BibitemOpen
  \bibfield  {author} {\bibinfo {author} {\bibfnamefont {J.}~\bibnamefont
  {Yu}}, \bibinfo {author} {\bibfnamefont {B.~A.}\ \bibnamefont {Foutty}},
  \bibinfo {author} {\bibfnamefont {Y.~H.}\ \bibnamefont {Kwan}}, \bibinfo
  {author} {\bibfnamefont {M.~E.}\ \bibnamefont {Barber}}, \bibinfo {author}
  {\bibfnamefont {K.}~\bibnamefont {Watanabe}}, \bibinfo {author}
  {\bibfnamefont {T.}~\bibnamefont {Taniguchi}}, \bibinfo {author}
  {\bibfnamefont {Z.-X.}\ \bibnamefont {Shen}}, \bibinfo {author}
  {\bibfnamefont {S.~A.}\ \bibnamefont {Parameswaran}},\ and\ \bibinfo {author}
  {\bibfnamefont {B.~E.}\ \bibnamefont {Feldman}},\ }\bibfield  {title} {\emph
  {\bibinfo {title} {Spin skyrmion gaps as signatures of strong-coupling
  insulators in magic-angle twisted bilayer graphene}},\ }\href
  {https://doi.org/10.1038/s41467-023-42275-6} {\bibfield  {journal} {\bibinfo
  {journal} {Nature Communications}\ }\textbf {\bibinfo {volume} {14}},\
  \bibinfo {pages} {6679} (\bibinfo {year} {2023})}\BibitemShut {NoStop}%
\bibitem [{\citenamefont {Morissette}\ \emph {et~al.}(2023)\citenamefont
  {Morissette}, \citenamefont {Lin}, \citenamefont {Sun}, \citenamefont
  {Zhang}, \citenamefont {Liu}, \citenamefont {Rhodes}, \citenamefont
  {Watanabe}, \citenamefont {Taniguchi}, \citenamefont {Hone}, \citenamefont
  {Pollanen}, \citenamefont {Scheurer}, \citenamefont {Lilly}, \citenamefont
  {Mounce},\ and\ \citenamefont {Li}}]{Morisette2022Hunds}%
  \BibitemOpen
  \bibfield  {author} {\bibinfo {author} {\bibfnamefont {E.}~\bibnamefont
  {Morissette}}, \bibinfo {author} {\bibfnamefont {J.-X.}\ \bibnamefont {Lin}},
  \bibinfo {author} {\bibfnamefont {D.}~\bibnamefont {Sun}}, \bibinfo {author}
  {\bibfnamefont {L.}~\bibnamefont {Zhang}}, \bibinfo {author} {\bibfnamefont
  {S.}~\bibnamefont {Liu}}, \bibinfo {author} {\bibfnamefont {D.}~\bibnamefont
  {Rhodes}}, \bibinfo {author} {\bibfnamefont {K.}~\bibnamefont {Watanabe}},
  \bibinfo {author} {\bibfnamefont {T.}~\bibnamefont {Taniguchi}}, \bibinfo
  {author} {\bibfnamefont {J.}~\bibnamefont {Hone}}, \bibinfo {author}
  {\bibfnamefont {J.}~\bibnamefont {Pollanen}}, \bibinfo {author}
  {\bibfnamefont {M.~S.}\ \bibnamefont {Scheurer}}, \bibinfo {author}
  {\bibfnamefont {M.}~\bibnamefont {Lilly}}, \bibinfo {author} {\bibfnamefont
  {A.}~\bibnamefont {Mounce}},\ and\ \bibinfo {author} {\bibfnamefont
  {J.~I.~A.}\ \bibnamefont {Li}},\ }\bibfield  {title} {\emph {\bibinfo {title}
  {Dirac revivals drive a resonance response in twisted bilayer graphene}},\
  }\href {https://doi.org/10.1038/s41567-023-02060-0} {\bibfield  {journal}
  {\bibinfo  {journal} {Nature Physics}\ }\textbf {\bibinfo {volume} {19}},\
  \bibinfo {pages} {1156} (\bibinfo {year} {2023})}\BibitemShut {NoStop}%
\bibitem [{\citenamefont {Tseng}\ \emph {et~al.}(2022)\citenamefont {Tseng},
  \citenamefont {Ma}, \citenamefont {Liu}, \citenamefont {Watanabe},
  \citenamefont {Taniguchi}, \citenamefont {Chu},\ and\ \citenamefont
  {Yankowitz}}]{Tseng2022nu2QAH}%
  \BibitemOpen
  \bibfield  {author} {\bibinfo {author} {\bibfnamefont {C.-C.}\ \bibnamefont
  {Tseng}}, \bibinfo {author} {\bibfnamefont {X.}~\bibnamefont {Ma}}, \bibinfo
  {author} {\bibfnamefont {Z.}~\bibnamefont {Liu}}, \bibinfo {author}
  {\bibfnamefont {K.}~\bibnamefont {Watanabe}}, \bibinfo {author}
  {\bibfnamefont {T.}~\bibnamefont {Taniguchi}}, \bibinfo {author}
  {\bibfnamefont {J.-H.}\ \bibnamefont {Chu}},\ and\ \bibinfo {author}
  {\bibfnamefont {M.}~\bibnamefont {Yankowitz}},\ }\bibfield  {title} {\emph
  {\bibinfo {title} {Anomalous hall effect at half filling in twisted bilayer
  graphene}},\ }\href {https://doi.org/10.1038/s41567-022-01697-7} {\bibfield
  {journal} {\bibinfo  {journal} {Nature Physics}\ }\textbf {\bibinfo {volume}
  {18}},\ \bibinfo {pages} {1038} (\bibinfo {year} {2022})}\BibitemShut
  {NoStop}%
\bibitem [{\citenamefont {Choi}\ \emph {et~al.}(2019)\citenamefont {Choi},
  \citenamefont {Kemmer}, \citenamefont {Peng}, \citenamefont {Thomson},
  \citenamefont {Arora}, \citenamefont {Polski}, \citenamefont {Zhang},
  \citenamefont {Ren}, \citenamefont {Alicea}, \citenamefont {Refael} \emph
  {et~al.}}]{Choi2019}%
  \BibitemOpen
  \bibfield  {author} {\bibinfo {author} {\bibfnamefont {Y.}~\bibnamefont
  {Choi}}, \bibinfo {author} {\bibfnamefont {J.}~\bibnamefont {Kemmer}},
  \bibinfo {author} {\bibfnamefont {Y.}~\bibnamefont {Peng}}, \bibinfo {author}
  {\bibfnamefont {A.}~\bibnamefont {Thomson}}, \bibinfo {author} {\bibfnamefont
  {H.}~\bibnamefont {Arora}}, \bibinfo {author} {\bibfnamefont
  {R.}~\bibnamefont {Polski}}, \bibinfo {author} {\bibfnamefont
  {Y.}~\bibnamefont {Zhang}}, \bibinfo {author} {\bibfnamefont
  {H.}~\bibnamefont {Ren}}, \bibinfo {author} {\bibfnamefont {J.}~\bibnamefont
  {Alicea}}, \bibinfo {author} {\bibfnamefont {G.}~\bibnamefont {Refael}},
  \emph {et~al.},\ }\bibfield  {title} {\emph {\bibinfo {title} {Electronic
  correlations in twisted bilayer graphene near the magic angle}},\ }\href@noop
  {} {\bibfield  {journal} {\bibinfo  {journal} {Nature Physics}\ }\textbf
  {\bibinfo {volume} {15}},\ \bibinfo {pages} {1174} (\bibinfo {year}
  {2019})}\BibitemShut {NoStop}%
\bibitem [{\citenamefont {Oh}\ \emph {et~al.}(2021)\citenamefont {Oh},
  \citenamefont {Nuckolls}, \citenamefont {Wong}, \citenamefont {Lee},
  \citenamefont {Liu}, \citenamefont {Watanabe}, \citenamefont {Taniguchi},\
  and\ \citenamefont {Yazdani}}]{Oh2021unconventional}%
  \BibitemOpen
  \bibfield  {author} {\bibinfo {author} {\bibfnamefont {M.}~\bibnamefont
  {Oh}}, \bibinfo {author} {\bibfnamefont {K.~P.}\ \bibnamefont {Nuckolls}},
  \bibinfo {author} {\bibfnamefont {D.}~\bibnamefont {Wong}}, \bibinfo {author}
  {\bibfnamefont {R.~L.}\ \bibnamefont {Lee}}, \bibinfo {author} {\bibfnamefont
  {X.}~\bibnamefont {Liu}}, \bibinfo {author} {\bibfnamefont {K.}~\bibnamefont
  {Watanabe}}, \bibinfo {author} {\bibfnamefont {T.}~\bibnamefont
  {Taniguchi}},\ and\ \bibinfo {author} {\bibfnamefont {A.}~\bibnamefont
  {Yazdani}},\ }\bibfield  {title} {\emph {\bibinfo {title} {Evidence for
  unconventional superconductivity in twisted bilayer graphene}},\ }\href
  {https://doi.org/10.1038/s41586-021-04121-x} {\bibfield  {journal} {\bibinfo
  {journal} {Nature}\ }\textbf {\bibinfo {volume} {600}},\ \bibinfo {pages}
  {240} (\bibinfo {year} {2021})}\BibitemShut {NoStop}%
\bibitem [{\citenamefont {Nuckolls}\ \emph {et~al.}(2020)\citenamefont
  {Nuckolls}, \citenamefont {Oh}, \citenamefont {Wong}, \citenamefont {Lian},
  \citenamefont {Watanabe}, \citenamefont {Taniguchi}, \citenamefont
  {Bernevig},\ and\ \citenamefont {Yazdani}}]{Nuckolls2020strongly}%
  \BibitemOpen
  \bibfield  {author} {\bibinfo {author} {\bibfnamefont {K.~P.}\ \bibnamefont
  {Nuckolls}}, \bibinfo {author} {\bibfnamefont {M.}~\bibnamefont {Oh}},
  \bibinfo {author} {\bibfnamefont {D.}~\bibnamefont {Wong}}, \bibinfo {author}
  {\bibfnamefont {B.}~\bibnamefont {Lian}}, \bibinfo {author} {\bibfnamefont
  {K.}~\bibnamefont {Watanabe}}, \bibinfo {author} {\bibfnamefont
  {T.}~\bibnamefont {Taniguchi}}, \bibinfo {author} {\bibfnamefont {B.~A.}\
  \bibnamefont {Bernevig}},\ and\ \bibinfo {author} {\bibfnamefont
  {A.}~\bibnamefont {Yazdani}},\ }\bibfield  {title} {\emph {\bibinfo {title}
  {Strongly correlated chern insulators in magic-angle twisted bilayer
  graphene}},\ }\href {https://doi.org/10.1038/s41586-020-3028-8} {\bibfield
  {journal} {\bibinfo  {journal} {Nature}\ }\textbf {\bibinfo {volume} {588}},\
  \bibinfo {pages} {610} (\bibinfo {year} {2020})}\BibitemShut {NoStop}%
\bibitem [{\citenamefont {Xie}\ \emph {et~al.}(2021)\citenamefont {Xie},
  \citenamefont {Pierce}, \citenamefont {Park}, \citenamefont {Parker},
  \citenamefont {Khalaf}, \citenamefont {Ledwith}, \citenamefont {Cao},
  \citenamefont {Lee}, \citenamefont {Chen}, \citenamefont {Forrester},
  \citenamefont {Watanabe}, \citenamefont {Taniguchi}, \citenamefont
  {Vishwanath}, \citenamefont {Jarillo-Herrero},\ and\ \citenamefont
  {Yacoby}}]{Xie2021fractional}%
  \BibitemOpen
  \bibfield  {author} {\bibinfo {author} {\bibfnamefont {Y.}~\bibnamefont
  {Xie}}, \bibinfo {author} {\bibfnamefont {A.~T.}\ \bibnamefont {Pierce}},
  \bibinfo {author} {\bibfnamefont {J.~M.}\ \bibnamefont {Park}}, \bibinfo
  {author} {\bibfnamefont {D.~E.}\ \bibnamefont {Parker}}, \bibinfo {author}
  {\bibfnamefont {E.}~\bibnamefont {Khalaf}}, \bibinfo {author} {\bibfnamefont
  {P.}~\bibnamefont {Ledwith}}, \bibinfo {author} {\bibfnamefont
  {Y.}~\bibnamefont {Cao}}, \bibinfo {author} {\bibfnamefont {S.~H.}\
  \bibnamefont {Lee}}, \bibinfo {author} {\bibfnamefont {S.}~\bibnamefont
  {Chen}}, \bibinfo {author} {\bibfnamefont {P.~R.}\ \bibnamefont {Forrester}},
  \bibinfo {author} {\bibfnamefont {K.}~\bibnamefont {Watanabe}}, \bibinfo
  {author} {\bibfnamefont {T.}~\bibnamefont {Taniguchi}}, \bibinfo {author}
  {\bibfnamefont {A.}~\bibnamefont {Vishwanath}}, \bibinfo {author}
  {\bibfnamefont {P.}~\bibnamefont {Jarillo-Herrero}},\ and\ \bibinfo {author}
  {\bibfnamefont {A.}~\bibnamefont {Yacoby}},\ }\bibfield  {title} {\emph
  {\bibinfo {title} {Fractional chern insulators in magic-angle twisted bilayer
  graphene}},\ }\href {https://doi.org/10.1038/s41586-021-04002-3} {\bibfield
  {journal} {\bibinfo  {journal} {Nature}\ }\textbf {\bibinfo {volume} {600}},\
  \bibinfo {pages} {439} (\bibinfo {year} {2021})}\BibitemShut {NoStop}%
\bibitem [{\citenamefont {D{\'i}ez-M{\'e}rida}\ \emph
  {et~al.}(2023)\citenamefont {D{\'i}ez-M{\'e}rida}, \citenamefont
  {D{\'i}ez-Carl{\'o}n}, \citenamefont {Yang}, \citenamefont {Xie},
  \citenamefont {Gao}, \citenamefont {Senior}, \citenamefont {Watanabe},
  \citenamefont {Taniguchi}, \citenamefont {Lu}, \citenamefont {Higginbotham},
  \citenamefont {Law},\ and\ \citenamefont {Efetov}}]{Diez-Merida2021diode}%
  \BibitemOpen
  \bibfield  {author} {\bibinfo {author} {\bibfnamefont {J.}~\bibnamefont
  {D{\'i}ez-M{\'e}rida}}, \bibinfo {author} {\bibfnamefont {A.}~\bibnamefont
  {D{\'i}ez-Carl{\'o}n}}, \bibinfo {author} {\bibfnamefont {S.~Y.}\
  \bibnamefont {Yang}}, \bibinfo {author} {\bibfnamefont {Y.-M.}\ \bibnamefont
  {Xie}}, \bibinfo {author} {\bibfnamefont {X.-J.}\ \bibnamefont {Gao}},
  \bibinfo {author} {\bibfnamefont {J.}~\bibnamefont {Senior}}, \bibinfo
  {author} {\bibfnamefont {K.}~\bibnamefont {Watanabe}}, \bibinfo {author}
  {\bibfnamefont {T.}~\bibnamefont {Taniguchi}}, \bibinfo {author}
  {\bibfnamefont {X.}~\bibnamefont {Lu}}, \bibinfo {author} {\bibfnamefont
  {A.~P.}\ \bibnamefont {Higginbotham}}, \bibinfo {author} {\bibfnamefont
  {K.~T.}\ \bibnamefont {Law}},\ and\ \bibinfo {author} {\bibfnamefont {D.~K.}\
  \bibnamefont {Efetov}},\ }\bibfield  {title} {\emph {\bibinfo {title}
  {Symmetry-broken josephson junctions and superconducting diodes in
  magic-angle twisted bilayer graphene}},\ }\href
  {https://doi.org/10.1038/s41467-023-38005-7} {\bibfield  {journal} {\bibinfo
  {journal} {Nature Communications}\ }\textbf {\bibinfo {volume} {14}},\
  \bibinfo {pages} {2396} (\bibinfo {year} {2023})}\BibitemShut {NoStop}%
\bibitem [{\citenamefont {Jiang}\ \emph {et~al.}(2019)\citenamefont {Jiang},
  \citenamefont {Lai}, \citenamefont {Watanabe}, \citenamefont {Taniguchi},
  \citenamefont {Haule}, \citenamefont {Mao},\ and\ \citenamefont
  {Andrei}}]{Jiang2019charge}%
  \BibitemOpen
  \bibfield  {author} {\bibinfo {author} {\bibfnamefont {Y.}~\bibnamefont
  {Jiang}}, \bibinfo {author} {\bibfnamefont {X.}~\bibnamefont {Lai}}, \bibinfo
  {author} {\bibfnamefont {K.}~\bibnamefont {Watanabe}}, \bibinfo {author}
  {\bibfnamefont {T.}~\bibnamefont {Taniguchi}}, \bibinfo {author}
  {\bibfnamefont {K.}~\bibnamefont {Haule}}, \bibinfo {author} {\bibfnamefont
  {J.}~\bibnamefont {Mao}},\ and\ \bibinfo {author} {\bibfnamefont {E.~Y.}\
  \bibnamefont {Andrei}},\ }\bibfield  {title} {\emph {\bibinfo {title} {Charge
  order and broken rotational symmetry in magic-angle twisted bilayer
  graphene}},\ }\href {https://doi.org/10.1038/s41586-019-1460-4} {\bibfield
  {journal} {\bibinfo  {journal} {Nature}\ }\textbf {\bibinfo {volume} {573}},\
  \bibinfo {pages} {91} (\bibinfo {year} {2019})}\BibitemShut {NoStop}%
\bibitem [{\citenamefont {Arora}\ \emph {et~al.}(2020)\citenamefont {Arora},
  \citenamefont {Polski}, \citenamefont {Zhang}, \citenamefont {Thomson},
  \citenamefont {Choi}, \citenamefont {Kim}, \citenamefont {Lin}, \citenamefont
  {Wilson}, \citenamefont {Xu}, \citenamefont {Chu}, \citenamefont {Watanabe},
  \citenamefont {Taniguchi}, \citenamefont {Alicea},\ and\ \citenamefont
  {Nadj-Perge}}]{Arora2020SC}%
  \BibitemOpen
  \bibfield  {author} {\bibinfo {author} {\bibfnamefont {H.~S.}\ \bibnamefont
  {Arora}}, \bibinfo {author} {\bibfnamefont {R.}~\bibnamefont {Polski}},
  \bibinfo {author} {\bibfnamefont {Y.}~\bibnamefont {Zhang}}, \bibinfo
  {author} {\bibfnamefont {A.}~\bibnamefont {Thomson}}, \bibinfo {author}
  {\bibfnamefont {Y.}~\bibnamefont {Choi}}, \bibinfo {author} {\bibfnamefont
  {H.}~\bibnamefont {Kim}}, \bibinfo {author} {\bibfnamefont {Z.}~\bibnamefont
  {Lin}}, \bibinfo {author} {\bibfnamefont {I.~Z.}\ \bibnamefont {Wilson}},
  \bibinfo {author} {\bibfnamefont {X.}~\bibnamefont {Xu}}, \bibinfo {author}
  {\bibfnamefont {J.-H.}\ \bibnamefont {Chu}}, \bibinfo {author} {\bibfnamefont
  {K.}~\bibnamefont {Watanabe}}, \bibinfo {author} {\bibfnamefont
  {T.}~\bibnamefont {Taniguchi}}, \bibinfo {author} {\bibfnamefont
  {J.}~\bibnamefont {Alicea}},\ and\ \bibinfo {author} {\bibfnamefont
  {S.}~\bibnamefont {Nadj-Perge}},\ }\bibfield  {title} {\emph {\bibinfo
  {title} {Superconductivity in metallic twisted bilayer graphene stabilized by
  {WSe}2}},\ }\href {https://doi.org/10.1038/s41586-020-2473-8} {\bibfield
  {journal} {\bibinfo  {journal} {Nature}\ }\textbf {\bibinfo {volume} {583}},\
  \bibinfo {pages} {379} (\bibinfo {year} {2020})}\BibitemShut {NoStop}%
\bibitem [{\citenamefont {Kerelsky}\ \emph {et~al.}(2019)\citenamefont
  {Kerelsky}, \citenamefont {McGilly}, \citenamefont {Kennes}, \citenamefont
  {Xian}, \citenamefont {Yankowitz}, \citenamefont {Chen}, \citenamefont
  {Watanabe}, \citenamefont {Taniguchi}, \citenamefont {Hone}, \citenamefont
  {Dean} \emph {et~al.}}]{Kerelsky2019}%
  \BibitemOpen
  \bibfield  {author} {\bibinfo {author} {\bibfnamefont {A.}~\bibnamefont
  {Kerelsky}}, \bibinfo {author} {\bibfnamefont {L.~J.}\ \bibnamefont
  {McGilly}}, \bibinfo {author} {\bibfnamefont {D.~M.}\ \bibnamefont {Kennes}},
  \bibinfo {author} {\bibfnamefont {L.}~\bibnamefont {Xian}}, \bibinfo {author}
  {\bibfnamefont {M.}~\bibnamefont {Yankowitz}}, \bibinfo {author}
  {\bibfnamefont {S.}~\bibnamefont {Chen}}, \bibinfo {author} {\bibfnamefont
  {K.}~\bibnamefont {Watanabe}}, \bibinfo {author} {\bibfnamefont
  {T.}~\bibnamefont {Taniguchi}}, \bibinfo {author} {\bibfnamefont
  {J.}~\bibnamefont {Hone}}, \bibinfo {author} {\bibfnamefont {C.}~\bibnamefont
  {Dean}}, \emph {et~al.},\ }\bibfield  {title} {\emph {\bibinfo {title}
  {Maximized electron interactions at the magic angle in twisted bilayer
  graphene}},\ }\href@noop {} {\bibfield  {journal} {\bibinfo  {journal}
  {Nature}\ }\textbf {\bibinfo {volume} {572}},\ \bibinfo {pages} {95}
  (\bibinfo {year} {2019})}\BibitemShut {NoStop}%
\bibitem [{\citenamefont {Choi}\ \emph {et~al.}(2021)\citenamefont {Choi},
  \citenamefont {Kim}, \citenamefont {Peng}, \citenamefont {Thomson},
  \citenamefont {Lewandowski}, \citenamefont {Polski}, \citenamefont {Zhang},
  \citenamefont {Arora}, \citenamefont {Watanabe}, \citenamefont {Taniguchi}
  \emph {et~al.}}]{choi2021correlation}%
  \BibitemOpen
  \bibfield  {author} {\bibinfo {author} {\bibfnamefont {Y.}~\bibnamefont
  {Choi}}, \bibinfo {author} {\bibfnamefont {H.}~\bibnamefont {Kim}}, \bibinfo
  {author} {\bibfnamefont {Y.}~\bibnamefont {Peng}}, \bibinfo {author}
  {\bibfnamefont {A.}~\bibnamefont {Thomson}}, \bibinfo {author} {\bibfnamefont
  {C.}~\bibnamefont {Lewandowski}}, \bibinfo {author} {\bibfnamefont
  {R.}~\bibnamefont {Polski}}, \bibinfo {author} {\bibfnamefont
  {Y.}~\bibnamefont {Zhang}}, \bibinfo {author} {\bibfnamefont {H.~S.}\
  \bibnamefont {Arora}}, \bibinfo {author} {\bibfnamefont {K.}~\bibnamefont
  {Watanabe}}, \bibinfo {author} {\bibfnamefont {T.}~\bibnamefont {Taniguchi}},
  \emph {et~al.},\ }\bibfield  {title} {\emph {\bibinfo {title}
  {Correlation-driven topological phases in magic-angle twisted bilayer
  graphene}},\ }\href@noop {} {\bibfield  {journal} {\bibinfo  {journal}
  {Nature}\ }\textbf {\bibinfo {volume} {589}},\ \bibinfo {pages} {536}
  (\bibinfo {year} {2021})}\BibitemShut {NoStop}%
\bibitem [{\citenamefont {Xie}\ \emph {et~al.}(2019)\citenamefont {Xie},
  \citenamefont {Lian}, \citenamefont {J{\"a}ck}, \citenamefont {Liu},
  \citenamefont {Chiu}, \citenamefont {Watanabe}, \citenamefont {Taniguchi},
  \citenamefont {Bernevig},\ and\ \citenamefont {Yazdani}}]{Xie2019stm}%
  \BibitemOpen
  \bibfield  {author} {\bibinfo {author} {\bibfnamefont {Y.}~\bibnamefont
  {Xie}}, \bibinfo {author} {\bibfnamefont {B.}~\bibnamefont {Lian}}, \bibinfo
  {author} {\bibfnamefont {B.}~\bibnamefont {J{\"a}ck}}, \bibinfo {author}
  {\bibfnamefont {X.}~\bibnamefont {Liu}}, \bibinfo {author} {\bibfnamefont
  {C.-L.}\ \bibnamefont {Chiu}}, \bibinfo {author} {\bibfnamefont
  {K.}~\bibnamefont {Watanabe}}, \bibinfo {author} {\bibfnamefont
  {T.}~\bibnamefont {Taniguchi}}, \bibinfo {author} {\bibfnamefont {B.~A.}\
  \bibnamefont {Bernevig}},\ and\ \bibinfo {author} {\bibfnamefont
  {A.}~\bibnamefont {Yazdani}},\ }\bibfield  {title} {\emph {\bibinfo {title}
  {Spectroscopic signatures of many-body correlations in magic angle twisted
  bilayer graphene}},\ }\href@noop {} {\bibfield  {journal} {\bibinfo
  {journal} {Nature}\ }\textbf {\bibinfo {volume} {572}},\ \bibinfo {pages}
  {101} (\bibinfo {year} {2019})}\BibitemShut {NoStop}%
\bibitem [{\citenamefont {Wong}\ \emph {et~al.}(2020)\citenamefont {Wong},
  \citenamefont {Nuckolls}, \citenamefont {Oh}, \citenamefont {Lian},
  \citenamefont {Xie}, \citenamefont {Jeon}, \citenamefont {Watanabe},
  \citenamefont {Taniguchi}, \citenamefont {Bernevig},\ and\ \citenamefont
  {Yazdani}}]{Wong_2020}%
  \BibitemOpen
  \bibfield  {author} {\bibinfo {author} {\bibfnamefont {D.}~\bibnamefont
  {Wong}}, \bibinfo {author} {\bibfnamefont {K.~P.}\ \bibnamefont {Nuckolls}},
  \bibinfo {author} {\bibfnamefont {M.}~\bibnamefont {Oh}}, \bibinfo {author}
  {\bibfnamefont {B.}~\bibnamefont {Lian}}, \bibinfo {author} {\bibfnamefont
  {Y.}~\bibnamefont {Xie}}, \bibinfo {author} {\bibfnamefont {S.}~\bibnamefont
  {Jeon}}, \bibinfo {author} {\bibfnamefont {K.}~\bibnamefont {Watanabe}},
  \bibinfo {author} {\bibfnamefont {T.}~\bibnamefont {Taniguchi}}, \bibinfo
  {author} {\bibfnamefont {B.~A.}\ \bibnamefont {Bernevig}},\ and\ \bibinfo
  {author} {\bibfnamefont {A.}~\bibnamefont {Yazdani}},\ }\bibfield  {title}
  {\emph {\bibinfo {title} {Cascade of electronic transitions in magic-angle
  twisted bilayer graphene}},\ }\href
  {https://doi.org/10.1038/s41586-020-2339-0} {\bibfield  {journal} {\bibinfo
  {journal} {Nature}\ }\textbf {\bibinfo {volume} {582}},\ \bibinfo {pages}
  {198} (\bibinfo {year} {2020})}\BibitemShut {NoStop}%
\bibitem [{\citenamefont {Pierce}\ \emph {et~al.}(2021)\citenamefont {Pierce},
  \citenamefont {Xie}, \citenamefont {Park}, \citenamefont {Khalaf},
  \citenamefont {Lee}, \citenamefont {Cao}, \citenamefont {Parker},
  \citenamefont {Forrester}, \citenamefont {Chen}, \citenamefont {Watanabe},
  \citenamefont {Taniguchi}, \citenamefont {Vishwanath}, \citenamefont
  {Jarillo-Herrero},\ and\ \citenamefont {Yacoby}}]{pierce2021unconventional}%
  \BibitemOpen
  \bibfield  {author} {\bibinfo {author} {\bibfnamefont {A.~T.}\ \bibnamefont
  {Pierce}}, \bibinfo {author} {\bibfnamefont {Y.}~\bibnamefont {Xie}},
  \bibinfo {author} {\bibfnamefont {J.~M.}\ \bibnamefont {Park}}, \bibinfo
  {author} {\bibfnamefont {E.}~\bibnamefont {Khalaf}}, \bibinfo {author}
  {\bibfnamefont {S.~H.}\ \bibnamefont {Lee}}, \bibinfo {author} {\bibfnamefont
  {Y.}~\bibnamefont {Cao}}, \bibinfo {author} {\bibfnamefont {D.~E.}\
  \bibnamefont {Parker}}, \bibinfo {author} {\bibfnamefont {P.~R.}\
  \bibnamefont {Forrester}}, \bibinfo {author} {\bibfnamefont {S.}~\bibnamefont
  {Chen}}, \bibinfo {author} {\bibfnamefont {K.}~\bibnamefont {Watanabe}},
  \bibinfo {author} {\bibfnamefont {T.}~\bibnamefont {Taniguchi}}, \bibinfo
  {author} {\bibfnamefont {A.}~\bibnamefont {Vishwanath}}, \bibinfo {author}
  {\bibfnamefont {P.}~\bibnamefont {Jarillo-Herrero}},\ and\ \bibinfo {author}
  {\bibfnamefont {A.}~\bibnamefont {Yacoby}},\ }\href@noop {} {\bibinfo {title}
  {Unconventional sequence of correlated chern insulators in magic-angle
  twisted bilayer graphene}} (\bibinfo {year} {2021}),\ \Eprint
  {https://arxiv.org/abs/2101.04123} {arXiv:2101.04123 [cond-mat.mes-hall]}
  \BibitemShut {NoStop}%
\bibitem [{\citenamefont {Tarnopolsky}\ \emph {et~al.}(2019)\citenamefont
  {Tarnopolsky}, \citenamefont {Kruchkov},\ and\ \citenamefont
  {Vishwanath}}]{tarnopolsky_origin_2019}%
  \BibitemOpen
  \bibfield  {author} {\bibinfo {author} {\bibfnamefont {G.}~\bibnamefont
  {Tarnopolsky}}, \bibinfo {author} {\bibfnamefont {A.~J.}\ \bibnamefont
  {Kruchkov}},\ and\ \bibinfo {author} {\bibfnamefont {A.}~\bibnamefont
  {Vishwanath}},\ }\bibfield  {title} {\emph {\bibinfo {title} {Origin of
  {Magic} {Angles} in {Twisted} {Bilayer} {Graphene}}},\ }\href
  {https://doi.org/10.1103/PhysRevLett.122.106405} {\bibfield  {journal}
  {\bibinfo  {journal} {Phys. Rev. Lett.}\ }\textbf {\bibinfo {volume} {122}},\
  \bibinfo {pages} {106405} (\bibinfo {year} {2019})},\ \bibinfo {note}
  {publisher: American Physical Society}\BibitemShut {NoStop}%
\bibitem [{\citenamefont {Bultinck}\ \emph {et~al.}(2020)\citenamefont
  {Bultinck}, \citenamefont {Khalaf}, \citenamefont {Liu}, \citenamefont
  {Chatterjee}, \citenamefont {Vishwanath},\ and\ \citenamefont
  {Zaletel}}]{bultinck_ground_2020}%
  \BibitemOpen
  \bibfield  {author} {\bibinfo {author} {\bibfnamefont {N.}~\bibnamefont
  {Bultinck}}, \bibinfo {author} {\bibfnamefont {E.}~\bibnamefont {Khalaf}},
  \bibinfo {author} {\bibfnamefont {S.}~\bibnamefont {Liu}}, \bibinfo {author}
  {\bibfnamefont {S.}~\bibnamefont {Chatterjee}}, \bibinfo {author}
  {\bibfnamefont {A.}~\bibnamefont {Vishwanath}},\ and\ \bibinfo {author}
  {\bibfnamefont {M.~P.}\ \bibnamefont {Zaletel}},\ }\bibfield  {title} {\emph
  {\bibinfo {title} {Ground {State} and {Hidden} {Symmetry} of {Magic}-{Angle}
  {Graphene} at {Even} {Integer} {Filling}}},\ }\href
  {https://doi.org/10.1103/PhysRevX.10.031034} {\bibfield  {journal} {\bibinfo
  {journal} {Phys. Rev. X}\ }\textbf {\bibinfo {volume} {10}},\ \bibinfo
  {pages} {031034} (\bibinfo {year} {2020})},\ \bibinfo {note} {publisher:
  American Physical Society}\BibitemShut {NoStop}%
\bibitem [{\citenamefont {Lian}\ \emph {et~al.}(2021)\citenamefont {Lian},
  \citenamefont {Song}, \citenamefont {Regnault}, \citenamefont {Efetov},
  \citenamefont {Yazdani},\ and\ \citenamefont {Bernevig}}]{TBG4}%
  \BibitemOpen
  \bibfield  {author} {\bibinfo {author} {\bibfnamefont {B.}~\bibnamefont
  {Lian}}, \bibinfo {author} {\bibfnamefont {Z.-D.}\ \bibnamefont {Song}},
  \bibinfo {author} {\bibfnamefont {N.}~\bibnamefont {Regnault}}, \bibinfo
  {author} {\bibfnamefont {D.~K.}\ \bibnamefont {Efetov}}, \bibinfo {author}
  {\bibfnamefont {A.}~\bibnamefont {Yazdani}},\ and\ \bibinfo {author}
  {\bibfnamefont {B.~A.}\ \bibnamefont {Bernevig}},\ }\bibfield  {title} {\emph
  {\bibinfo {title} {Twisted bilayer graphene. iv. exact insulator ground
  states and phase diagram}},\ }\href
  {https://doi.org/10.1103/PhysRevB.103.205414} {\bibfield  {journal} {\bibinfo
   {journal} {Phys. Rev. B}\ }\textbf {\bibinfo {volume} {103}},\ \bibinfo
  {pages} {205414} (\bibinfo {year} {2021})}\BibitemShut {NoStop}%
\bibitem [{\citenamefont {Kwan}\ \emph {et~al.}(2021)\citenamefont {Kwan},
  \citenamefont {Wagner}, \citenamefont {Soejima}, \citenamefont {Zaletel},
  \citenamefont {Simon}, \citenamefont {Parameswaran},\ and\ \citenamefont
  {Bultinck}}]{kwan_kekule_2021}%
  \BibitemOpen
  \bibfield  {author} {\bibinfo {author} {\bibfnamefont {Y.~H.}\ \bibnamefont
  {Kwan}}, \bibinfo {author} {\bibfnamefont {G.}~\bibnamefont {Wagner}},
  \bibinfo {author} {\bibfnamefont {T.}~\bibnamefont {Soejima}}, \bibinfo
  {author} {\bibfnamefont {M.~P.}\ \bibnamefont {Zaletel}}, \bibinfo {author}
  {\bibfnamefont {S.~H.}\ \bibnamefont {Simon}}, \bibinfo {author}
  {\bibfnamefont {S.~A.}\ \bibnamefont {Parameswaran}},\ and\ \bibinfo {author}
  {\bibfnamefont {N.}~\bibnamefont {Bultinck}},\ }\bibfield  {title} {\emph
  {\bibinfo {title} {Kekul\'e spiral order at all nonzero integer fillings in
  twisted bilayer graphene}},\ }\href
  {https://doi.org/10.1103/PhysRevX.11.041063} {\bibfield  {journal} {\bibinfo
  {journal} {Phys. Rev. X}\ }\textbf {\bibinfo {volume} {11}},\ \bibinfo
  {pages} {041063} (\bibinfo {year} {2021})}\BibitemShut {NoStop}%
\bibitem [{\citenamefont {Soejima}\ \emph {et~al.}(2020)\citenamefont
  {Soejima}, \citenamefont {Parker}, \citenamefont {Bultinck}, \citenamefont
  {Hauschild},\ and\ \citenamefont {Zaletel}}]{SoejimaDMRG}%
  \BibitemOpen
  \bibfield  {author} {\bibinfo {author} {\bibfnamefont {T.}~\bibnamefont
  {Soejima}}, \bibinfo {author} {\bibfnamefont {D.~E.}\ \bibnamefont {Parker}},
  \bibinfo {author} {\bibfnamefont {N.}~\bibnamefont {Bultinck}}, \bibinfo
  {author} {\bibfnamefont {J.}~\bibnamefont {Hauschild}},\ and\ \bibinfo
  {author} {\bibfnamefont {M.~P.}\ \bibnamefont {Zaletel}},\ }\bibfield
  {title} {\emph {\bibinfo {title} {Efficient simulation of moir\'e materials
  using the density matrix renormalization group}},\ }\href
  {https://doi.org/10.1103/PhysRevB.102.205111} {\bibfield  {journal} {\bibinfo
   {journal} {Phys. Rev. B}\ }\textbf {\bibinfo {volume} {102}},\ \bibinfo
  {pages} {205111} (\bibinfo {year} {2020})}\BibitemShut {NoStop}%
\bibitem [{\citenamefont {Kwan}\ \emph {et~al.}(2024)\citenamefont {Kwan},
  \citenamefont {Wang}, \citenamefont {Wagner}, \citenamefont {Simon},
  \citenamefont {Parameswaran},\ and\ \citenamefont
  {Bultinck}}]{kwan2024texturedexcitoninsulators}%
  \BibitemOpen
  \bibfield  {author} {\bibinfo {author} {\bibfnamefont {Y.~H.}\ \bibnamefont
  {Kwan}}, \bibinfo {author} {\bibfnamefont {Z.}~\bibnamefont {Wang}}, \bibinfo
  {author} {\bibfnamefont {G.}~\bibnamefont {Wagner}}, \bibinfo {author}
  {\bibfnamefont {S.~H.}\ \bibnamefont {Simon}}, \bibinfo {author}
  {\bibfnamefont {S.~A.}\ \bibnamefont {Parameswaran}},\ and\ \bibinfo {author}
  {\bibfnamefont {N.}~\bibnamefont {Bultinck}},\ }\href
  {https://arxiv.org/abs/2406.15343} {\bibinfo {title} {Textured exciton
  insulators}} (\bibinfo {year} {2024}),\ \Eprint
  {https://arxiv.org/abs/2406.15343} {arXiv:2406.15343 [cond-mat.str-el]}
  \BibitemShut {NoStop}%
\bibitem [{\citenamefont {Wang}\ \emph {et~al.}(2024)\citenamefont {Wang},
  \citenamefont {Kwan}, \citenamefont {Wagner}, \citenamefont {Simon},
  \citenamefont {Bultinck},\ and\ \citenamefont
  {Parameswaran}}]{wang2024cherntexturedexcitoninsulatorsvalley}%
  \BibitemOpen
  \bibfield  {author} {\bibinfo {author} {\bibfnamefont {Z.}~\bibnamefont
  {Wang}}, \bibinfo {author} {\bibfnamefont {Y.~H.}\ \bibnamefont {Kwan}},
  \bibinfo {author} {\bibfnamefont {G.}~\bibnamefont {Wagner}}, \bibinfo
  {author} {\bibfnamefont {S.~H.}\ \bibnamefont {Simon}}, \bibinfo {author}
  {\bibfnamefont {N.}~\bibnamefont {Bultinck}},\ and\ \bibinfo {author}
  {\bibfnamefont {S.~A.}\ \bibnamefont {Parameswaran}},\ }\href
  {https://arxiv.org/abs/2406.15342} {\bibinfo {title} {Chern-textured exciton
  insulators with valley spiral order in moir\'e materials}} (\bibinfo {year}
  {2024}),\ \Eprint {https://arxiv.org/abs/2406.15342} {arXiv:2406.15342
  [cond-mat.str-el]} \BibitemShut {NoStop}%
\bibitem [{\citenamefont {Mendez-Valderrama}\ \emph {et~al.}(2024)\citenamefont
  {Mendez-Valderrama}, \citenamefont {Mao},\ and\ \citenamefont
  {Chowdhury}}]{opticalsumrule2023}%
  \BibitemOpen
  \bibfield  {author} {\bibinfo {author} {\bibfnamefont {J.~F.}\ \bibnamefont
  {Mendez-Valderrama}}, \bibinfo {author} {\bibfnamefont {D.}~\bibnamefont
  {Mao}},\ and\ \bibinfo {author} {\bibfnamefont {D.}~\bibnamefont
  {Chowdhury}},\ }\bibfield  {title} {\emph {\bibinfo {title} {Low-energy
  optical sum rule in moir\'e graphene}},\ }\href
  {https://doi.org/10.1103/PhysRevLett.133.196501} {\bibfield  {journal}
  {\bibinfo  {journal} {Phys. Rev. Lett.}\ }\textbf {\bibinfo {volume} {133}},\
  \bibinfo {pages} {196501} (\bibinfo {year} {2024})}\BibitemShut {NoStop}%
\bibitem [{\citenamefont {Tang}\ \emph {et~al.}(2011)\citenamefont {Tang},
  \citenamefont {Mei},\ and\ \citenamefont {Wen}}]{XGW11}%
  \BibitemOpen
  \bibfield  {author} {\bibinfo {author} {\bibfnamefont {E.}~\bibnamefont
  {Tang}}, \bibinfo {author} {\bibfnamefont {J.-W.}\ \bibnamefont {Mei}},\ and\
  \bibinfo {author} {\bibfnamefont {X.-G.}\ \bibnamefont {Wen}},\ }\bibfield
  {title} {\emph {\bibinfo {title} {High-temperature fractional quantum hall
  states}},\ }\href {https://doi.org/10.1103/PhysRevLett.106.236802} {\bibfield
   {journal} {\bibinfo  {journal} {Phys. Rev. Lett.}\ }\textbf {\bibinfo
  {volume} {106}},\ \bibinfo {pages} {236802} (\bibinfo {year}
  {2011})}\BibitemShut {NoStop}%
\bibitem [{\citenamefont {Neupert}\ \emph {et~al.}(2011)\citenamefont
  {Neupert}, \citenamefont {Santos}, \citenamefont {Chamon},\ and\
  \citenamefont {Mudry}}]{TN11}%
  \BibitemOpen
  \bibfield  {author} {\bibinfo {author} {\bibfnamefont {T.}~\bibnamefont
  {Neupert}}, \bibinfo {author} {\bibfnamefont {L.}~\bibnamefont {Santos}},
  \bibinfo {author} {\bibfnamefont {C.}~\bibnamefont {Chamon}},\ and\ \bibinfo
  {author} {\bibfnamefont {C.}~\bibnamefont {Mudry}},\ }\bibfield  {title}
  {\emph {\bibinfo {title} {Fractional quantum hall states at zero magnetic
  field}},\ }\href {https://doi.org/10.1103/PhysRevLett.106.236804} {\bibfield
  {journal} {\bibinfo  {journal} {Phys. Rev. Lett.}\ }\textbf {\bibinfo
  {volume} {106}},\ \bibinfo {pages} {236804} (\bibinfo {year}
  {2011})}\BibitemShut {NoStop}%
\bibitem [{\citenamefont {Regnault}\ and\ \citenamefont
  {Bernevig}(2011)}]{BAB11}%
  \BibitemOpen
  \bibfield  {author} {\bibinfo {author} {\bibfnamefont {N.}~\bibnamefont
  {Regnault}}\ and\ \bibinfo {author} {\bibfnamefont {B.~A.}\ \bibnamefont
  {Bernevig}},\ }\bibfield  {title} {\emph {\bibinfo {title} {Fractional chern
  insulator}},\ }\href {https://doi.org/10.1103/PhysRevX.1.021014} {\bibfield
  {journal} {\bibinfo  {journal} {Phys. Rev. X}\ }\textbf {\bibinfo {volume}
  {1}},\ \bibinfo {pages} {021014} (\bibinfo {year} {2011})}\BibitemShut
  {NoStop}%
\bibitem [{\citenamefont {Sheng}\ \emph {et~al.}(2011)\citenamefont {Sheng},
  \citenamefont {Gu}, \citenamefont {Sun},\ and\ \citenamefont
  {Sheng}}]{Sheng2011}%
  \BibitemOpen
  \bibfield  {author} {\bibinfo {author} {\bibfnamefont {D.~N.}\ \bibnamefont
  {Sheng}}, \bibinfo {author} {\bibfnamefont {Z.-C.}\ \bibnamefont {Gu}},
  \bibinfo {author} {\bibfnamefont {K.}~\bibnamefont {Sun}},\ and\ \bibinfo
  {author} {\bibfnamefont {L.}~\bibnamefont {Sheng}},\ }\bibfield  {title}
  {\emph {\bibinfo {title} {Fractional quantum hall effect in the absence of
  landau levels}},\ }\href {https://doi.org/10.1038/ncomms1380} {\bibfield
  {journal} {\bibinfo  {journal} {Nature Communications}\ }\textbf {\bibinfo
  {volume} {2}},\ \bibinfo {pages} {389} (\bibinfo {year} {2011})}\BibitemShut
  {NoStop}%
\bibitem [{\citenamefont {Ju}\ \emph {et~al.}(2024)\citenamefont {Ju},
  \citenamefont {MacDonald}, \citenamefont {Mak}, \citenamefont {Shan},\ and\
  \citenamefont {Xu}}]{FQAHrev}%
  \BibitemOpen
  \bibfield  {author} {\bibinfo {author} {\bibfnamefont {L.}~\bibnamefont
  {Ju}}, \bibinfo {author} {\bibfnamefont {A.~H.}\ \bibnamefont {MacDonald}},
  \bibinfo {author} {\bibfnamefont {K.~F.}\ \bibnamefont {Mak}}, \bibinfo
  {author} {\bibfnamefont {J.}~\bibnamefont {Shan}},\ and\ \bibinfo {author}
  {\bibfnamefont {X.}~\bibnamefont {Xu}},\ }\bibfield  {title} {\emph {\bibinfo
  {title} {The fractional quantum anomalous hall effect}},\ }\href
  {https://doi.org/10.1038/s41578-024-00694-x} {\bibfield  {journal} {\bibinfo
  {journal} {Nature Reviews Materials}\ }\textbf {\bibinfo {volume} {9}},\
  \bibinfo {pages} {455} (\bibinfo {year} {2024})}\BibitemShut {NoStop}%
\bibitem [{\citenamefont {Cai}\ \emph {et~al.}(2023)\citenamefont {Cai},
  \citenamefont {Anderson}, \citenamefont {Wang}, \citenamefont {Zhang},
  \citenamefont {Liu}, \citenamefont {Holtzmann}, \citenamefont {Zhang},
  \citenamefont {Fan}, \citenamefont {Taniguchi}, \citenamefont {Watanabe},
  \citenamefont {Ran}, \citenamefont {Cao}, \citenamefont {Fu}, \citenamefont
  {Xiao}, \citenamefont {Yao},\ and\ \citenamefont {Xu}}]{xiaodong}%
  \BibitemOpen
  \bibfield  {author} {\bibinfo {author} {\bibfnamefont {J.}~\bibnamefont
  {Cai}}, \bibinfo {author} {\bibfnamefont {E.}~\bibnamefont {Anderson}},
  \bibinfo {author} {\bibfnamefont {C.}~\bibnamefont {Wang}}, \bibinfo {author}
  {\bibfnamefont {X.}~\bibnamefont {Zhang}}, \bibinfo {author} {\bibfnamefont
  {X.}~\bibnamefont {Liu}}, \bibinfo {author} {\bibfnamefont {W.}~\bibnamefont
  {Holtzmann}}, \bibinfo {author} {\bibfnamefont {Y.}~\bibnamefont {Zhang}},
  \bibinfo {author} {\bibfnamefont {F.}~\bibnamefont {Fan}}, \bibinfo {author}
  {\bibfnamefont {T.}~\bibnamefont {Taniguchi}}, \bibinfo {author}
  {\bibfnamefont {K.}~\bibnamefont {Watanabe}}, \bibinfo {author}
  {\bibfnamefont {Y.}~\bibnamefont {Ran}}, \bibinfo {author} {\bibfnamefont
  {T.}~\bibnamefont {Cao}}, \bibinfo {author} {\bibfnamefont {L.}~\bibnamefont
  {Fu}}, \bibinfo {author} {\bibfnamefont {D.}~\bibnamefont {Xiao}}, \bibinfo
  {author} {\bibfnamefont {W.}~\bibnamefont {Yao}},\ and\ \bibinfo {author}
  {\bibfnamefont {X.}~\bibnamefont {Xu}},\ }\bibfield  {title} {\emph {\bibinfo
  {title} {Signatures of fractional quantum anomalous hall states in twisted
  mote2}},\ }\href {https://doi.org/10.1038/s41586-023-06289-w} {\bibfield
  {journal} {\bibinfo  {journal} {Nature}\ }\textbf {\bibinfo {volume} {622}},\
  \bibinfo {pages} {63} (\bibinfo {year} {2023})}\BibitemShut {NoStop}%
\bibitem [{\citenamefont {Zeng}\ \emph {et~al.}(2023)\citenamefont {Zeng},
  \citenamefont {Xia}, \citenamefont {Kang}, \citenamefont {Zhu}, \citenamefont
  {Kn{\"u}ppel}, \citenamefont {Vaswani}, \citenamefont {Watanabe},
  \citenamefont {Taniguchi}, \citenamefont {Mak},\ and\ \citenamefont
  {Shan}}]{FaiFCI}%
  \BibitemOpen
  \bibfield  {author} {\bibinfo {author} {\bibfnamefont {Y.}~\bibnamefont
  {Zeng}}, \bibinfo {author} {\bibfnamefont {Z.}~\bibnamefont {Xia}}, \bibinfo
  {author} {\bibfnamefont {K.}~\bibnamefont {Kang}}, \bibinfo {author}
  {\bibfnamefont {J.}~\bibnamefont {Zhu}}, \bibinfo {author} {\bibfnamefont
  {P.}~\bibnamefont {Kn{\"u}ppel}}, \bibinfo {author} {\bibfnamefont
  {C.}~\bibnamefont {Vaswani}}, \bibinfo {author} {\bibfnamefont
  {K.}~\bibnamefont {Watanabe}}, \bibinfo {author} {\bibfnamefont
  {T.}~\bibnamefont {Taniguchi}}, \bibinfo {author} {\bibfnamefont {K.~F.}\
  \bibnamefont {Mak}},\ and\ \bibinfo {author} {\bibfnamefont {J.}~\bibnamefont
  {Shan}},\ }\bibfield  {title} {\emph {\bibinfo {title} {Thermodynamic
  evidence of fractional chern insulator in moir{\'e} mote2}},\ }\href
  {https://doi.org/10.1038/s41586-023-06452-3} {\bibfield  {journal} {\bibinfo
  {journal} {Nature}\ }\textbf {\bibinfo {volume} {622}},\ \bibinfo {pages}
  {69} (\bibinfo {year} {2023})}\BibitemShut {NoStop}%
\bibitem [{\citenamefont {Xu}\ \emph {et~al.}(2023)\citenamefont {Xu},
  \citenamefont {Sun}, \citenamefont {Jia}, \citenamefont {Liu}, \citenamefont
  {Xu}, \citenamefont {Li}, \citenamefont {Gu}, \citenamefont {Watanabe},
  \citenamefont {Taniguchi}, \citenamefont {Tong}, \citenamefont {Jia},
  \citenamefont {Shi}, \citenamefont {Jiang}, \citenamefont {Zhang},
  \citenamefont {Liu},\ and\ \citenamefont {Li}}]{tingxin}%
  \BibitemOpen
  \bibfield  {author} {\bibinfo {author} {\bibfnamefont {F.}~\bibnamefont
  {Xu}}, \bibinfo {author} {\bibfnamefont {Z.}~\bibnamefont {Sun}}, \bibinfo
  {author} {\bibfnamefont {T.}~\bibnamefont {Jia}}, \bibinfo {author}
  {\bibfnamefont {C.}~\bibnamefont {Liu}}, \bibinfo {author} {\bibfnamefont
  {C.}~\bibnamefont {Xu}}, \bibinfo {author} {\bibfnamefont {C.}~\bibnamefont
  {Li}}, \bibinfo {author} {\bibfnamefont {Y.}~\bibnamefont {Gu}}, \bibinfo
  {author} {\bibfnamefont {K.}~\bibnamefont {Watanabe}}, \bibinfo {author}
  {\bibfnamefont {T.}~\bibnamefont {Taniguchi}}, \bibinfo {author}
  {\bibfnamefont {B.}~\bibnamefont {Tong}}, \bibinfo {author} {\bibfnamefont
  {J.}~\bibnamefont {Jia}}, \bibinfo {author} {\bibfnamefont {Z.}~\bibnamefont
  {Shi}}, \bibinfo {author} {\bibfnamefont {S.}~\bibnamefont {Jiang}}, \bibinfo
  {author} {\bibfnamefont {Y.}~\bibnamefont {Zhang}}, \bibinfo {author}
  {\bibfnamefont {X.}~\bibnamefont {Liu}},\ and\ \bibinfo {author}
  {\bibfnamefont {T.}~\bibnamefont {Li}},\ }\bibfield  {title} {\emph {\bibinfo
  {title} {Observation of integer and fractional quantum anomalous hall effects
  in twisted bilayer ${\mathrm{mote}}_{2}$}},\ }\href
  {https://doi.org/10.1103/PhysRevX.13.031037} {\bibfield  {journal} {\bibinfo
  {journal} {Phys. Rev. X}\ }\textbf {\bibinfo {volume} {13}},\ \bibinfo
  {pages} {031037} (\bibinfo {year} {2023})}\BibitemShut {NoStop}%
\bibitem [{\citenamefont {Lu}\ \emph {et~al.}(2024)\citenamefont {Lu},
  \citenamefont {Han}, \citenamefont {Yao}, \citenamefont {Reddy},
  \citenamefont {Yang}, \citenamefont {Seo}, \citenamefont {Watanabe},
  \citenamefont {Taniguchi}, \citenamefont {Fu},\ and\ \citenamefont
  {Ju}}]{LongJu}%
  \BibitemOpen
  \bibfield  {author} {\bibinfo {author} {\bibfnamefont {Z.}~\bibnamefont
  {Lu}}, \bibinfo {author} {\bibfnamefont {T.}~\bibnamefont {Han}}, \bibinfo
  {author} {\bibfnamefont {Y.}~\bibnamefont {Yao}}, \bibinfo {author}
  {\bibfnamefont {A.~P.}\ \bibnamefont {Reddy}}, \bibinfo {author}
  {\bibfnamefont {J.}~\bibnamefont {Yang}}, \bibinfo {author} {\bibfnamefont
  {J.}~\bibnamefont {Seo}}, \bibinfo {author} {\bibfnamefont {K.}~\bibnamefont
  {Watanabe}}, \bibinfo {author} {\bibfnamefont {T.}~\bibnamefont {Taniguchi}},
  \bibinfo {author} {\bibfnamefont {L.}~\bibnamefont {Fu}},\ and\ \bibinfo
  {author} {\bibfnamefont {L.}~\bibnamefont {Ju}},\ }\bibfield  {title} {\emph
  {\bibinfo {title} {Fractional quantum anomalous hall effect in multilayer
  graphene}},\ }\href {https://doi.org/10.1038/s41586-023-07010-7} {\bibfield
  {journal} {\bibinfo  {journal} {Nature}\ }\textbf {\bibinfo {volume} {626}},\
  \bibinfo {pages} {759} (\bibinfo {year} {2024})}\BibitemShut {NoStop}%
\bibitem [{\citenamefont {Repellin}\ and\ \citenamefont
  {Senthil}(2020)}]{cecile2020chern}%
  \BibitemOpen
  \bibfield  {author} {\bibinfo {author} {\bibfnamefont {C.}~\bibnamefont
  {Repellin}}\ and\ \bibinfo {author} {\bibfnamefont {T.}~\bibnamefont
  {Senthil}},\ }\bibfield  {title} {\emph {\bibinfo {title} {Chern bands of
  twisted bilayer graphene: Fractional chern insulators and spin phase
  transition}},\ }\href {https://doi.org/10.1103/PhysRevResearch.2.023238}
  {\bibfield  {journal} {\bibinfo  {journal} {Phys. Rev. Res.}\ }\textbf
  {\bibinfo {volume} {2}},\ \bibinfo {pages} {023238} (\bibinfo {year}
  {2020})}\BibitemShut {NoStop}%
\bibitem [{\citenamefont {Li}\ \emph {et~al.}(2021{\natexlab{a}})\citenamefont
  {Li}, \citenamefont {Kumar}, \citenamefont {Sun},\ and\ \citenamefont
  {Lin}}]{Li2021spontaneous}%
  \BibitemOpen
  \bibfield  {author} {\bibinfo {author} {\bibfnamefont {H.}~\bibnamefont
  {Li}}, \bibinfo {author} {\bibfnamefont {U.}~\bibnamefont {Kumar}}, \bibinfo
  {author} {\bibfnamefont {K.}~\bibnamefont {Sun}},\ and\ \bibinfo {author}
  {\bibfnamefont {S.-Z.}\ \bibnamefont {Lin}},\ }\bibfield  {title} {\emph
  {\bibinfo {title} {Spontaneous fractional chern insulators in transition
  metal dichalcogenide moir\'e superlattices}},\ }\href
  {https://doi.org/10.1103/PhysRevResearch.3.L032070} {\bibfield  {journal}
  {\bibinfo  {journal} {Phys. Rev. Res.}\ }\textbf {\bibinfo {volume} {3}},\
  \bibinfo {pages} {L032070} (\bibinfo {year}
  {2021}{\natexlab{a}})}\BibitemShut {NoStop}%
\bibitem [{\citenamefont {Wilhelm}\ \emph {et~al.}(2021)\citenamefont
  {Wilhelm}, \citenamefont {Lang},\ and\ \citenamefont
  {L\"auchli}}]{wilhelm2021}%
  \BibitemOpen
  \bibfield  {author} {\bibinfo {author} {\bibfnamefont {P.}~\bibnamefont
  {Wilhelm}}, \bibinfo {author} {\bibfnamefont {T.~C.}\ \bibnamefont {Lang}},\
  and\ \bibinfo {author} {\bibfnamefont {A.~M.}\ \bibnamefont {L\"auchli}},\
  }\bibfield  {title} {\emph {\bibinfo {title} {Interplay of fractional chern
  insulator and charge density wave phases in twisted bilayer graphene}},\
  }\href {https://doi.org/10.1103/PhysRevB.103.125406} {\bibfield  {journal}
  {\bibinfo  {journal} {Phys. Rev. B}\ }\textbf {\bibinfo {volume} {103}},\
  \bibinfo {pages} {125406} (\bibinfo {year} {2021})}\BibitemShut {NoStop}%
\bibitem [{\citenamefont {Liu}\ and\ \citenamefont
  {Bergholtz}(2024)}]{liu2024fci}%
  \BibitemOpen
  \bibfield  {author} {\bibinfo {author} {\bibfnamefont {Z.}~\bibnamefont
  {Liu}}\ and\ \bibinfo {author} {\bibfnamefont {E.~J.}\ \bibnamefont
  {Bergholtz}},\ }in\ \href
  {https://doi.org/https://doi.org/10.1016/B978-0-323-90800-9.00136-0} {\emph
  {\bibinfo {booktitle} {Encyclopedia of Condensed Matter Physics (Second
  Edition)}}},\ \bibinfo {editor} {edited by\ \bibinfo {editor} {\bibfnamefont
  {T.}~\bibnamefont {Chakraborty}}}\ (\bibinfo  {publisher} {Academic Press},\
  \bibinfo {address} {Oxford},\ \bibinfo {year} {2024})\ \bibinfo {edition}
  {second edition}\ ed.,\ pp.\ \bibinfo {pages} {515--538}\BibitemShut
  {NoStop}%
\bibitem [{\citenamefont {Ledwith}\ \emph {et~al.}(2023)\citenamefont
  {Ledwith}, \citenamefont {Vishwanath},\ and\ \citenamefont
  {Parker}}]{vortexability}%
  \BibitemOpen
  \bibfield  {author} {\bibinfo {author} {\bibfnamefont {P.~J.}\ \bibnamefont
  {Ledwith}}, \bibinfo {author} {\bibfnamefont {A.}~\bibnamefont
  {Vishwanath}},\ and\ \bibinfo {author} {\bibfnamefont {D.~E.}\ \bibnamefont
  {Parker}},\ }\bibfield  {title} {\emph {\bibinfo {title} {Vortexability: A
  unifying criterion for ideal fractional chern insulators}},\ }\href
  {https://doi.org/10.1103/PhysRevB.108.205144} {\bibfield  {journal} {\bibinfo
   {journal} {Phys. Rev. B}\ }\textbf {\bibinfo {volume} {108}},\ \bibinfo
  {pages} {205144} (\bibinfo {year} {2023})}\BibitemShut {NoStop}%
\bibitem [{\citenamefont {Kourtis}\ \emph {et~al.}(2014)\citenamefont
  {Kourtis}, \citenamefont {Neupert}, \citenamefont {Chamon},\ and\
  \citenamefont {Mudry}}]{Kourtis2014fractional}%
  \BibitemOpen
  \bibfield  {author} {\bibinfo {author} {\bibfnamefont {S.}~\bibnamefont
  {Kourtis}}, \bibinfo {author} {\bibfnamefont {T.}~\bibnamefont {Neupert}},
  \bibinfo {author} {\bibfnamefont {C.}~\bibnamefont {Chamon}},\ and\ \bibinfo
  {author} {\bibfnamefont {C.}~\bibnamefont {Mudry}},\ }\bibfield  {title}
  {\emph {\bibinfo {title} {Fractional chern insulators with strong
  interactions that far exceed band gaps}},\ }\href
  {https://doi.org/10.1103/PhysRevLett.112.126806} {\bibfield  {journal}
  {\bibinfo  {journal} {Phys. Rev. Lett.}\ }\textbf {\bibinfo {volume} {112}},\
  \bibinfo {pages} {126806} (\bibinfo {year} {2014})}\BibitemShut {NoStop}%
\bibitem [{\citenamefont {Grushin}\ \emph {et~al.}(2015)\citenamefont
  {Grushin}, \citenamefont {Motruk}, \citenamefont {Zaletel},\ and\
  \citenamefont {Pollmann}}]{grushin15}%
  \BibitemOpen
  \bibfield  {author} {\bibinfo {author} {\bibfnamefont {A.~G.}\ \bibnamefont
  {Grushin}}, \bibinfo {author} {\bibfnamefont {J.}~\bibnamefont {Motruk}},
  \bibinfo {author} {\bibfnamefont {M.~P.}\ \bibnamefont {Zaletel}},\ and\
  \bibinfo {author} {\bibfnamefont {F.}~\bibnamefont {Pollmann}},\ }\bibfield
  {title} {\emph {\bibinfo {title} {Characterization and stability of a
  fermionic $\ensuremath{\nu}=1/3$ fractional chern insulator}},\ }\href
  {https://doi.org/10.1103/PhysRevB.91.035136} {\bibfield  {journal} {\bibinfo
  {journal} {Phys. Rev. B}\ }\textbf {\bibinfo {volume} {91}},\ \bibinfo
  {pages} {035136} (\bibinfo {year} {2015})}\BibitemShut {NoStop}%
\bibitem [{\citenamefont {Sharma}\ \emph {et~al.}(2024)\citenamefont {Sharma},
  \citenamefont {Peng},\ and\ \citenamefont {Sheng}}]{sharma2024}%
  \BibitemOpen
  \bibfield  {author} {\bibinfo {author} {\bibfnamefont {P.}~\bibnamefont
  {Sharma}}, \bibinfo {author} {\bibfnamefont {Y.}~\bibnamefont {Peng}},\ and\
  \bibinfo {author} {\bibfnamefont {D.~N.}\ \bibnamefont {Sheng}},\ }\bibfield
  {title} {\emph {\bibinfo {title} {Topological quantum phase transitions
  driven by a displacement field in twisted ${\mathrm{mote}}_{2}$ bilayers}},\
  }\href {https://doi.org/10.1103/PhysRevB.110.125142} {\bibfield  {journal}
  {\bibinfo  {journal} {Phys. Rev. B}\ }\textbf {\bibinfo {volume} {110}},\
  \bibinfo {pages} {125142} (\bibinfo {year} {2024})}\BibitemShut {NoStop}%
\bibitem [{\citenamefont {Yu}\ \emph {et~al.}(2024)\citenamefont {Yu},
  \citenamefont {Herzog-Arbeitman}, \citenamefont {Wang}, \citenamefont
  {Vafek}, \citenamefont {Bernevig},\ and\ \citenamefont
  {Regnault}}]{yu2024fractional}%
  \BibitemOpen
  \bibfield  {author} {\bibinfo {author} {\bibfnamefont {J.}~\bibnamefont
  {Yu}}, \bibinfo {author} {\bibfnamefont {J.}~\bibnamefont
  {Herzog-Arbeitman}}, \bibinfo {author} {\bibfnamefont {M.}~\bibnamefont
  {Wang}}, \bibinfo {author} {\bibfnamefont {O.}~\bibnamefont {Vafek}},
  \bibinfo {author} {\bibfnamefont {B.~A.}\ \bibnamefont {Bernevig}},\ and\
  \bibinfo {author} {\bibfnamefont {N.}~\bibnamefont {Regnault}},\ }\bibfield
  {title} {\emph {\bibinfo {title} {Fractional chern insulators versus
  nonmagnetic states in twisted bilayer ${\mathrm{mote}}_{2}$}},\ }\href
  {https://doi.org/10.1103/PhysRevB.109.045147} {\bibfield  {journal} {\bibinfo
   {journal} {Phys. Rev. B}\ }\textbf {\bibinfo {volume} {109}},\ \bibinfo
  {pages} {045147} (\bibinfo {year} {2024})}\BibitemShut {NoStop}%
\bibitem [{\citenamefont {Sheffer}\ and\ \citenamefont
  {Stern}(2021)}]{sheffer2021}%
  \BibitemOpen
  \bibfield  {author} {\bibinfo {author} {\bibfnamefont {Y.}~\bibnamefont
  {Sheffer}}\ and\ \bibinfo {author} {\bibfnamefont {A.}~\bibnamefont
  {Stern}},\ }\bibfield  {title} {\emph {\bibinfo {title} {Chiral magic-angle
  twisted bilayer graphene in a magnetic field: Landau level correspondence,
  exact wave functions, and fractional chern insulators}},\ }\href
  {https://doi.org/10.1103/PhysRevB.104.L121405} {\bibfield  {journal}
  {\bibinfo  {journal} {Phys. Rev. B}\ }\textbf {\bibinfo {volume} {104}},\
  \bibinfo {pages} {L121405} (\bibinfo {year} {2021})}\BibitemShut {NoStop}%
\bibitem [{\citenamefont {Wang}\ \emph {et~al.}(2021)\citenamefont {Wang},
  \citenamefont {Cano}, \citenamefont {Millis}, \citenamefont {Liu},\ and\
  \citenamefont {Yang}}]{wangcano}%
  \BibitemOpen
  \bibfield  {author} {\bibinfo {author} {\bibfnamefont {J.}~\bibnamefont
  {Wang}}, \bibinfo {author} {\bibfnamefont {J.}~\bibnamefont {Cano}}, \bibinfo
  {author} {\bibfnamefont {A.~J.}\ \bibnamefont {Millis}}, \bibinfo {author}
  {\bibfnamefont {Z.}~\bibnamefont {Liu}},\ and\ \bibinfo {author}
  {\bibfnamefont {B.}~\bibnamefont {Yang}},\ }\bibfield  {title} {\emph
  {\bibinfo {title} {Exact landau level description of geometry and interaction
  in a flatband}},\ }\href {https://doi.org/10.1103/PhysRevLett.127.246403}
  {\bibfield  {journal} {\bibinfo  {journal} {Phys. Rev. Lett.}\ }\textbf
  {\bibinfo {volume} {127}},\ \bibinfo {pages} {246403} (\bibinfo {year}
  {2021})}\BibitemShut {NoStop}%
\bibitem [{\citenamefont {Dong}\ \emph {et~al.}(2023)\citenamefont {Dong},
  \citenamefont {Ledwith}, \citenamefont {Khalaf}, \citenamefont {Lee},\ and\
  \citenamefont {Vishwanath}}]{dong23manybody}%
  \BibitemOpen
  \bibfield  {author} {\bibinfo {author} {\bibfnamefont {J.}~\bibnamefont
  {Dong}}, \bibinfo {author} {\bibfnamefont {P.~J.}\ \bibnamefont {Ledwith}},
  \bibinfo {author} {\bibfnamefont {E.}~\bibnamefont {Khalaf}}, \bibinfo
  {author} {\bibfnamefont {J.~Y.}\ \bibnamefont {Lee}},\ and\ \bibinfo {author}
  {\bibfnamefont {A.}~\bibnamefont {Vishwanath}},\ }\bibfield  {title} {\emph
  {\bibinfo {title} {Many-body ground states from decomposition of ideal higher
  chern bands: Applications to chirally twisted graphene multilayers}},\ }\href
  {https://doi.org/10.1103/PhysRevResearch.5.023166} {\bibfield  {journal}
  {\bibinfo  {journal} {Phys. Rev. Res.}\ }\textbf {\bibinfo {volume} {5}},\
  \bibinfo {pages} {023166} (\bibinfo {year} {2023})}\BibitemShut {NoStop}%
\bibitem [{\citenamefont {Shi}\ \emph {et~al.}(2024)\citenamefont {Shi},
  \citenamefont {Morales-Dur\'an}, \citenamefont {Khalaf},\ and\ \citenamefont
  {MacDonald}}]{shi2024prb}%
  \BibitemOpen
  \bibfield  {author} {\bibinfo {author} {\bibfnamefont {J.}~\bibnamefont
  {Shi}}, \bibinfo {author} {\bibfnamefont {N.}~\bibnamefont
  {Morales-Dur\'an}}, \bibinfo {author} {\bibfnamefont {E.}~\bibnamefont
  {Khalaf}},\ and\ \bibinfo {author} {\bibfnamefont {A.~H.}\ \bibnamefont
  {MacDonald}},\ }\bibfield  {title} {\emph {\bibinfo {title} {Adiabatic
  approximation and aharonov-casher bands in twisted homobilayer transition
  metal dichalcogenides}},\ }\href
  {https://doi.org/10.1103/PhysRevB.110.035130} {\bibfield  {journal} {\bibinfo
   {journal} {Phys. Rev. B}\ }\textbf {\bibinfo {volume} {110}},\ \bibinfo
  {pages} {035130} (\bibinfo {year} {2024})}\BibitemShut {NoStop}%
\bibitem [{\citenamefont {Morales-Dur\'an}\ \emph {et~al.}(2024)\citenamefont
  {Morales-Dur\'an}, \citenamefont {Wei}, \citenamefont {Shi},\ and\
  \citenamefont {MacDonald}}]{nicolas}%
  \BibitemOpen
  \bibfield  {author} {\bibinfo {author} {\bibfnamefont {N.}~\bibnamefont
  {Morales-Dur\'an}}, \bibinfo {author} {\bibfnamefont {N.}~\bibnamefont
  {Wei}}, \bibinfo {author} {\bibfnamefont {J.}~\bibnamefont {Shi}},\ and\
  \bibinfo {author} {\bibfnamefont {A.~H.}\ \bibnamefont {MacDonald}},\
  }\bibfield  {title} {\emph {\bibinfo {title} {Magic angles and fractional
  chern insulators in twisted homobilayer transition metal dichalcogenides}},\
  }\href {https://doi.org/10.1103/PhysRevLett.132.096602} {\bibfield  {journal}
  {\bibinfo  {journal} {Phys. Rev. Lett.}\ }\textbf {\bibinfo {volume} {132}},\
  \bibinfo {pages} {096602} (\bibinfo {year} {2024})}\BibitemShut {NoStop}%
\bibitem [{\citenamefont {{Wolf}}\ \emph {et~al.}(2024)\citenamefont {{Wolf}},
  \citenamefont {{Chao}}, \citenamefont {{MacDonald}},\ and\ \citenamefont
  {{Su}}}]{WolfAMD}%
  \BibitemOpen
  \bibfield  {author} {\bibinfo {author} {\bibfnamefont {T.~M.~R.}\
  \bibnamefont {{Wolf}}}, \bibinfo {author} {\bibfnamefont {Y.-C.}\
  \bibnamefont {{Chao}}}, \bibinfo {author} {\bibfnamefont {A.~H.}\
  \bibnamefont {{MacDonald}}},\ and\ \bibinfo {author} {\bibfnamefont {J.~J.}\
  \bibnamefont {{Su}}},\ }\bibfield  {title} {\emph {\bibinfo {title}
  {{Intraband collective excitations in fractional Chern insulators are
  dark}}},\ }\href {https://doi.org/10.48550/arXiv.2406.10709} {\bibfield
  {journal} {\bibinfo  {journal} {arXiv e-prints}\ ,\ \bibinfo {eid}
  {arXiv:2406.10709}} (\bibinfo {year} {2024})},\ \Eprint
  {https://arxiv.org/abs/2406.10709} {arXiv:2406.10709 [cond-mat.str-el]}
  \BibitemShut {NoStop}%
\bibitem [{\citenamefont {{Abouelkomsan}}\ \emph {et~al.}(2024)\citenamefont
  {{Abouelkomsan}}, \citenamefont {{Paul}}, \citenamefont {{Stern}},\ and\
  \citenamefont {{Fu}}}]{ASLF24}%
  \BibitemOpen
  \bibfield  {author} {\bibinfo {author} {\bibfnamefont {A.}~\bibnamefont
  {{Abouelkomsan}}}, \bibinfo {author} {\bibfnamefont {N.}~\bibnamefont
  {{Paul}}}, \bibinfo {author} {\bibfnamefont {A.}~\bibnamefont {{Stern}}},\
  and\ \bibinfo {author} {\bibfnamefont {L.}~\bibnamefont {{Fu}}},\ }\bibfield
  {title} {\emph {\bibinfo {title} {{Compressible quantum matter with vanishing
  Drude weight}}},\ }\href {https://doi.org/10.48550/arXiv.2403.14747}
  {\bibfield  {journal} {\bibinfo  {journal} {arXiv e-prints}\ ,\ \bibinfo
  {eid} {arXiv:2403.14747}} (\bibinfo {year} {2024})},\ \Eprint
  {https://arxiv.org/abs/2403.14747} {arXiv:2403.14747 [cond-mat.str-el]}
  \BibitemShut {NoStop}%
\bibitem [{\citenamefont {Kohn}(1961)}]{kohn1961}%
  \BibitemOpen
  \bibfield  {author} {\bibinfo {author} {\bibfnamefont {W.}~\bibnamefont
  {Kohn}},\ }\bibfield  {title} {\emph {\bibinfo {title} {Cyclotron resonance
  and de haas-van alphen oscillations of an interacting electron gas}},\ }\href
  {https://doi.org/10.1103/PhysRev.123.1242} {\bibfield  {journal} {\bibinfo
  {journal} {Phys. Rev.}\ }\textbf {\bibinfo {volume} {123}},\ \bibinfo {pages}
  {1242} (\bibinfo {year} {1961})}\BibitemShut {NoStop}%
\bibitem [{\citenamefont {Yip}(1989)}]{yip89}%
  \BibitemOpen
  \bibfield  {author} {\bibinfo {author} {\bibfnamefont {S.-K.}\ \bibnamefont
  {Yip}},\ }\bibfield  {title} {\emph {\bibinfo {title} {Extensions of kohn's
  theorem}},\ }\href {https://doi.org/10.1103/PhysRevB.40.3682} {\bibfield
  {journal} {\bibinfo  {journal} {Phys. Rev. B}\ }\textbf {\bibinfo {volume}
  {40}},\ \bibinfo {pages} {3682} (\bibinfo {year} {1989})}\BibitemShut
  {NoStop}%
\bibitem [{\citenamefont {Halperin}\ \emph {et~al.}(1993)\citenamefont
  {Halperin}, \citenamefont {Lee},\ and\ \citenamefont {Read}}]{HLR}%
  \BibitemOpen
  \bibfield  {author} {\bibinfo {author} {\bibfnamefont {B.~I.}\ \bibnamefont
  {Halperin}}, \bibinfo {author} {\bibfnamefont {P.~A.}\ \bibnamefont {Lee}},\
  and\ \bibinfo {author} {\bibfnamefont {N.}~\bibnamefont {Read}},\ }\bibfield
  {title} {\emph {\bibinfo {title} {Theory of the half-filled landau level}},\
  }\href {https://doi.org/10.1103/PhysRevB.47.7312} {\bibfield  {journal}
  {\bibinfo  {journal} {Phys. Rev. B}\ }\textbf {\bibinfo {volume} {47}},\
  \bibinfo {pages} {7312} (\bibinfo {year} {1993})}\BibitemShut {NoStop}%
\bibitem [{\citenamefont {Girvin}\ \emph {et~al.}(1986)\citenamefont {Girvin},
  \citenamefont {MacDonald},\ and\ \citenamefont {Platzman}}]{GMP}%
  \BibitemOpen
  \bibfield  {author} {\bibinfo {author} {\bibfnamefont {S.~M.}\ \bibnamefont
  {Girvin}}, \bibinfo {author} {\bibfnamefont {A.~H.}\ \bibnamefont
  {MacDonald}},\ and\ \bibinfo {author} {\bibfnamefont {P.~M.}\ \bibnamefont
  {Platzman}},\ }\bibfield  {title} {\emph {\bibinfo {title} {Magneto-roton
  theory of collective excitations in the fractional quantum hall effect}},\
  }\href {https://doi.org/10.1103/PhysRevB.33.2481} {\bibfield  {journal}
  {\bibinfo  {journal} {Phys. Rev. B}\ }\textbf {\bibinfo {volume} {33}},\
  \bibinfo {pages} {2481} (\bibinfo {year} {1986})}\BibitemShut {NoStop}%
\bibitem [{\citenamefont {Haldane}(2011)}]{HaldaneSF}%
  \BibitemOpen
  \bibfield  {author} {\bibinfo {author} {\bibfnamefont {F.~D.~M.}\
  \bibnamefont {Haldane}},\ }\bibfield  {title} {\emph {\bibinfo {title}
  {Geometrical description of the fractional quantum hall effect}},\ }\href
  {https://doi.org/10.1103/PhysRevLett.107.116801} {\bibfield  {journal}
  {\bibinfo  {journal} {Phys. Rev. Lett.}\ }\textbf {\bibinfo {volume} {107}},\
  \bibinfo {pages} {116801} (\bibinfo {year} {2011})}\BibitemShut {NoStop}%
\bibitem [{\citenamefont {Read}(1998)}]{read}%
  \BibitemOpen
  \bibfield  {author} {\bibinfo {author} {\bibfnamefont {N.}~\bibnamefont
  {Read}},\ }\bibfield  {title} {\emph {\bibinfo {title} {Lowest-landau-level
  theory of the quantum hall effect: The fermi-liquid-like state of bosons at
  filling factor one}},\ }\href {https://doi.org/10.1103/PhysRevB.58.16262}
  {\bibfield  {journal} {\bibinfo  {journal} {Phys. Rev. B}\ }\textbf {\bibinfo
  {volume} {58}},\ \bibinfo {pages} {16262} (\bibinfo {year}
  {1998})}\BibitemShut {NoStop}%
\bibitem [{\citenamefont {Kumar}\ and\ \citenamefont
  {Haldane}(2022)}]{prashant}%
  \BibitemOpen
  \bibfield  {author} {\bibinfo {author} {\bibfnamefont {P.}~\bibnamefont
  {Kumar}}\ and\ \bibinfo {author} {\bibfnamefont {F.~D.~M.}\ \bibnamefont
  {Haldane}},\ }\bibfield  {title} {\emph {\bibinfo {title} {Neutral
  excitations of quantum hall states: A density matrix renormalization group
  study}},\ }\href {https://doi.org/10.1103/PhysRevB.106.075116} {\bibfield
  {journal} {\bibinfo  {journal} {Phys. Rev. B}\ }\textbf {\bibinfo {volume}
  {106}},\ \bibinfo {pages} {075116} (\bibinfo {year} {2022})}\BibitemShut
  {NoStop}%
\bibitem [{\citenamefont {Antoniou}\ and\ \citenamefont
  {MacDonald}(1992)}]{AMdis}%
  \BibitemOpen
  \bibfield  {author} {\bibinfo {author} {\bibfnamefont {D.}~\bibnamefont
  {Antoniou}}\ and\ \bibinfo {author} {\bibfnamefont {A.~H.}\ \bibnamefont
  {MacDonald}},\ }\bibfield  {title} {\emph {\bibinfo {title} {Magnetoplasmons
  and cyclotron resonance in disordered two-dimensional electronic systems}},\
  }\href {https://doi.org/10.1103/PhysRevB.46.15225} {\bibfield  {journal}
  {\bibinfo  {journal} {Phys. Rev. B}\ }\textbf {\bibinfo {volume} {46}},\
  \bibinfo {pages} {15225} (\bibinfo {year} {1992})}\BibitemShut {NoStop}%
\bibitem [{\citenamefont {Wu}\ and\ \citenamefont
  {MacDonald}(2016)}]{wu2016moire}%
  \BibitemOpen
  \bibfield  {author} {\bibinfo {author} {\bibfnamefont {F.}~\bibnamefont
  {Wu}}\ and\ \bibinfo {author} {\bibfnamefont {A.~H.}\ \bibnamefont
  {MacDonald}},\ }\bibfield  {title} {\emph {\bibinfo {title} {Moir\'e assisted
  fractional quantum hall state spectroscopy}},\ }\href
  {https://doi.org/10.1103/PhysRevB.94.241108} {\bibfield  {journal} {\bibinfo
  {journal} {Phys. Rev. B}\ }\textbf {\bibinfo {volume} {94}},\ \bibinfo
  {pages} {241108} (\bibinfo {year} {2016})}\BibitemShut {NoStop}%
\bibitem [{\citenamefont {Dong}\ \emph {et~al.}(2022)\citenamefont {Dong},
  \citenamefont {Wang},\ and\ \citenamefont
  {Fu}}]{dong2022diracelectronperiodicmagnetic}%
  \BibitemOpen
  \bibfield  {author} {\bibinfo {author} {\bibfnamefont {J.}~\bibnamefont
  {Dong}}, \bibinfo {author} {\bibfnamefont {J.}~\bibnamefont {Wang}},\ and\
  \bibinfo {author} {\bibfnamefont {L.}~\bibnamefont {Fu}},\ }\href
  {https://arxiv.org/abs/2208.10516} {\bibinfo {title} {Dirac electron under
  periodic magnetic field: Platform for fractional chern insulator and
  generalized wigner crystal}} (\bibinfo {year} {2022}),\ \Eprint
  {https://arxiv.org/abs/2208.10516} {arXiv:2208.10516 [cond-mat.mes-hall]}
  \BibitemShut {NoStop}%
\bibitem [{\citenamefont {Estienne}\ \emph {et~al.}(2023)\citenamefont
  {Estienne}, \citenamefont {Regnault},\ and\ \citenamefont
  {Cr\'epel}}]{estienne2023}%
  \BibitemOpen
  \bibfield  {author} {\bibinfo {author} {\bibfnamefont {B.}~\bibnamefont
  {Estienne}}, \bibinfo {author} {\bibfnamefont {N.}~\bibnamefont {Regnault}},\
  and\ \bibinfo {author} {\bibfnamefont {V.}~\bibnamefont {Cr\'epel}},\
  }\bibfield  {title} {\emph {\bibinfo {title} {Ideal chern bands as landau
  levels in curved space}},\ }\href
  {https://doi.org/10.1103/PhysRevResearch.5.L032048} {\bibfield  {journal}
  {\bibinfo  {journal} {Phys. Rev. Res.}\ }\textbf {\bibinfo {volume} {5}},\
  \bibinfo {pages} {L032048} (\bibinfo {year} {2023})}\BibitemShut {NoStop}%
\bibitem [{\citenamefont {Li}\ and\ \citenamefont
  {Wu}(2024)}]{li2024variationalmappingchernbands}%
  \BibitemOpen
  \bibfield  {author} {\bibinfo {author} {\bibfnamefont {B.}~\bibnamefont
  {Li}}\ and\ \bibinfo {author} {\bibfnamefont {F.}~\bibnamefont {Wu}},\ }\href
  {https://arxiv.org/abs/2405.20307} {\bibinfo {title} {Variational mapping of
  chern bands to landau levels: Application to fractional chern insulators in
  twisted mote$_2$}} (\bibinfo {year} {2024}),\ \Eprint
  {https://arxiv.org/abs/2405.20307} {arXiv:2405.20307 [cond-mat.mes-hall]}
  \BibitemShut {NoStop}%
\bibitem [{\citenamefont {Aharonov}\ and\ \citenamefont
  {Casher}(1979)}]{aharonov1979ground}%
  \BibitemOpen
  \bibfield  {author} {\bibinfo {author} {\bibfnamefont {Y.}~\bibnamefont
  {Aharonov}}\ and\ \bibinfo {author} {\bibfnamefont {A.}~\bibnamefont
  {Casher}},\ }\bibfield  {title} {\emph {\bibinfo {title} {Ground state of a
  spin-\textonehalf{} charged particle in a two-dimensional magnetic field}},\
  }\href {https://doi.org/10.1103/PhysRevA.19.2461} {\bibfield  {journal}
  {\bibinfo  {journal} {Phys. Rev. A}\ }\textbf {\bibinfo {volume} {19}},\
  \bibinfo {pages} {2461} (\bibinfo {year} {1979})}\BibitemShut {NoStop}%
\bibitem [{\citenamefont {Dubrovin}\ and\ \citenamefont
  {Novikov}(1980)}]{dubrovin1980ground}%
  \BibitemOpen
  \bibfield  {author} {\bibinfo {author} {\bibfnamefont {B.}~\bibnamefont
  {Dubrovin}}\ and\ \bibinfo {author} {\bibfnamefont {S.}~\bibnamefont
  {Novikov}},\ }\bibfield  {title} {\emph {\bibinfo {title} {Ground states of a
  two-dimensional electron in a periodic magnetic field}},\ }\href@noop {}
  {\bibfield  {journal} {\bibinfo  {journal} {Journal of Experimental and
  Theoretical Physics - J EXP THEOR PHYS}\ }\textbf {\bibinfo {volume} {52}}
  (\bibinfo {year} {1980})}\BibitemShut {NoStop}%
\bibitem [{\citenamefont {Mao}\ and\ \citenamefont
  {Chowdhury}(2024)}]{mao2024upper}%
  \BibitemOpen
  \bibfield  {author} {\bibinfo {author} {\bibfnamefont {D.}~\bibnamefont
  {Mao}}\ and\ \bibinfo {author} {\bibfnamefont {D.}~\bibnamefont
  {Chowdhury}},\ }\bibfield  {title} {\emph {\bibinfo {title} {Upper bounds on
  superconducting and excitonic phase stiffness for interacting isolated narrow
  bands}},\ }\href {https://doi.org/10.1103/PhysRevB.109.024507} {\bibfield
  {journal} {\bibinfo  {journal} {Phys. Rev. B}\ }\textbf {\bibinfo {volume}
  {109}},\ \bibinfo {pages} {024507} (\bibinfo {year} {2024})}\BibitemShut
  {NoStop}%
\bibitem [{si()}]{si}%
  \BibitemOpen
  \href@noop {} {}\bibinfo {note} {See supplementary material for additional
  details}\BibitemShut {NoStop}%
\bibitem [{\citenamefont {Wang}\ \emph {et~al.}(2023)\citenamefont {Wang},
  \citenamefont {Klevtsov},\ and\ \citenamefont {Liu}}]{wang2023origin}%
  \BibitemOpen
  \bibfield  {author} {\bibinfo {author} {\bibfnamefont {J.}~\bibnamefont
  {Wang}}, \bibinfo {author} {\bibfnamefont {S.}~\bibnamefont {Klevtsov}},\
  and\ \bibinfo {author} {\bibfnamefont {Z.}~\bibnamefont {Liu}},\ }\bibfield
  {title} {\emph {\bibinfo {title} {Origin of model fractional chern insulators
  in all topological ideal flatbands: Explicit color-entangled wave function
  and exact density algebra}},\ }\href@noop {} {\bibfield  {journal} {\bibinfo
  {journal} {Physical Review Research}\ }\textbf {\bibinfo {volume} {5}},\
  \bibinfo {pages} {023167} (\bibinfo {year} {2023})}\BibitemShut {NoStop}%
\bibitem [{\citenamefont {Wu}\ \emph {et~al.}(2019)\citenamefont {Wu},
  \citenamefont {Lovorn}, \citenamefont {Tutuc}, \citenamefont {Martin},\ and\
  \citenamefont {MacDonald}}]{wu2019topological}%
  \BibitemOpen
  \bibfield  {author} {\bibinfo {author} {\bibfnamefont {F.}~\bibnamefont
  {Wu}}, \bibinfo {author} {\bibfnamefont {T.}~\bibnamefont {Lovorn}}, \bibinfo
  {author} {\bibfnamefont {E.}~\bibnamefont {Tutuc}}, \bibinfo {author}
  {\bibfnamefont {I.}~\bibnamefont {Martin}},\ and\ \bibinfo {author}
  {\bibfnamefont {A.~H.}\ \bibnamefont {MacDonald}},\ }\bibfield  {title}
  {\emph {\bibinfo {title} {Topological insulators in twisted transition metal
  dichalcogenide homobilayers}},\ }\href
  {https://doi.org/10.1103/PhysRevLett.122.086402} {\bibfield  {journal}
  {\bibinfo  {journal} {Phys. Rev. Lett.}\ }\textbf {\bibinfo {volume} {122}},\
  \bibinfo {pages} {086402} (\bibinfo {year} {2019})}\BibitemShut {NoStop}%
\bibitem [{\citenamefont {Can}\ \emph {et~al.}(2014)\citenamefont {Can},
  \citenamefont {Laskin},\ and\ \citenamefont {Wiegmann}}]{can2014prl}%
  \BibitemOpen
  \bibfield  {author} {\bibinfo {author} {\bibfnamefont {T.}~\bibnamefont
  {Can}}, \bibinfo {author} {\bibfnamefont {M.}~\bibnamefont {Laskin}},\ and\
  \bibinfo {author} {\bibfnamefont {P.}~\bibnamefont {Wiegmann}},\ }\bibfield
  {title} {\emph {\bibinfo {title} {Fractional quantum hall effect in a curved
  space: Gravitational anomaly and electromagnetic response}},\ }\href
  {https://doi.org/10.1103/PhysRevLett.113.046803} {\bibfield  {journal}
  {\bibinfo  {journal} {Phys. Rev. Lett.}\ }\textbf {\bibinfo {volume} {113}},\
  \bibinfo {pages} {046803} (\bibinfo {year} {2014})}\BibitemShut {NoStop}%
\bibitem [{\citenamefont {{Onishi}}\ and\ \citenamefont
  {{Fu}}(2024{\natexlab{b}})}]{OnishiFu24c}%
  \BibitemOpen
  \bibfield  {author} {\bibinfo {author} {\bibfnamefont {Y.}~\bibnamefont
  {{Onishi}}}\ and\ \bibinfo {author} {\bibfnamefont {L.}~\bibnamefont
  {{Fu}}},\ }\bibfield  {title} {\emph {\bibinfo {title} {{Topological bound on
  structure factor}}},\ }\href {https://doi.org/10.48550/arXiv.2406.18654}
  {\bibfield  {journal} {\bibinfo  {journal} {arXiv e-prints}\ ,\ \bibinfo
  {eid} {arXiv:2406.18654}} (\bibinfo {year} {2024}{\natexlab{b}})},\ \Eprint
  {https://arxiv.org/abs/2406.18654} {arXiv:2406.18654 [cond-mat.str-el]}
  \BibitemShut {NoStop}%
\bibitem [{\citenamefont {Zaklama}\ \emph {et~al.}(2025)\citenamefont
  {Zaklama}, \citenamefont {Luo},\ and\ \citenamefont
  {Fu}}]{zaklama2025structurefactortopologicalbound}%
  \BibitemOpen
  \bibfield  {author} {\bibinfo {author} {\bibfnamefont {T.}~\bibnamefont
  {Zaklama}}, \bibinfo {author} {\bibfnamefont {D.}~\bibnamefont {Luo}},\ and\
  \bibinfo {author} {\bibfnamefont {L.}~\bibnamefont {Fu}},\ }\href
  {https://arxiv.org/abs/2411.03496} {\bibinfo {title} {Structure factor and
  topological bound of twisted bilayer semiconductors at fractional fillings}}
  (\bibinfo {year} {2025}),\ \Eprint {https://arxiv.org/abs/2411.03496}
  {arXiv:2411.03496 [cond-mat.str-el]} \BibitemShut {NoStop}%
\bibitem [{\citenamefont {Hofmann}\ \emph {et~al.}(2023)\citenamefont
  {Hofmann}, \citenamefont {Berg},\ and\ \citenamefont
  {Chowdhury}}]{chiralPRL}%
  \BibitemOpen
  \bibfield  {author} {\bibinfo {author} {\bibfnamefont {J.~S.}\ \bibnamefont
  {Hofmann}}, \bibinfo {author} {\bibfnamefont {E.}~\bibnamefont {Berg}},\ and\
  \bibinfo {author} {\bibfnamefont {D.}~\bibnamefont {Chowdhury}},\ }\bibfield
  {title} {\emph {\bibinfo {title} {Superconductivity, charge density wave, and
  supersolidity in flat bands with a tunable quantum metric}},\ }\href
  {https://doi.org/10.1103/PhysRevLett.130.226001} {\bibfield  {journal}
  {\bibinfo  {journal} {Phys. Rev. Lett.}\ }\textbf {\bibinfo {volume} {130}},\
  \bibinfo {pages} {226001} (\bibinfo {year} {2023})}\BibitemShut {NoStop}%
\bibitem [{\citenamefont {Watanabe}\ and\ \citenamefont
  {Oshikawa}(2020)}]{oshikawaNL}%
  \BibitemOpen
  \bibfield  {author} {\bibinfo {author} {\bibfnamefont {H.}~\bibnamefont
  {Watanabe}}\ and\ \bibinfo {author} {\bibfnamefont {M.}~\bibnamefont
  {Oshikawa}},\ }\bibfield  {title} {\emph {\bibinfo {title} {Generalized
  $f$-sum rules and kohn formulas on nonlinear conductivities}},\ }\href
  {https://doi.org/10.1103/PhysRevB.102.165137} {\bibfield  {journal} {\bibinfo
   {journal} {Phys. Rev. B}\ }\textbf {\bibinfo {volume} {102}},\ \bibinfo
  {pages} {165137} (\bibinfo {year} {2020})}\BibitemShut {NoStop}%
\bibitem [{\citenamefont {Broholm}\ \emph {et~al.}(2020)\citenamefont
  {Broholm}, \citenamefont {Cava}, \citenamefont {Kivelson}, \citenamefont
  {Nocera}, \citenamefont {Norman},\ and\ \citenamefont {Senthil}}]{QSL_rev}%
  \BibitemOpen
  \bibfield  {author} {\bibinfo {author} {\bibfnamefont {C.}~\bibnamefont
  {Broholm}}, \bibinfo {author} {\bibfnamefont {R.~J.}\ \bibnamefont {Cava}},
  \bibinfo {author} {\bibfnamefont {S.~A.}\ \bibnamefont {Kivelson}}, \bibinfo
  {author} {\bibfnamefont {D.~G.}\ \bibnamefont {Nocera}}, \bibinfo {author}
  {\bibfnamefont {M.~R.}\ \bibnamefont {Norman}},\ and\ \bibinfo {author}
  {\bibfnamefont {T.}~\bibnamefont {Senthil}},\ }\bibfield  {title} {\emph
  {\bibinfo {title} {Quantum spin liquids}},\ }\href
  {https://doi.org/10.1126/science.aay0668} {\bibfield  {journal} {\bibinfo
  {journal} {Science}\ }\textbf {\bibinfo {volume} {367}},\ \bibinfo {pages}
  {eaay0668} (\bibinfo {year} {2020})},\ \Eprint
  {https://arxiv.org/abs/https://www.science.org/doi/pdf/10.1126/science.aay0668}
  {https://www.science.org/doi/pdf/10.1126/science.aay0668} \BibitemShut
  {NoStop}%
\bibitem [{\citenamefont {Ng}\ and\ \citenamefont {Lee}(2007)}]{Ng}%
  \BibitemOpen
  \bibfield  {author} {\bibinfo {author} {\bibfnamefont {T.-K.}\ \bibnamefont
  {Ng}}\ and\ \bibinfo {author} {\bibfnamefont {P.~A.}\ \bibnamefont {Lee}},\
  }\bibfield  {title} {\emph {\bibinfo {title} {Power-law conductivity inside
  the mott gap: Application to
  $\ensuremath{\kappa}\mathrm{\text{\ensuremath{-}}}(\mathrm{BEDT}\mathrm{\text{\ensuremath{-}}}\mathrm{TTF}{)}_{2}{\mathrm{cu}}_{2}(\mathrm{CN}{)}_{3}$}},\
  }\href {https://doi.org/10.1103/PhysRevLett.99.156402} {\bibfield  {journal}
  {\bibinfo  {journal} {Phys. Rev. Lett.}\ }\textbf {\bibinfo {volume} {99}},\
  \bibinfo {pages} {156402} (\bibinfo {year} {2007})}\BibitemShut {NoStop}%
\bibitem [{\citenamefont {Potter}\ \emph {et~al.}(2013)\citenamefont {Potter},
  \citenamefont {Senthil},\ and\ \citenamefont {Lee}}]{potter}%
  \BibitemOpen
  \bibfield  {author} {\bibinfo {author} {\bibfnamefont {A.~C.}\ \bibnamefont
  {Potter}}, \bibinfo {author} {\bibfnamefont {T.}~\bibnamefont {Senthil}},\
  and\ \bibinfo {author} {\bibfnamefont {P.~A.}\ \bibnamefont {Lee}},\
  }\bibfield  {title} {\emph {\bibinfo {title} {Mechanisms for sub-gap optical
  conductivity in herbertsmithite}},\ }\href
  {https://doi.org/10.1103/PhysRevB.87.245106} {\bibfield  {journal} {\bibinfo
  {journal} {Phys. Rev. B}\ }\textbf {\bibinfo {volume} {87}},\ \bibinfo
  {pages} {245106} (\bibinfo {year} {2013})}\BibitemShut {NoStop}%
\bibitem [{\citenamefont {Huh}\ \emph {et~al.}(2013)\citenamefont {Huh},
  \citenamefont {Punk},\ and\ \citenamefont {Sachdev}}]{huh}%
  \BibitemOpen
  \bibfield  {author} {\bibinfo {author} {\bibfnamefont {Y.}~\bibnamefont
  {Huh}}, \bibinfo {author} {\bibfnamefont {M.}~\bibnamefont {Punk}},\ and\
  \bibinfo {author} {\bibfnamefont {S.}~\bibnamefont {Sachdev}},\ }\bibfield
  {title} {\emph {\bibinfo {title} {Optical conductivity of visons in ${Z}_{2}$
  spin liquids close to a valence bond solid transition on the kagome
  lattice}},\ }\href {https://doi.org/10.1103/PhysRevB.87.235108} {\bibfield
  {journal} {\bibinfo  {journal} {Phys. Rev. B}\ }\textbf {\bibinfo {volume}
  {87}},\ \bibinfo {pages} {235108} (\bibinfo {year} {2013})}\BibitemShut
  {NoStop}%
\bibitem [{\citenamefont {Ioffe}\ and\ \citenamefont {Larkin}(1989)}]{IL}%
  \BibitemOpen
  \bibfield  {author} {\bibinfo {author} {\bibfnamefont {L.~B.}\ \bibnamefont
  {Ioffe}}\ and\ \bibinfo {author} {\bibfnamefont {A.~I.}\ \bibnamefont
  {Larkin}},\ }\bibfield  {title} {\emph {\bibinfo {title} {Gapless fermions
  and gauge fields in dielectrics}},\ }\href
  {https://doi.org/10.1103/PhysRevB.39.8988} {\bibfield  {journal} {\bibinfo
  {journal} {Phys. Rev. B}\ }\textbf {\bibinfo {volume} {39}},\ \bibinfo
  {pages} {8988} (\bibinfo {year} {1989})}\BibitemShut {NoStop}%
\bibitem [{\citenamefont {Li}\ \emph {et~al.}(2021{\natexlab{b}})\citenamefont
  {Li}, \citenamefont {Jiang}, \citenamefont {Li}, \citenamefont {Zhang},
  \citenamefont {Kang}, \citenamefont {Zhu}, \citenamefont {Watanabe},
  \citenamefont {Taniguchi}, \citenamefont {Chowdhury}, \citenamefont {Fu},
  \citenamefont {Shan},\ and\ \citenamefont {Mak}}]{FaiMIT}%
  \BibitemOpen
  \bibfield  {author} {\bibinfo {author} {\bibfnamefont {T.}~\bibnamefont
  {Li}}, \bibinfo {author} {\bibfnamefont {S.}~\bibnamefont {Jiang}}, \bibinfo
  {author} {\bibfnamefont {L.}~\bibnamefont {Li}}, \bibinfo {author}
  {\bibfnamefont {Y.}~\bibnamefont {Zhang}}, \bibinfo {author} {\bibfnamefont
  {K.}~\bibnamefont {Kang}}, \bibinfo {author} {\bibfnamefont {J.}~\bibnamefont
  {Zhu}}, \bibinfo {author} {\bibfnamefont {K.}~\bibnamefont {Watanabe}},
  \bibinfo {author} {\bibfnamefont {T.}~\bibnamefont {Taniguchi}}, \bibinfo
  {author} {\bibfnamefont {D.}~\bibnamefont {Chowdhury}}, \bibinfo {author}
  {\bibfnamefont {L.}~\bibnamefont {Fu}}, \bibinfo {author} {\bibfnamefont
  {J.}~\bibnamefont {Shan}},\ and\ \bibinfo {author} {\bibfnamefont {K.~F.}\
  \bibnamefont {Mak}},\ }\bibfield  {title} {\emph {\bibinfo {title}
  {Continuous {Mott} transition in semiconductor moir\'{e} superlattices}},\
  }\href {https://doi.org/10.1038/s41586-021-03853-0} {\bibfield  {journal}
  {\bibinfo  {journal} {Nature}\ }\textbf {\bibinfo {volume} {597}},\ \bibinfo
  {pages} {350} (\bibinfo {year} {2021}{\natexlab{b}})}\BibitemShut {NoStop}%
\bibitem [{\citenamefont {Costa~de Almeida}\ and\ \citenamefont
  {Hauke}(2021)}]{hauke21}%
  \BibitemOpen
  \bibfield  {author} {\bibinfo {author} {\bibfnamefont {R.}~\bibnamefont
  {Costa~de Almeida}}\ and\ \bibinfo {author} {\bibfnamefont {P.}~\bibnamefont
  {Hauke}},\ }\bibfield  {title} {\emph {\bibinfo {title} {From entanglement
  certification with quench dynamics to multipartite entanglement of
  interacting fermions}},\ }\href
  {https://doi.org/10.1103/PhysRevResearch.3.L032051} {\bibfield  {journal}
  {\bibinfo  {journal} {Phys. Rev. Res.}\ }\textbf {\bibinfo {volume} {3}},\
  \bibinfo {pages} {L032051} (\bibinfo {year} {2021})}\BibitemShut {NoStop}%
\bibitem [{\citenamefont {{Balut}}\ \emph {et~al.}(2024)\citenamefont
  {{Balut}}, \citenamefont {{Bradlyn}},\ and\ \citenamefont
  {{Abbamonte}}}]{BB24}%
  \BibitemOpen
  \bibfield  {author} {\bibinfo {author} {\bibfnamefont {D.}~\bibnamefont
  {{Balut}}}, \bibinfo {author} {\bibfnamefont {B.}~\bibnamefont {{Bradlyn}}},\
  and\ \bibinfo {author} {\bibfnamefont {P.}~\bibnamefont {{Abbamonte}}},\
  }\bibfield  {title} {\emph {\bibinfo {title} {{Quantum entanglement and
  quantum geometry measured with inelastic X-ray scattering}}},\ }\href
  {https://doi.org/10.48550/arXiv.2409.15583} {\bibfield  {journal} {\bibinfo
  {journal} {arXiv e-prints}\ ,\ \bibinfo {eid} {arXiv:2409.15583}} (\bibinfo
  {year} {2024})},\ \Eprint {https://arxiv.org/abs/2409.15583}
  {arXiv:2409.15583 [cond-mat.mes-hall]} \BibitemShut {NoStop}%
\bibitem [{\citenamefont {Canc\'es}\ and\ \citenamefont
  {Le~Bris}(2000)}]{ODAalgo}%
  \BibitemOpen
  \bibfield  {author} {\bibinfo {author} {\bibfnamefont {E.}~\bibnamefont
  {Canc\'es}}\ and\ \bibinfo {author} {\bibfnamefont {C.}~\bibnamefont
  {Le~Bris}},\ }\bibfield  {title} {\emph {\bibinfo {title} {Can we outperform
  the diis approach for electronic structure calculations?}},\ }\href
  {https://doi.org/https://doi.org/10.1002/1097-461X(2000)79:2<82::AID-QUA3>3.0.CO;2-I}
  {\bibfield  {journal} {\bibinfo  {journal} {International Journal of Quantum
  Chemistry}\ }\textbf {\bibinfo {volume} {79}},\ \bibinfo {pages} {82}
  (\bibinfo {year} {2000})}\BibitemShut {NoStop}%
\bibitem [{\citenamefont {Parker}\ \emph {et~al.}(2021)\citenamefont {Parker},
  \citenamefont {Ledwith}, \citenamefont {Khalaf}, \citenamefont {Soejima},
  \citenamefont {Hauschild}, \citenamefont {Xie}, \citenamefont {Pierce},
  \citenamefont {Zaletel}, \citenamefont {Yacoby},\ and\ \citenamefont
  {Vishwanath}}]{parker2021fieldtuned}%
  \BibitemOpen
  \bibfield  {author} {\bibinfo {author} {\bibfnamefont {D.}~\bibnamefont
  {Parker}}, \bibinfo {author} {\bibfnamefont {P.}~\bibnamefont {Ledwith}},
  \bibinfo {author} {\bibfnamefont {E.}~\bibnamefont {Khalaf}}, \bibinfo
  {author} {\bibfnamefont {T.}~\bibnamefont {Soejima}}, \bibinfo {author}
  {\bibfnamefont {J.}~\bibnamefont {Hauschild}}, \bibinfo {author}
  {\bibfnamefont {Y.}~\bibnamefont {Xie}}, \bibinfo {author} {\bibfnamefont
  {A.}~\bibnamefont {Pierce}}, \bibinfo {author} {\bibfnamefont {M.~P.}\
  \bibnamefont {Zaletel}}, \bibinfo {author} {\bibfnamefont {A.}~\bibnamefont
  {Yacoby}},\ and\ \bibinfo {author} {\bibfnamefont {A.}~\bibnamefont
  {Vishwanath}},\ }\href {https://arxiv.org/abs/2112.13837} {\bibinfo {title}
  {Field-tuned and zero-field fractional chern insulators in magic angle
  graphene}} (\bibinfo {year} {2021}),\ \Eprint
  {https://arxiv.org/abs/2112.13837} {arXiv:2112.13837 [cond-mat.str-el]}
  \BibitemShut {NoStop}%
\bibitem [{\citenamefont {Jain}(1989)}]{Jain1989}%
  \BibitemOpen
  \bibfield  {author} {\bibinfo {author} {\bibfnamefont {J.~K.}\ \bibnamefont
  {Jain}},\ }\bibfield  {title} {\emph {\bibinfo {title} {Incompressible
  quantum hall states}},\ }\href {https://doi.org/10.1103/PhysRevB.40.8079}
  {\bibfield  {journal} {\bibinfo  {journal} {Phys. Rev. B}\ }\textbf {\bibinfo
  {volume} {40}},\ \bibinfo {pages} {8079} (\bibinfo {year}
  {1989})}\BibitemShut {NoStop}%
\bibitem [{\citenamefont {Jain}(1992)}]{Jain1992}%
  \BibitemOpen
  \bibfield  {author} {\bibinfo {author} {\bibfnamefont {J.~K.}\ \bibnamefont
  {Jain}},\ }\bibfield  {title} {\emph {\bibinfo {title} {Microscopic theory of
  the fractional quantum hall effect}},\ }\href
  {https://doi.org/10.1080/00018739200101483} {\bibfield  {journal} {\bibinfo
  {journal} {Advances in Physics}\ }\textbf {\bibinfo {volume} {41}},\ \bibinfo
  {pages} {105} (\bibinfo {year} {1992})},\ \Eprint
  {https://arxiv.org/abs/https://doi.org/10.1080/00018739200101483}
  {https://doi.org/10.1080/00018739200101483} \BibitemShut {NoStop}%
\bibitem [{\citenamefont {Wen}(1992)}]{Wen1992}%
  \BibitemOpen
  \bibfield  {author} {\bibinfo {author} {\bibfnamefont {X.-G.}\ \bibnamefont
  {Wen}},\ }\bibfield  {title} {\emph {\bibinfo {title} {{Theory of the edge
  states in fractional quantum Hall effects}}},\ }\href
  {https://doi.org/10.1142/S0217979292000840} {\bibfield  {journal} {\bibinfo
  {journal} {Int. J. Mod. Phys. B}\ }\textbf {\bibinfo {volume} {6}},\ \bibinfo
  {pages} {1711} (\bibinfo {year} {1992})}\BibitemShut {NoStop}%
\end{thebibliography}%
\begin{widetext}
    \appendix
    \section{Lehmann representation for optical sum-rules}
\label{sec:Kubo}
In this appendix, we express the partial sum rule in Eq.~\ref{PSR} for the longitudinal conductivity at temperature $T \ll E_{\rm{gap}}$ and $\Lambda$ inside the bandgap to the remote bands. In the spectral representation,  the real part of the longitudinal conductivity can be written as,
\begin{equation}
    \begin{split}
       \tn{ Re} [\sigma_{\tn{xx}}^{\tn{eff}}(q_x\rightarrow 0, \omega)] = \frac{\pi e^2}{\omega} \sum_{m,n} \frac{e^{-\beta E_n} - e^{-\beta E_m}}{Z}\langle n| j_x^{\tn{eff}}(q_x\rightarrow 0) |m\rangle \langle m| j_x^{\tn{eff}}(-q_x\rightarrow 0) |n\rangle \delta(\omega - E_m + E_n),
    \end{split}
\end{equation}
where $Z$ is the partition function. If we perform an integral over $\omega$, at temperature $T \ll E_{\rm{gap}}$, for $|\omega| \lesssim E_{\rm{gap}}$, we only need to consider the many-body states $|n\rangle$ and $|m\rangle$ that belong to the low-energy projected Hilbert space $\mathbb{H}$ with eigen-energies $E_n,~E_m$, and therefore,
\begin{equation}
    \begin{split}
        \int_{0}^\Lambda d\omega~ \frac{\tn{Re} [\sigma^{\tn{eff}}_{\tn{xx}}(q_x\rightarrow 0, \omega)]}{\omega} &\approx \pi e^2 \sum_{m,n\in \mathbb{H}} \langle n| j_x^{\tn{eff}}(q_x\rightarrow 0) |m\rangle \langle m| j_x^{\tn{eff}}(-q_x\rightarrow 0) |n\rangle  \frac{e^{-\beta E_n} - e^{-\beta E_m}}{Z(E_m - E_n)^2} \theta(E_m-E_n)\\
        &= - \pi e^2 \sum_{m,n\in \mathbb{H}} \langle n| [\hat{X}, H] |m\rangle \langle m| [\hat{X}, H] |n\rangle \frac{e^{-\beta E_n} - e^{-\beta E_m}}{Z(E_m - E_n)^2}\theta(E_m-E_n)\\
        &= \pi e^2 \sum_{m,n\in \mathbb{H}} \langle n| \hat{X} |m\rangle \langle m| \hat{X} |n\rangle \frac{e^{-\beta E_n} - e^{-\beta E_m}}{Z} \theta(E_m-E_n). 
    \end{split}
\end{equation}
Importantly, since the many-body states are restricted only to $\mathbb{H}$, the matrix-elements above are associated with the projected position coordinate, $\hat{X}\rightarrow\hat{\overline{X}}$. 

\section{Details of Hartree-Fock calculations for MATBG}
\label{app:HFTBG}
For all calculations in the symmetry broken insulators of TBG, we start by determining the ground state by self-consistent Hartree-Fock using the optimal damping algorithm \cite{ODAalgo} to speed up convergence, and initialize the Hartree-Fock Projector $\mathcal{P}$ by filling up the states of a random quadratic Hamiltonian. This is done in parallel using 128 random seeds, and the minimum-energy self-consistent solution is selected to be the ground-state. For all the calculations we used the ``average" subtraction scheme \cite{parker2021fieldtuned,kwan_kekule_2021} with $\theta=1.08$ and $\kappa=0.75$. For the results in Fig.~\ref{fig:Obar}a, the calculations for the bound, $\overline{\mathbb{S}}/\Delta$, are taken from \cite{opticalsumrule2023} which where performed on a $12\times12$ system due to the complexity involved in calculating $\overline{\mathbb{S}}$ which involves various derivatives of projectors whose dimension scales with the plane-wave cutoff of the continuum model. For details on the calculation of $\overline{\mathbb{S}}$ see the supplementary information of \cite{opticalsumrule2023}. The calculations of the MBPQM and the structure factor in Fig.~\ref{fig:Obar} were performed on a $24\times24$ system. For $\varepsilon>0.1$, the Hartree-Fock ground state is found to be a spin-unpolarized IKS close to a momentum $\vec{q}_{IKS}=2\vec{M}/3$ (within the resolution of our grid ), with $\vec{M}$ being the M-point of the MBZ \cite{kwan_kekule_2021}. For $\varepsilon<0.1$, the solution is instead a spin-polarized KIVC state. To achieve lower energy solutions that improve the convergence of $\mathrm{tr}[G_{\mathrm{xx}}(\vec{k})]$, we ran the HF loop starting from the ground state of either a spin diagonal, or a spin-independent random Hamiltonian for $\varepsilon<0.1$, and $\varepsilon>0.1$, respectively.  Once the ground-state reduced density matrix is obtained, the projected spectral weight is computed using the projected structure factor (quantum weight), as detailed in Appendix \ref{sec:SF}. We explicitly avoid the region of the phase diagram near the first order transition between KIVC and IKS since the exact location of the transition is dependent on the subtraction scheme, and instead we focused primarily deep in the two phases. Near the transition point, the indirect gap on the KIVC side is also reduced drastically from its value at zero strain (with a strong tendency towards semi-metallic behavior).

\section{Projected quantum weight for interacting Slater determinant states}
\label{sec:SF}
Since the projected quantum weight is related to the long wavelength behavior of the projected structure factor, we can evaluate it directly by simply expanding $\overline{\mathcal{S}}(\q)$ around small $\q$. For a Slater determinant state, such as the Hartree-Fock mean-field solution for TBG with projected Coulomb interactions at integer fillings, the projected structure factor can then be easily obtained via Wick's theorem:
\begin{subequations}
\beq
\overline{\mathcal{S}}(\q) &=& \sum_{\k,\k'} \langle c_{\k,\alpha}^\dag c_{\k - \q,\beta} \lambda_{\alpha\beta}(\k,\q) c_{\k',\alpha'}^\dag c_{\k'+\q, \beta'} \lambda_{\alpha' \beta'}(\k', -\q)\rangle - \langle c_{\k,\alpha}^\dag c_{\k - \q,\beta} \lambda_{\alpha\beta}(\k,\q) \rangle\langle c_{\k',\alpha'}^\dag c_{\k'+\q, \beta'} \lambda_{\alpha' \beta'}(\k', -\q) \rangle\nn\\
&=& \sum_{\k,\k'} \langle c_{\k,\alpha}^\dag c_{\k'+\q, \beta'}  \rangle \langle c_{\k - \q,\beta} c_{\k',\alpha'}^\dag \rangle \lambda_{\alpha\beta}(\k,\q) \lambda_{\alpha' \beta'}(\k', -\q)\\
&=& \sum_{\k,\k'} \tn{Tr}\left\{ \hat{\lambda}(\k,\q)\left\{\mathbf{1} \delta_{\k',\k-\q}- [\mathcal{P} (\k',\k-\q)]^T \right\} \hat{\lambda}(\k',-\q) [\mathcal{P}(\k,\k'+\q)]^T \right\},
\eeq
\end{subequations}
where the trace is taken over the band, orbital and spin indices of the projected bands, respectively, and $\mathcal{P}(\k,\k')$ is the one-particle reduced density matrix, whose matrix elements are given by
\beq
\left[\mathcal{P}(\k,\k')\right]_{\alpha\beta} = \langle c_{\k,\alpha}^\dag c_{\k',\beta} \rangle.
\eeq

Note that for mean-field states that are generalized ferromagnetic (FM) states in the orbital basis at each $\k$ point, the structure factor can be considerably simplified; this includes e.g. the IVC ordering in TBG, as well as the IKS states. For these generalized FM states, the wavefunction is an eigenstate of the projected particle number operator at ``each" $\k$ point, namely,
\beq
c_{\k, \alpha}^\dag c_{\k,\alpha} |\Psi_{FM} \rangle = \nu |\Psi_{FM} \rangle, 
\label{eq:g-FM}
\eeq
where $\nu$ is the filling factor. For IKS, we simply replace $\k$ by $\tilde{\k} = \k - \vec{Q}\tau_z/2$, where $\vec{Q}$ is the ordering wave vector for IKS and $\tau_z$ denotes the Pauli matrix in valley space.

From Eq.\ref{eq:g-FM}, the following holds, 
\beq
\langle \Psi_{FM} | c_{\k, \alpha}^\dag c_{\k,\alpha} c_{\k',\beta}^\dag c_{\k'',\gamma}| \Psi_{FM} \rangle - \langle \Psi_{FM} | c_{\k, \alpha}^\dag c_{\k,\alpha}| \Psi_{FM} \rangle \langle \Psi_{FM} | c_{\k',\beta}^\dag c_{\k'',\gamma}| \Psi_{FM} \rangle = 0\\
\Leftrightarrow \langle \Psi_{FM} |c_{\k, \alpha}^\dag c_{\k'',\gamma}| \Psi_{FM} \rangle \langle \Psi_{FM} | c_{\k,\alpha} c_{\k',\beta}^\dag | \Psi_{FM} \rangle = 0,
\label{eq:slater_projector}
\eeq
where $\k', \beta$ and  $\k'', \gamma$ are arbitrary.

Also, for the generalized FM states (excluding IKS),
\beq
\mathcal{P}(\k,\k') \equiv \delta_{\k,\k'} \hat{P}_{\k},
\eeq
while for IKS, we have
\beq
\mathcal{P}(\k,\k') \equiv \delta_{\k,\tilde{\k} + \vec{Q}\tau_z/2} \delta_{\k',\tilde{\k} + \vec{Q}\tau_z/2} \hat{P}_{\tilde{\k}}.
\eeq

Therefore, from Eq.\ref{eq:slater_projector} we have $(\mathbf{1}- \hat{P}_\k) \hat{P}_\k = 0$ and $(\mathbf{1}- \hat{P}_{\tilde{\k}}) \hat{P}_{\tilde{\k}} = 0$.

From the above equations, it is readily seen that for the generalized FM states in TBG, the terms in $\overline{\mathcal{S}}(\q)$ vanish if $\alpha = \beta$ or $\alpha' = \beta'$ and the corresponding $\q = 0$ for the fermion operators.

Therefore, when we expand around $\q$, the leading order  terms in $\q$ vanish. For the $q^2-$ contribution, the terms involving double derivatives of the form-factor vanishes. To conclude, for generalized FM states,
\beq
\overline{\mathcal{S}}(\q) &=& q_\mu q_\nu \sum_{\k} \tn{Tr}\left\{ \left[\overrightarrow{\partial_\mu} - i\hat{\mathcal{A}}_{\k,\mu} \right]\left\{\mathbf{1} - [\hat{P}_\k]^T \right\} \left[\overleftarrow{\partial_\nu}+ i\hat{\mathcal{A}}_{\k',\nu} \right][\hat{P}_\k]^T \right\},
\label{eq: S_slater}
\eeq
where the arrows over the derivatives denote the direction of taking derivative, i.e., they all act on the  $\left\{\mathbf{1} - [\hat{P}_\k]^T \right\} $ term. For the IKS states, note that the form factor is always diagonal in valley, we arrive at the same expression as in Eq.\ref{eq: S_slater} but with $\k$ replaced by $\tilde{\k}$. 

Alternatively, we can calculate the same $\overline{\mathcal{S}}(\q)$ in a new basis where the HF Hamiltonian is diagonalized. $H_{MF} = \sum_{\k,m} \tilde{\epsilon}_{\k,m} \tilde{c}_{m,\k}^\dag \tilde{c}_{m,\k}$, where $m$ labels the HF bands. The HF approximation is done by projecting the Hamiltonian to the low-energy, with the same energy cut-off that $\overline{\mathcal{S}}(\q)$ is evaluated within.

In the new basis,
\begin{subequations}
\beq
\overline{\mathcal{S}}(\q)
&=& \sum_{\k,\k'} \langle \tilde{c}_{m,\k}^\dag \tilde{c}_{ n', \k'+\q}  \rangle \langle \tilde{c}_{n, \k - \q} \tilde{c}_{m', \k'}^\dag \rangle \tilde{\lambda}_{mn}(\k,\q) \tilde{\lambda}_{m' n'}(\k', -\q)\\
&=& \sum_{\k,m\in \text{occ}, n\notin\text{occ}} \tilde{\lambda}_{mn}(\k,\q) \tilde{\lambda}_{n m}(\k-\q, -\q)\\
&=& q_\mu q_\nu \sum_{\k, m\in \text{occ}} \langle \partial_\mu \tilde{u}_{m,\k} |(\mathcal{P}_{\text{IR}}(\k)- \mathcal{P}_{\text{occ}}(\k))|\partial_\nu \tilde{u}_{m,\k} \rangle + (\mu \leftrightarrow \nu),
\eeq
\end{subequations}
where $\mathcal{P}_{\text{IR}}(\k)) = \sum_{m \in \text{IR}} |\tilde{u}_{m,\k} \rangle \langle \tilde{u}_{m,\k}|$ sums over all the low-energy bands that the HF Hamiltonian keeps, and $\mathcal{P}_{\text{occ}}(\k)) = \sum_{m \in \text{occ}} |\tilde{u}_{m,\k} \rangle \langle \tilde{u}_{m,\k}| $ is the projection to the occupied HF bands. For IKS states, the above expression holds by replacing $\vec{k}$ with $\tilde{\vec{k}}$. Note that the above formula can also be generalized to Slater determinant of CDW insulators, by considering the Bloch wavefunction in the reduced BZ of the enlarged unitcell.

Therefore, we can define a ``projected" quantum geometric tensor,
\beq
G_{\mu\nu}^{nm}\left(\boldsymbol{k}\right)= \langle \partial_\mu \tilde{u}_{m,\k} |(\mathcal{P}_{\text{IR}}(\k)- \mathcal{P}_{\text{occ}}(\k))|\partial_\nu \tilde{u}_{n,\k} \rangle, 
\label{eqn:Gproj}
\eeq
which takes into account only the ``low-energy" part contribution to the quantum geometry and is related to the fluctuation of the ``projected" position operator. 

To evaluate the projected quantum geometric tensor, we perform a finite difference evaluation of Eq.~\ref{eqn:Gproj}. First we evaluate the metric by setting $m=n$ and $\mu=\nu=\mathrm{x}$. In this case, we approximate the derivative of the states by $|\partial_{k_{\nu}}u_{n\boldsymbol{k}} \rangle \approx \left[ |u_{n\boldsymbol{k}+\delta k_{\mathrm{x}} }\rangle - |u_{n\boldsymbol{k}} \rangle\right]/\delta k_{\mathrm{x}}$, which results in the following expression for the metric in terms of the finite differences of the form factor \cite{kwan2024texturedexcitoninsulators}: 

\begin{equation}
    G^{nn}_{\mathrm{xx}}\left(\boldsymbol{k}\right)=\frac{1}{\delta k_{\mathrm{x}}^{2}}\left(\sum_{\ell \in \mathrm{IR} } \left|\langle u_{n\boldsymbol{k}+\delta k_{\mathrm{x}}}|u_{\ell\boldsymbol{k}}\rangle\right|^{2}-\sum_{\ell \in \mathrm{occ} } \left|\langle u_{n\boldsymbol{k}+\delta k_{\mathrm{x}}}|u_{\ell\boldsymbol{k}}\rangle\right|^{2}\right).
\end{equation}

When evaluating this expression in the Hartree-Fock ground states of TBG, we calculate  $G^{nn}_{\mathrm{xx}}$ by preforming a unitary transformation on the form factors of TBG determined by the eigenstates of the self-consistent Hartree-Fock Hamiltonian. Since the intervalley form factor is parametrically smaller than the intravalley form facter at small angles, the rotation neglects valley mixing.Finally, when comparing to the projected quantum weight, we compute the trace over occupied orbitals in the symmetry broken phase, given by:
\begin{equation}
    \mathrm{Tr}[G_{\mathrm{xx}}] = \frac{1}{V}\sum_{\boldsymbol{k}}\sum_{\ell \in \mathrm{occ} }G_{\mathrm{xx}}^{\ell \ell}\left(\boldsymbol{k}\right).
\end{equation}
\section{Emergent guiding center coordinate in LLL under periodic magnetic field}
\label{app:perB}
Recall that for a spinless LLL under a uniform magnetic field, we can choose a set of orthonormal basis $| n \rangle$ on the plane, 
\beq
\psi_n(\vec{r}) = \langle \vec{r} | n \rangle = R_n^{-1} z^n e^{-\frac{|z|^2}{4}},
\eeq
where $R_n = \sqrt{\pi 2^{n+1} n!} $ is the normalization factor.

Therefore, in the $| n \rangle$ basis, the matrix elements of the projected position operator can be written as,
\beq
\langle l | \hat{\overline{X}} |n \rangle = \frac{1}{\sqrt{2}}\left[\delta_{l,n+1} \sqrt{n+1} + \delta_{l,n-1} \sqrt{n} \right]\\
\langle l | \hat{\overline{Y}} |n \rangle = \frac{1}{\sqrt{2} i }\left[\delta_{l,n+1} \sqrt{n+1} - \delta_{l,n-1} \sqrt{n} \right].
\eeq
Hence, we can write $\hat{\overline{X}}$ acting on the wavefunctions as,
\beq
\hat{\overline{X}} \psi_n(\vec{r}) = \frac{1}{2}\left[z + 2 \partial_{z} + \frac{\bar{z}}{2} \right] \varphi_n(\vec{r}),
\eeq
and similarly for $\hat{\overline{Y}}$.

Equivalently, we can write $\hat{\overline{X}}$ in the second quantized language. We define the creation operator $c_n^\dag$ as creating a particle in state $|n\rangle$,
\beq
\hat{\overline{X}} = \frac{1}{\sqrt{2}}\sum_n \left(\sqrt{n+1} c_{n+1}^\dag c_n + \sqrt{n} c_{n-1}^\dag c_n \right).
\eeq

For a completely filled LL denoted $|\Psi_{\nu = 1}\rangle$, the projected position operator annihilates the state in the thermodynamic limit due to the full occupation of the single-particle orbitals of the LL. For a finite system with open boundary conditions, we need to truncate the Hilbert space to $|n_{\text{max}}\rangle = |N_{\phi} -1 \rangle$, where $2\pi N_{\phi}$ is the total magnetic flux, such that there is one orbital per magnetic flux; the thermodynamic limit is approached for $N\rightarrow \infty$. With the truncation for a finite particle number, we can justify that $\hat{\overline{X}} |\Psi_{\nu = 1}\rangle = 0$, which is also the Laughlin state with $m=1$. In fact, for any Laughlin wavefunction, $\hat{\overline{X}} |\Psi_{L}\rangle = 0 $ holds. We will come back to this in more detail using parton construction when we discuss the generalized Laughlin states under a periodic magnetic field.

Let us now consider the LLL under a periodic magnetic field. We can still define a set of complete and independent eigenvectors $|n_s \rangle$,
\beq
 \psi_{n_s}(\vec{r}) = \langle \vec{r} | n_s \rangle = R_n^{-1} z^n e^{-\frac{|z|^2}{4} - \tilde{\varphi}(\vec{r})},
\eeq
where $\tilde{\varphi}(\vec{r}) = - \frac{\phi_1}{|G|^2} \sum_j e^{i \vec{k}_j \cdot \vec{r}}$ and $|G|$ is the norm of the reciprocal lattice vector $\vec{k}_j$'s. For a non-vanishing $\phi_1$, the $|n_s\rangle$ are not normalized and different $|n_s\rangle$'s are non-orthogonal.

Since $|n_s\rangle$'s are independent and form a complete basis, we can still define a projector $\mathcal{P}$,
\beq
\mathcal{P} = \sum_{l,n} |l_s\rangle (\hat{O}^{-1})_{l n} \langle n_s|,
\eeq
where $\hat{O}_{ln} = \langle l_s| n_s\rangle $ is the overlapping matrix. Since $\hat{O}$ is Hermitian by definition, as long as $\hat{O}$ is positive definite, we can always diagonalize $\hat{O}$ as,
\beq
\left(\hat{O}^{-1}\right)^T = \hat{A}^\dag \hat{A}.
\eeq

We can find $\hat{A}$ perturbatively in $\phi_1$ and
\beq
|\tilde{n} \rangle  = \sum_l \hat{A}_{n l}|l_s\rangle = |n\rangle + \frac{\phi_1}{|G|^2} \sum_j (1- |l\rangle\langle l|) e^{i \vec{k}_j \cdot \vec{r}}|n\rangle + O(\phi_1^2),
\eeq
defines an orthonormal basis and the projector can be written as $\mathcal{P} = \sum_n |\tilde{n}\rangle \langle\tilde{n}|$.

Now let us consider the projected density operators: 
\beq
\hat{\overline{\rho}}_{\vec{q}} = \mathcal{P} e^{i \vec{q} \cdot \vec{r}} \mathcal{P} &=& \sum_{l,n}|\tilde{l} \rangle \langle \tilde{l} |  e^{i \vec{q} \cdot \vec{r}} |\tilde{n}\rangle \langle\tilde{n}|\\
&=& \sum_{l,n}|\tilde{l} \rangle \left(\hat{\overline{\rho}}_{\vec{q}}^0 \right)_{ln} \langle\tilde{n}|+ \frac{\phi_1}{|G|^2} \sum_{l,n,j}|\tilde{l} \rangle \left[ 2 \hat{\overline{\rho}}_{\vec{q}+\vec{k}_j}^0 - \hat{\overline{\rho}}_{\vec{q}}^0 \hat{\overline{\rho}}_{\vec{k}_j}^0 - \hat{\overline{\rho}}_{\vec{k}_j}^0 \hat{\overline{\rho}}_{\vec{q}}^0\right]_{ln}\langle\tilde{n}| + O(\phi_1^2)\\
&\equiv& \hat{\overline{\rho}}_{\vec{q}}^0 + \frac{\phi_1}{|G|^2} \sum_j \hat{\overline{\rho}}_{\vec{q}+\vec{k}_j}^0 \left[ 2  - e^{\frac{\bar{q} k_j}{2}}- e^{\frac{q \bar{k}_j}{2}} \right] + O(\phi_1^2),
\eeq
where $\left(\hat{\overline{\rho}}_{\vec{q}}^0 \right)_{ln} \equiv \langle l | e^{i \vec{q} \cdot \vec{r}}  |n\rangle$ is the matrix element of the density operator of the LLL under uniform field, which satisfies the GMP algebra, $\left(\hat{\overline{\rho}}_{\vec{q}_1}^0 \hat{\overline{\rho}}_{\vec{q}_2}^0\right)_{ln} = e^{\frac{\bar{q}_1 q_2}{2}} \left(\hat{\overline{\rho}}_{\vec{q}_1 + \vec{q}_2}^0\right)_{ln}$. Therefore $\sum_{l,n}|\tilde{l} \rangle \left(\hat{\overline{\rho}}_{\vec{q}}^0 \right)_{ln} \langle\tilde{n}| \equiv \hat{\overline{\rho}}_{\vec{q}}^0$ are operators acting on the states of the LLL under {\it periodic} magnetic field that also satisfy GMP algebra. Hence, in the $\vec{q} \rightarrow 0$ limit, we call the linear in $\vec{q}$ term derived from $\hat{\overline{\rho}}_{\vec{q}}^0$ the ``emergent" guiding center coordinates $\hat{\overline{X}}_{\mu,0}$ since they satisfy the same algebra as the guiding center coordinates of the LLL under uniform field (for an equivalent definition written in momentum space basis, see Ref.~\cite{wang2023origin}). In the $|\tilde{n}\rangle$ basis, we have,
\beq
\hat{\overline{X}}_{0} =  \sum_{l,n}|\tilde{l} \rangle \langle l | \hat{\overline{X}} | n \rangle \langle\tilde{n}| = \frac{1}{\sqrt{2}} \sum_n \left[\sqrt{n+1} | \widetilde{n+1} \rangle \langle \widetilde{n}| + \sqrt{n} | \widetilde{n-1} \rangle \langle \widetilde{n}|\right].
\eeq
Therefore, if we consider a many-body state at $\nu =1$ of the LLL under periodic magnetic field,
\beq
|\Psi_{\nu=1} \rangle = \tilde{R}^{-1} \prod_{n = 0}^{n_{\text{max}}}\tilde{c}_{n}^\dag |0\rangle,
\eeq
where $\tilde{c}_{n}^\dag$ creates particle with orbital $|\tilde{n} \rangle$ and $\tilde{R}$ is the normalization factor. It is readily seen that $\hat{\overline{X}}_{0} |\Psi_{\nu=1} \rangle = 0$. Now let us expand $\tilde{c}_{n}$ in the $|n_s\rangle$ basis. In this case, due to the uniqueness of the $\nu = 1$ state, we also have,
\beq
|\Psi_{\nu=1} \rangle = R^{-1} \prod_{n = 0}^{n_{\text{max}}} c_{n_s}^\dag |0\rangle,
\eeq
where $c_{n_s}^\dag$ creates particle with orbital $|n_s \rangle$ and $R$ is another normalization factor. Hence in real space, we can write the wavefunction as,
\beq
\Psi_{\nu=1} (\vec{r}_1,...,\vec{r}_N) = R^{-1} \prod_{i<j} (z_i - z_j) e^{-\sum_i \frac{|z_i|^2}{4} - \tilde{\varphi}(\vec{r}_i)}.
\eeq

Now let us discuss the generalized Laughlin states,
\beq
\Psi_{gL} (\vec{r}_1,...,\vec{r}_N) = R_g^{-1} \prod_{i<j} (z_i - z_j)^m e^{-\sum_i \frac{|z_i|^2}{4} - \tilde{\varphi}(\vec{r}_i)}.
\eeq
Again, one can view $\Psi_{gLL} $ as $m$ copies of $\Psi_{\nu=1}$, with particle number $N = \frac{N_\phi}{m}$ and with magnetic field $\frac{B(\vec{r})}{m}$, namely,
\beq
\Psi_{gL} (\vec{r}_1,...,\vec{r}_N) = R_g^{-1} \int\prod_{i = 1}^N \prod_{l = 1}^m d\vec{r}_{i,\alpha_l} \delta(\vec{r}_{i} - \vec{r}_{i,\alpha_l}) \prod_{i<j} (z_{i,\alpha_l} - z_{j,\alpha_l}) e^{-\sum_i \frac{|z_{i,\alpha_l}|^2}{4 m } - \frac{\tilde{\varphi}(\vec{r}_{i,\alpha_l})}{m}},
\eeq
where $\alpha_j$ labels different copies.

This treatment of the Laughlin state is equivalent to a parton construction, where the physical fermion is fractionalized into $m$ species of fermions \cite{Jain1989, Jain1992, Wen1992}. To be explicit, we can write the generalized Laughlin wave function in terms of the parton fields as,
\beq
| \Psi_{gLL}\rangle \propto \mathcal{P}_g \prod_{l = 1}^m \prod_{n = 0}^{n_{\text{max}}} \tilde{c}_{\alpha_l,n}^\dag |0\rangle,
\eeq
where the $\propto$ means the LHS and RHS can differ by a normalization constant. We enlarge the Hilbert space to $m$ copies of LLL, labelled by $\alpha_l$, each of which has $n_{\text{max}} = \frac{N_\phi}{m}-1$, and $\mathcal{P}_g$ introduces the Guzwiller projection to the physical Hilbert space,
\beq
\langle \vec{X}|\mathcal{P}_g |\vec{X}_{\alpha_1},... \vec{X}_{\alpha_m} \rangle = \prod_{l = 1}^m \delta(\vec{X} - \vec{X}_{\alpha_l}),
\eeq
where $\vec{X}$ is a shorthand notation for the coordinates of $N$ particles, $\vec{x}_1,...\vec{x}_N$.

In order to evaluate the action of $\hat{\overline{X}}_{0}$ on the generalized Laughlin state, we introduce an auxillary operator $\hat{\widetilde{X}}_{0}$ acting on the enlarged Hilbert space such that
\beq
\hat{\overline{X}}_{0} = \mathcal{P}_g \hat{\widetilde{X}}_{0}.
\eeq
Note that the choice of $\hat{\widetilde{X}}_{0}$ is not unique due to the gauge redundancy in the parton construction. A convenient choice for our purpose is to let,
\beq
\hat{\widetilde{X}}_{0} = \frac{1}{\sqrt{2}} \sum_n \left[\sqrt{n+1} \tilde{c}_{\alpha_1, n+1}^\dag \tilde{c}_{\alpha_1, n } + \sqrt{n} \tilde{c}_{\alpha_1, n-1}^\dag \tilde{c}_{\alpha_1, n } \right]\otimes \prod_{l \neq 1} \mathbb{I}_{\alpha_l},
\eeq
where $\hat{\widetilde{X}}_{0}$ only acts non-trivially on $\alpha_1$ but acts trivially on all the other species.

Since $\hat{\widetilde{X}}_{0} \prod_{n = 0}^{n_{\text{max}}} \tilde{c}_{\alpha_1,n}^\dag |0\rangle = 0 $ due to the full occupation of the orbitals of the $\alpha_1$ species. We conclude that $\hat{\overline{X}}_{0}  |\Psi_{gLL}\rangle = 0$. Therefore, we can view $\hat{\overline{X}}_{0}$ (and $\hat{\overline{Y}}_{0}$) as ``emergent" guiding center coordinates and the generalized Laughlin state on the plane being their eigenstate with eigenvalue $0$. 
\end{widetext}

\end{document}